\documentclass[a4paper,11pt]{article}
\pdfoutput=1 

\usepackage{jcappub} 
\usepackage{subcaption}
\usepackage{ctable}
\usepackage{array}
\usepackage{multirow}
\usepackage{listings}

\definecolor{mygreen}{rgb}{0,0.6,0}
\definecolor{mygray}{rgb}{0.5,0.5,0.5}
\definecolor{mymauve}{rgb}{0.58,0,0.82}

\definecolor{purple}{rgb}{0.78,0.18,0.77}

\title{Efficient exploration of cosmology dependence in the EFT of LSS}

\author[1]{Matteo~Cataneo,}
\author[2,3]{Simon~Foreman,}
\author[2,3]{Leonardo~Senatore}

\affiliation[1]{Dark Cosmology Centre, Niels Bohr Institute, University of Copenhagen, Juliane Maries Vej 30, 2100 Copenhagen, Denmark}
\affiliation[2]{Stanford Institute for Theoretical Physics, Stanford University, Stanford, CA 94306}
\affiliation[3]{Kavli Institute for Particle Astrophysics and Cosmology, SLAC and Stanford University, Menlo Park, CA 94025}

\emailAdd{matteoc@dark-cosmology.dk}

\newcommand{\epstarget}{\epsilon_\text{target}}
\newcommand{\epsref}{\epsilon_\text{ref}}
\newcommand{\epsdelta}{\epsilon_\Delta}

\newcommand{\invMpc}{\,h\, {\rm Mpc}^{-1}\,}

\def\d{{\partial}}
\newcommand{\lp}{\left(}
\newcommand{\rp}{\right)}

\def\co{c_{s  (1)}^2}
\def\ct{c_{s  (2)}^2}
\def\knl{{k_{\rm NL}}}

\newcommand{\vk}{\boldsymbol{k}}
\newcommand{\vq}{\boldsymbol{q}}

\def\poneloop{P_\text{1-loop}}
\def\ptwoloop{P_\text{2-loop}}
\def\ptwoloopfull{P_\text{2-loop}^\text{(full)}}

\def\poneloopcs{P_\text{1-loop}^{(c_{\rm s})}}

\newcommand{\codename}{\texttt{CosmoEFT}}
\newcommand{\website}{\url{http://web.stanford.edu/~senatore/}}

\definecolor{mygreen}{rgb}{0,0.6,0}

\abstract{The most effective use of data from current and upcoming large scale structure~(LSS) and CMB observations requires the ability to predict the clustering of LSS with very high precision. The  Effective Field Theory of Large Scale Structure (EFTofLSS) provides an instrument for performing analytical computations of LSS observables with the required precision in the mildly nonlinear regime. In this paper, we develop efficient implementations of these computations that allow for an exploration of their dependence on cosmological parameters. They are based on two ideas. First, once an observable has been computed with high precision for a reference cosmology, for a new cosmology the same can be easily obtained with comparable precision just by adding the difference in that observable, evaluated with much less precision.  Second, most cosmologies of interest are sufficiently close to the Planck best-fit cosmology that observables can be obtained from a Taylor expansion around the reference cosmology.  These ideas are implemented for the matter power spectrum at two loops and are released as public codes. When applied to cosmologies that are within 3$\sigma$ of the Planck best-fit model, the first method evaluates the power spectrum in a few minutes on a laptop, with results that have 1\% or better precision, while with the Taylor expansion the same quantity is instantly generated with similar precision. The ideas and codes we present may easily be extended for other applications or higher-precision results. 
}

\begin{document}
\maketitle
\flushbottom


\section{Introduction} \label{sec:intro}

Ongoing and future sky surveys, such as the extended Baryon Oscillation Spectroscopic Survey (eBOSS)~\cite{2015arXiv150804473D}, the Large Synoptic Survey Telescope (LSST)~\cite{2009arXiv0912.0201L}, the Dark-Energy Spectroscopic Instrument (DESI)~\cite{2013arXiv1308.0847L} and Euclid~\cite{2010arXiv1001.0061R}, as well as current and next generation CMB experiments, such as the South Pole Telescope (SPT)~\cite{Ruhl:2004kv} and the Atacama Cosmology Telescope~(ACT)~\cite{Thornton:2016wjq}, will measure the statistical quantities of the  cosmological large-scale structure with percent/sub-percent precision~\cite{2013arXiv1312.5490P}. One way to compare our predictions to this wealth of data is to evolve structure formation down to highly non-linear scales with large-volume, high-resolution simulations. However, this approach is computationally expensive requiring millions of CPU hours, which makes impractical the investigation of cosmological parameter dependences of survey observables and their covariances. 

The reliance on numerical simulation has been largely motivated by the lack of a satisfactory analytic approach. Until recently, analytic techniques were not capable of going beyond tree-level computations in a well-defined fashion, which limited their applicability to very low wavenumbers or motivated the introduction of additional assumptions or approximations with errors that are impossible to estimate or correct within these approaches to perturbation theory. This situation has radically changed in the last few years, with the development of the so-called Effective Field Theory of Cosmological Large Scale Structures (EFTofLSS)~\cite{Baumann:2010tm,Carrasco:2012cv,Porto:2013qua,Senatore:2014via}, which has successfully shown its consistency and its capability to predict~\footnote{In this context, by ``predict" we mean to say that the EFTofLSS predicts the functional form of corrections to linear-theory calculations in the mildly nonlinear regime. Some of these corrections come with free coefficients whose values are not predicted within the EFTofLSS, but once these coefficients have been fixed (by matching to measurements of a certain observable, either from real data or simulations), the theory can indeed be used to calculate other observables or different configurations of the same obeservable at the same perturbative order with no residual freedom. It is in this sense (the standard sense of effective field theories) that the theory is predictive.} correlation functions of large scale structure quantities within the mildly non-linear regime~\cite{Baumann:2010tm,Carrasco:2012cv,Carrasco:2013sva,Carrasco:2013mua,Pajer:2013jj,Carroll:2013oxa,Porto:2013qua,Mercolli:2013bsa,Senatore:2014via,Angulo:2014tfa,Baldauf:2014qfa,Senatore:2014eva,Senatore:2014vja,Lewandowski:2014rca,Mirbabayi:2014zca,Foreman:2015uva,Angulo:2015eqa,McQuinn:2015tva,Assassi:2015jqa,Baldauf:2015tla,Baldauf:2015xfa,Foreman:2015lca,Baldauf:2015aha,Baldauf:2015zga,Bertolini:2015fya,Bertolini:2016bmt}. The EFTofLSS has been applied to the description of the dark matter two-point function~\cite{Carrasco:2012cv,Carrasco:2013mua,Senatore:2014via,Foreman:2015lca,Baldauf:2015aha}, three-point function~\cite{Angulo:2014tfa,Baldauf:2014qfa}, as well as four-point function (which includes the covariance of the power spectrum)~\cite{Bertolini:2015fya,Bertolini:2016bmt}.
The dark matter momentum power spectrum~\cite{Senatore:2014via,Baldauf:2015aha}, displacement field~\cite{Baldauf:2014qfa}, and vorticity slope~\cite{Carrasco:2013mua,Hahn:2014lca} have also been derived within this framework. Notably, despite the complexity of the physical processes involved on small scales, also the effects of baryons on the power spectrum in the perturbative regime can be naturally incorporated~\cite{Lewandowski:2014rca}. The extension of the EFTofLSS to biased tracers has been carried out in~\cite{Senatore:2014vja}, and used for the power spectrum and bispectrum in~\cite{Angulo:2015eqa}. Further applications include redshift space distortions~\cite{Senatore:2014vja,Lewandowski:2015ziq}, and the impact of primordial non-Gaussianity on large scale structure observables~\cite{Angulo:2015eqa,Assassi:2015jqa,Assassi:2015fma,Lewandowski:2015ziq}.

Several advantages result from this working analytic approach, as for example a simple and intuitive (yet rigorous) understanding of the physics governing the clustering of large scale structures, or the possibility of estimating the errors in our computations. In this work we introduce and implement a suite of computationally efficient algorithms to allow the exploration of a considerable volume of parameter space within the EFTofLSS.

Our idea is rather straightforward, and it is based on combining the analytical insight gained from the EFTofLSS with the fact that, after the Planck satellite, our knowledge of the cosmological parameters of our universe is quite accurate, which implies that only small differences among the observables are currently relevant.
Suppose we wish to compute the various observables as we change the cosmological parameters with a certain precision. Also, suppose that we have computed these observables for a certain reference cosmology, for example the Planck best fit one, with the desired precision. Now, in the EFTofLSS, observables are computed as multi-dimensional convolutions of  linear power spectra weighted by some kernels. Therefore, the difference in the computation between two different cosmologies lies just in the difference of the linear power spectra. Therefore, instead of performing these integrations directly for each cosmology, we can simply integrate the difference of the integrands (we will actually integrate the difference of rescaled integrands, which fits our purpose even better). In this way, the integral that we need to compute is very small, and therefore, in order to obtain a given precision on the final result, it can be computed with a smaller relative precision. (A similar idea has been applied to perturbation theory calculations in a different context in~\cite{Taruya:2012ut}.)

We implement this construction for the two-loop matter power spectrum in Sec.~\ref{sec:Difference}. With the integration techniques we use in this paper, we find that obtaining this quantity for all cosmologies within the $3\sigma$ contour around the Planck best fit cosmology with better than 1\% accuracy takes a few minutes on a laptop. This makes the cost of computing several cosmologies lighter by about a couple of orders of magnitude~\footnote{Using the same idea as presented here, the running time, and the gain, might be even improved by using more sophisticated integration techniques.}. We make two codes that perform this computation publicly available. The first, {\codename}, evaluates the required integrals using the approach described above. The second, \texttt{ResumEFT}, implements an ``IR-resummation" procedure that accounts for the leading effects of the long-wavelength displacements on the power spectrum at the wavenumbers of interest (see Sec.~\ref{sec:resumeft} for a brief description, or~\cite{Senatore:2014via,Angulo:2014tfa} for details).

Again, given that we are interested in relatively nearby cosmologies, it appears to be unnecessary to compute correlation functions for each cosmology directly. Instead, it should be enough to compute the Taylor expansion of the relevant quantities, such as the convolution integrals and the counterterms of the EFTofLSS, around the Planck cosmology. This can be done by evaluation of a relatively small number of cosmologies to compute the derivatives with respect the cosmological parameters (which is made easier by the use of {\codename}), so that the correlation functions for all the additional cosmologies can be simply read off the Taylor expansion, with negligible cost in time. We implement this procedure for the two-loop dark matter power spectrum in Sec.~\ref{sec:Taylor}, and we make publicly available a Mathematica notebook, \texttt{TaylorEFT}, that allows to read the dark matter power spectrum from the Taylor expansion. 

In Sec.~\ref{sec:Results} we perform additional checks on both \texttt{TaylorEFT} and the {\codename} procedures, finding that in general the power spectrum for all cosmologies within the $3\sigma$ contour of the Planck best fit cosmology can be reproduced within 1\% accuracy. Better precision can be achieved by adjusting the precision requirements and the order of the Taylor expansion. We also perform a first, preliminary, exploration of the cosmology dependence of the parameters present in the EFTofLSS which encode the effect of short distance physics at long distances, such as the well-known speed of sound.

The public codes we release with this work were designed to be useful within the context of a flat $\Lambda$CDM cosmology with parameters close to those of the latest Planck best-fit values, but we emphasize that the methods on which these codes are based can also be fruitfully applied in other contexts, for example in extensions to the base cosmological model. The future implementation of these methods in those other contexts has the potential to greatly reduce the need for large suites of expensive simulations.  In particular, the EFT parameters can in principle be extracted from measurements of short-distance correlations made on small, specially-designed simulations~\cite{Carrasco:2012cv}, eliminating the need to measure the entire functional forms of various statistical observables from large simulation boxes. This may allow for a more efficient use of computing time in running only the types of simulations necessary to extract these parameters. In fact, it may also be possible to determine these parameters entirely from observations, such that simulations would only be needed (in this setting) for checks of data analysis methods and pipelines, rather than as theoretical inputs.


\section{{\codename}:\\ Efficient exploration of cosmology-dependence in the EFTofLSS} \label{sec:Difference}

The current state-of-the-art (two-loop) calculation of the matter power spectrum in the EFTofLSS involves five-dimensional loop integrals. Several ways to make the integration more efficient have been proposed~\cite{Taruya:2012ut,Sherwin:2012nh,Blas:2013aba,Bertolini:2015fya,Schmittfull:2016jsw,McEwen:2016fjn}.
 Given a fixed computational cost for computing the predictions of the EFTofLSS for one cosmology, there remains the question if there is an efficient way to use this result in order not to pay the same price to obtain the result for any cosmology of interest. In Sec.~\ref{sec:DiffExplain} we present a re-organization of the perturbation theory integrals that greatly reduces the cost of exploring the cosmological parameter space once the result for a given cosmology is known. We give the details of its implementation in Sec.~\ref{sec:DiffImp}, and evaluate its performance in Sec.~\ref{sec:DiffChecks}. This new method is applied in the publicly available code {\codename}~\footnote{\website}, released with this work.

\subsection{Integrating differences between cosmologies} \label{sec:DiffExplain}

After the release of Planck's constraints on cosmological parameters~\cite{Ade:2015xua}, the cosmologies of most interest for future studies will likely be those that have only mild departures from the Planck best-fit model. This motivates the following decomposition of the loop integrals we wish to calculate:
\begin{equation}\label{eq:target_cosmo}
P_{\alpha}^{\text{target}}(k) = P_{\alpha}^{\text{ref}}(k) + \Delta P_{\alpha}(k)\ ,
\end{equation}
where $\alpha$ denotes a particular loop integral, and ``target'' and ``ref'' refer to the desired cosmology and a Planck-\text{like} fiducial (``reference'') cosmology respectively. The difference $\Delta P_{\alpha}(k)$ can trivially be written as the integral of the difference between the reference and target integrands:
\begin{equation}\label{eq:diff_integration}
\Delta P_{\alpha}(k) 
\equiv 
\int \text{d}^3 \vq_1\ldots \text{d}^3 \vq_n  \, \left[ P_{\alpha,\text{integrand}}^{\text{target}}(\vk,\vq_1,\ldots,\vq_n) 	
-  P_{\alpha,\text{integrand}}^{\text{ref}}(\vk,\vq_1,\ldots,\vq_n) \right]\ .
\end{equation}
Instead of computing the full $P_{\alpha}^{\text{target}}$ integral separately for each target cosmology, we can precompute $P_{\alpha}^{\text{ref}}$ once, and then only calculate $\Delta P_{\alpha}(k)$ for each target cosmology. If $P_{\alpha}^{\text{ref}}$ is precomputed with very high precision, then similar precision on the result for $P_{\alpha}^{\text{target}}$ can be achieved with a much lower requirement on the precision of $\Delta P_{\alpha}$. This incurs a significant reduction of the computational cost of running for several target cosmologies. We stress that this applies independently of the numerical technique used to compute the integrals (we will test our code using the IR-safe MonteCarlo integration of~\cite{Carrasco:2013sva}, but our results apply unaltered to any other potentially more efficient integration techniques~\cite{Taruya:2012ut,Sherwin:2012nh,Blas:2013aba,Bertolini:2015fya,Schmittfull:2016jsw,McEwen:2016fjn}): once the prediction for a reference cosmology has been computed, for the remaining target cosmologies one can directly compute the difference using much lower numerical precision.

One can estimate the relationship between the precision requirements on each term in Eq.~\eqref{eq:target_cosmo} using basic error propagation. Assuming the numerical integration of $P_{\alpha}^{\text{ref}}$ and $P_{\alpha}^{\text{target}}$
gives  uncorrelated errors, we have
\begin{equation}\label{eq:target_variance}
\sigma_{\Delta}^2 = \sigma_{\text{ref}}^2 + \sigma_\text{target}^2\ ,
\end{equation}
where $\sigma_{\text{target}}$ and $\sigma_{\text{ref}}$ are the integration errors of the target and reference cosmology loop integrals, and $\sigma_{\Delta}$ is the uncertainty in the integration~\eqref{eq:diff_integration}. Defining the corresponding relative errors
\begin{equation}
\epstarget \equiv \frac{\sigma_{\text{target}}}{|P_{\alpha}^{\text{target}}|}\ , \quad
\epsref \equiv \frac{\sigma_{\text{ref}}}{|P_{\alpha}^{\text{ref}}|}\ , \quad
\epsdelta \equiv \frac{\sigma_{\Delta}}{|\Delta P_{\alpha}|}\ ,
\end{equation}
Eq. \eqref{eq:target_variance} can be rewritten as
\begin{equation}
\label{eq:target_prec}
\epsdelta = \left| \frac{\Delta P_{\alpha}}{P_{\alpha}^{\text{target}}} \right|^{-1}
	\sqrt{ \epstarget^2 + \epsref^2 \lp \frac{P_{\alpha}^{\text{ref}}}{P_{\alpha}^{\text{target}}} \rp^2 }
	\approx \left| \frac{\Delta P_{\alpha}}{P_{\alpha}^{\text{target}}} \right|^{-1} \epstarget\ ,
\end{equation}
where the approximation is valid because $|P_{\alpha}^{\text{ref}}/P_{\alpha}^{\text{target}}| \sim \mathcal{O}(1)$ and $\epstarget\gtrsim \epsref$ (in other words, on the target cosmology we will always request at best similar precision to what we have used for the reference cosmology).

At this point, we can further reduce the difference between the target and reference cosmology by taking advantage of the linearity of Eq.~\eqref{eq:diff_integration}. In fact, we can exactly account for the difference caused by the primordial amplitude of scalar fluctuations $A_s$ with the following modified difference,
\begin{eqnarray}\label{eq:adj_DeltaPloop}
&&\Delta \tilde{P}_{\alpha}(k) \equiv\\ \nonumber
&&\quad\int \text{d}^3 \vq_1\ldots \text{d}^3 \vq_n   \, 
	\left[ P_{\alpha,\text{integrand}}^{\text{target}}(\vk,\vq_1,\ldots,\vq_n) 
	-  \left( \frac{A_s^{\text{target}}}{A_s^{\text{ref}}} \right)^{L+1} 
	P_{\alpha,\text{integrand}}^{\text{ref}}(\vk,\vq_1,\ldots,\vq_n) \right],
\end{eqnarray}
for any $L$-loop integral and adjusting Eq.~\eqref{eq:target_cosmo} accordingly. Hence, Eq.~\eqref{eq:target_prec} can be rewritten as
\begin{equation}
\label{eq:target_prec_modified}
\epsdelta \approx \left| \frac{\Delta \tilde{P}_{\alpha}}{P_{\alpha}^{\text{target}}} \right|^{-1} \epstarget\ ,
\end{equation}
and this equation can then be used to determine the  value of $\epsdelta$ that will yield a certain desired value for $\epstarget$.

\begin{figure}[t]
    \centering
    \begin{subfigure}[b]{0.48\columnwidth}
        \includegraphics[width=\columnwidth]{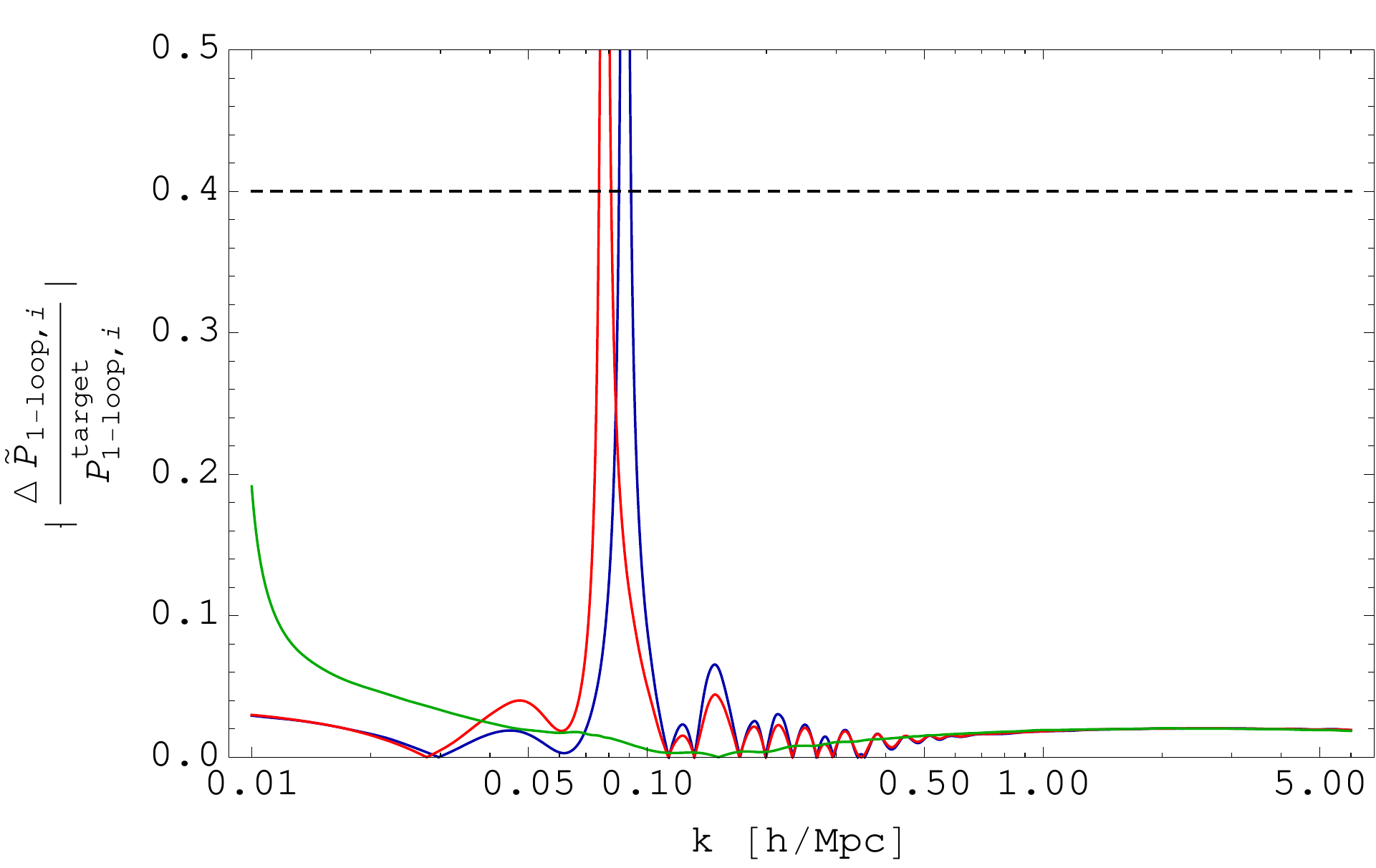}
    \end{subfigure}
    ~ 
    \begin{subfigure}[b]{0.48\columnwidth}
        \includegraphics[width=\columnwidth]{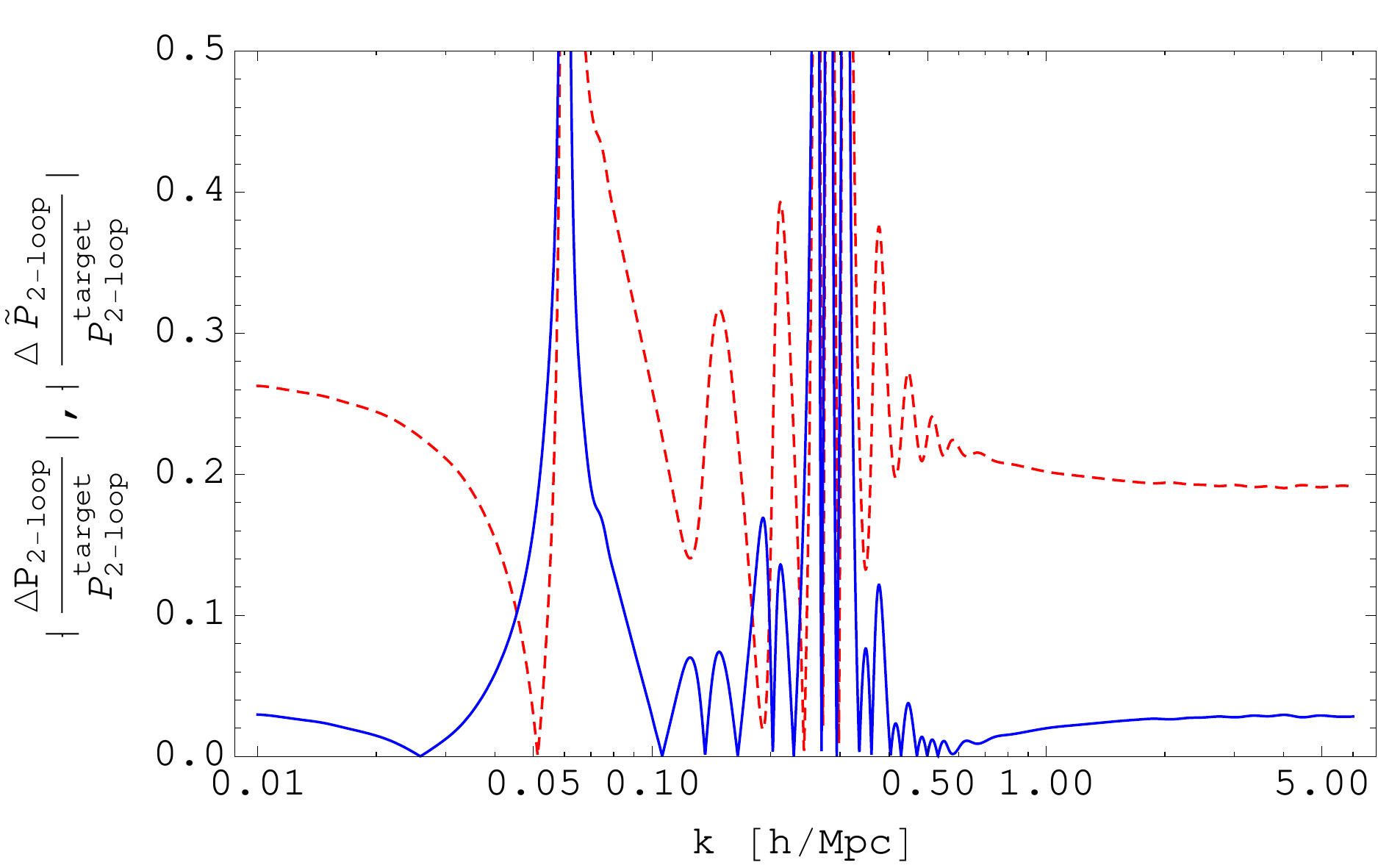}
    \end{subfigure}
    \caption{\textit{Left:} The ratio $\Delta \tilde{P}_\alpha(k)/P_\alpha^\text{target}$ for the one-loop terms that enter the two-loop EFTofLSS matter power spectrum (see Sec.~\ref{sec:Results} for more details):  $\poneloop$ (\textit{blue}), $\poneloopcs$ (\textit{red}) and $\poneloop^{(\text{quad},1)}$ (\textit{green}). $\Delta\tilde{P}_\alpha$ is given by Eq.~\eqref{eq:adj_DeltaPloop} with $L=1$, and we show results for the \texttt{cosmo\_5} test cosmology (given in Table~\ref{tab:Cosmo_list}). The black dashed line is the actual $k$-independent conservative value used in {\codename}  to set the relative precision required for one-loop integrations (see Eq.~\eqref{eq:target_prec_modified}). \textit{Right:} The same ratio for $\ptwoloop$, again for the \texttt{cosmo\_5} cosmology. The dashed red curve uses Eq.~\eqref{eq:diff_integration}, while the solid blue curve uses the adjusted form, Eq.~\eqref{eq:adj_DeltaPloop}, with $\ptwoloop^{\text{target/ref}}$ evaluated through direct integration. For general cosmologies, Eq.~\eqref{eq:adj_DeltaPloop} shows a similar improvement over Eq.~\eqref{eq:diff_integration} in  removing most of the difference associated with the cosmological parameter~$A_s$.}
\label{fig:DP1loop_DP2loop}
\end{figure}

Given their low dimensionality, one-loop integrals are calculated very efficiently and we can simply replace the $k$-dependent ratio in Eq.~\eqref{eq:target_prec_modified} with a constant that guarantees sub-percent accuracy over the entire range of scales we are interested in. We find that using $\left| \Delta \tilde{P}_{\text{1-loop}}/\poneloop^{\text{target}} \right| = 0.4$ always overestimates the real ratio (except at zero-crossing); an example of this is shown in the left panel of Fig.~\ref{fig:DP1loop_DP2loop} (We use $0.4$ to enable the code to handle cosmologies for which the  ratio $\left| \Delta \tilde{P}_{\text{1-loop}}/\poneloop^{\text{target}}\right|$ will likely be higher than that shown in Fig.~\ref{fig:DP1loop_DP2loop}; we have checked that using $0.2$ instead of $0.4$ does not appreciably affect the code's performance.). When fixing $\epstarget = 0.5\%$  and $\left| \Delta \tilde{P}_{\text{1-loop}}/\poneloop^{\text{target}} \right| = 0.4$, Eq.~\eqref{eq:target_prec_modified} sets $\epsdelta = 1.25\%$ for all wavenumbers. Due to our overestimation of the true value of $\left| \Delta \tilde{P}_{\text{1-loop}}/\poneloop^{\text{target}} \right|$, this will result in a target precision even better than 0.5\%; in Sec.~\ref{sec:DiffChecks}, we will show this explicitly for several test cosmologies.

A similar procedure works for two-loop integrals, which are, on the other hand, more computationally intensive. The right panel of Fig.~\ref{fig:DP1loop_DP2loop} illustrates the effectiveness of Eq.~\eqref{eq:adj_DeltaPloop} in removing most of the difference associated with the cosmological parameters. This figure also illustrates that, if we can obtain a good estimation for $\Delta\tilde{P}_\text{2-loop}/P_\text{2-loop}^\text{target}$ as a function of $k$, we can then adjust $\epsdelta$ to obtain the same $\epstarget$ at each $k$, possibly incurring a great reduction of computational expense. Obviously, we cannot use $\ptwoloop^{\text{target}}$ to calculate the ratio in Eq.~\eqref{eq:target_prec_modified}, since this is the final goal of the computation. However, the main contributor to the difference $\Delta\tilde{P}_\text{2-loop}$ is the difference between the linear power spectra that enter into the expressions for $P_{\text{2-loop}}^\text{target}$ and $P_{\text{2-loop}}^\text{ref}$. These power spectra appear in the integrands of each quantity, but since these integrands will be dominated by the region where all internal momenta are of order $k$, we can approximate the difference between $P_{\text{2-loop}}^\text{target}$ and $P_{\text{2-loop}}^\text{ref}$ by replacing each one with the appropriate power of the linear power spectrum $P_{11}(k)$, obtaining~\footnote{For higher loops one can trivially adjust this estimate.}: 
\begin{equation}
\label{eq:ratioestimate}
\left| \frac{\Delta \tilde{P}_{\text{2-loop}}}{P_{\text{2-loop}}^{\text{target}}} \right| \approx
	\left| 1 - \lp \frac{A_s^{\rm target}}{A_s^{\rm ref}} \rp^{3} 
	\lp \frac{P_{11}^{\rm ref}(k)}{P_{11}^{\rm target}(k)} \rp^{3} \right| .
\end{equation}

\begin{figure}[t]
\begin{center}
\includegraphics[width=0.48\columnwidth]{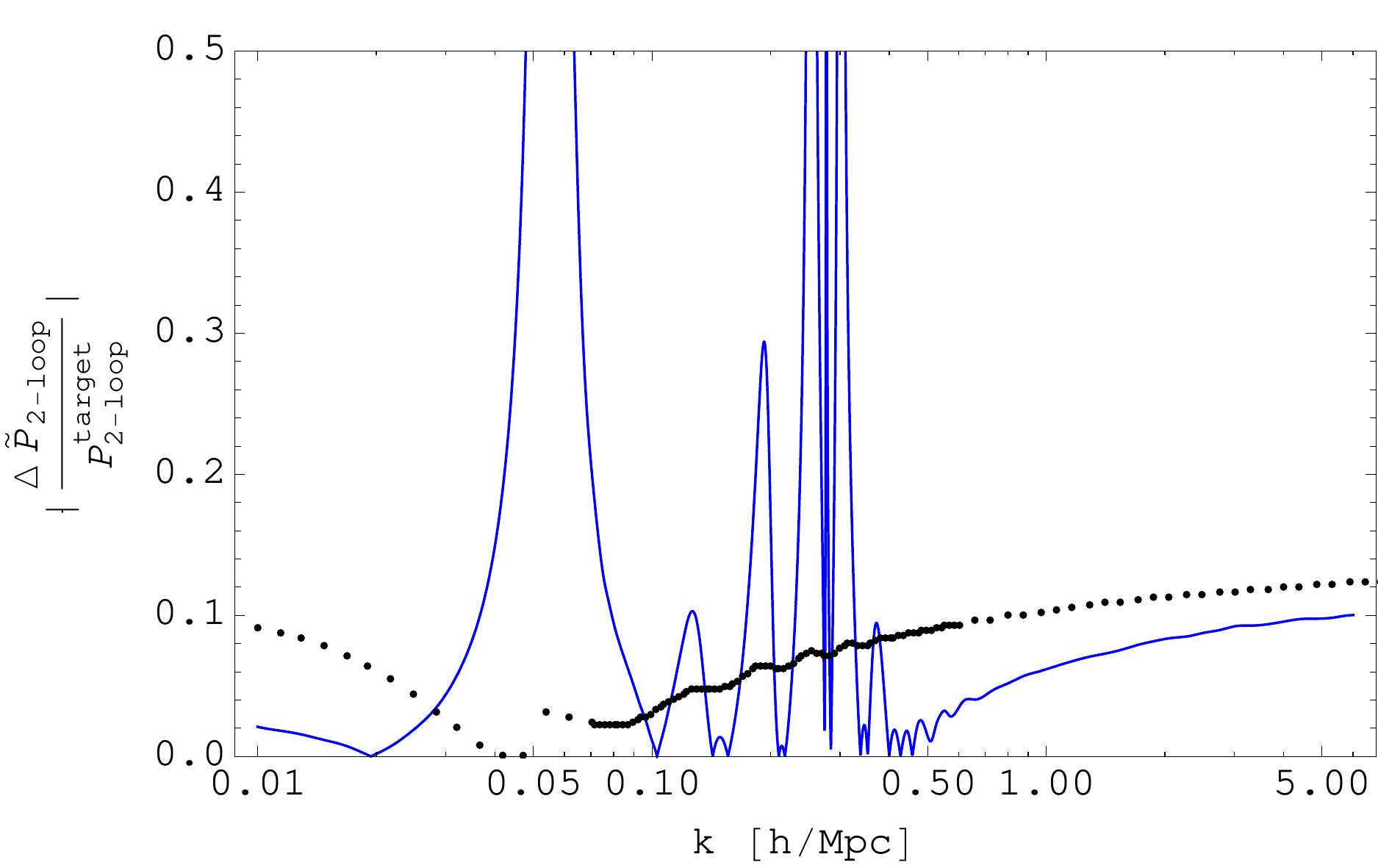}
\quad
\includegraphics[width=0.48\columnwidth]{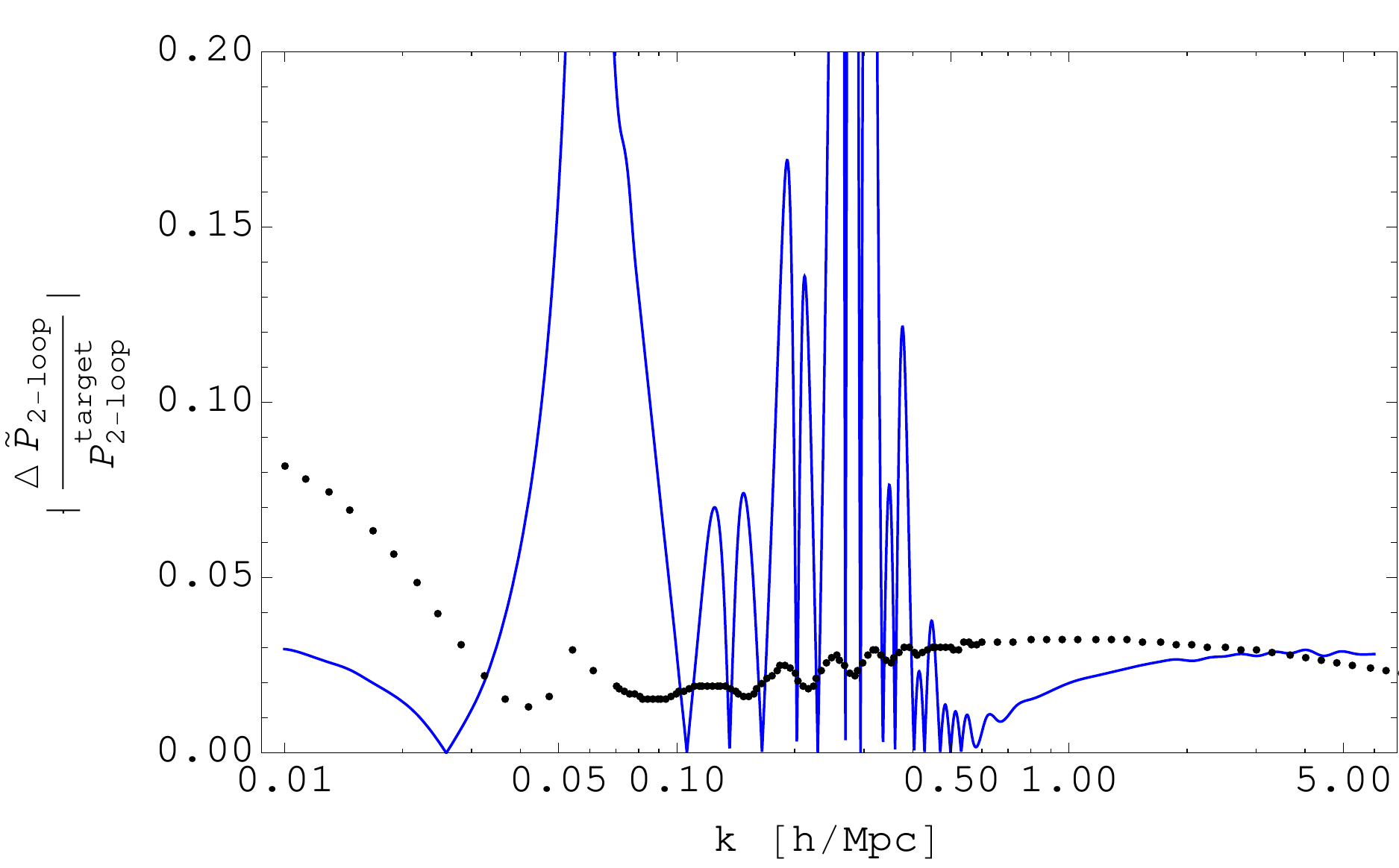}
\end{center}
\caption{ Comparison of estimate (Eq.~\eqref{eq:ratioestimate}, with the modifications described in the main text; \textit{black points}) and exact calculation of $|\Delta \tilde{P}_\text{2-loop} / \ptwoloop^\text{target}|$ (\textit{blue lines}) for two test cosmologies, \texttt{cosmo\_1} (\textit{left}) and \texttt{cosmo\_5} (\textit{right}), given in Table~\ref{tab:Cosmo_list}. On average, for $k\gtrsim 0.5\invMpc$, the estimate slightly over-predicts the exact calculation, but this only means that the precision requested for the integration of $\Delta \tilde{P}_\text{2-loop}$ is slightly more conservative than necessary. For lower wavenumbers, the estimate is less precise, but, as we describe in the text, the required precision is also lower.  Also, the estimate has the desirable feature of automatically limiting the precision requested close to the zero-crossings of $\ptwoloop^\text{target}(k)$, by setting a ceiling on the value of $|\Delta \tilde{P}_\text{2-loop} / \ptwoloop^\text{target}|$ in Eq.~\eqref{eq:ratioestimate}.
}
\label{fig:adjusted_ratio_comparison}
\end{figure}
However, this estimate makes predictions that are systematically in antiphase with BAO of the exact calculations of $|\Delta \tilde{P}_\text{2-loop}/\ptwoloop^\text{target}|$. Therefore, for $0.05 \invMpc < k < 0.5\invMpc$ we smooth the estimated adjusted ratios with a top-hat window function,
\begin{equation}\label{eq:adj_ratio_smooth}
\left| \frac{\Delta \tilde{P}_\text{2-loop}(k)}{P_\text{2-loop}^{\text{target}}(k)} \right|_\text{smooth} = 
	\frac{1}{\Delta k}\int_{k-\Delta k/2}^{k+\Delta k/2} \left| \frac{\Delta \tilde{P}_\text{2-loop}(k^\prime)}{P_\text{2-loop}^{\text{target}}(k^\prime)} \right| 
	\, \text{d}k^\prime\ ,
\end{equation}
where we set the window function width to $\Delta k = 0.1  \invMpc$, roughly equal to twice the BAO period in $k$-space.
The resulting estimate allows us to  use Eq.~\eqref{eq:target_prec_modified} to fix $\epsdelta$ as a function of $k$. In Fig.~\ref{fig:adjusted_ratio_comparison}, we compare this estimate to the exact calculation of $|\Delta \tilde{P}_\text{2-loop} / \ptwoloop^\text{target}|$ in two test cases. The estimate is slightly higher than the exact calculation for $k \gtrsim 0.5 \invMpc$, leading to conservative precision requirements on the integral evaluation. On the other hand, it fails to capture the exact relative ratio for smaller $k$'s. As we shall explain below, this is of marginal concern, since we only aim for global subpercent precision on the final matter power spectrum predictions, and $P_\text{2-loop}$ is very small at small wavenumbers. Moreover, to save computational time where $\Delta \tilde{P}_\text{2-loop} \approx 0$, we include the additional requirement $\sigma_\Delta=1.5({\rm Mpc}/h)^3$ for the absolute precision on the two-loop integral derived from Eq.~\eqref{eq:adj_DeltaPloop} (the typical values of $P(k)$ are of order a thousand in this units). This is chosen by the integration routine whenever it turns out to be less stringent than the relative precision $\epsdelta$. Note also that $\ptwoloop(k)$ typically possesses two zero crossings, one around $k\sim 0.05\invMpc$ and a second for $k\sim 0.3\invMpc$. Around these points, we do not require high precision on the evaluation of $\ptwoloop(k)$, and this is also accounted for by our estimates, which automatically impose a ceiling on the value of $|\Delta \tilde{P}_\text{2-loop} / \ptwoloop^\text{target}|$ (and therefore a floor on $\epsdelta$) in the relevant regions.

 \begin{figure*}[t]
\begin{center}
\includegraphics[width=0.48\columnwidth]{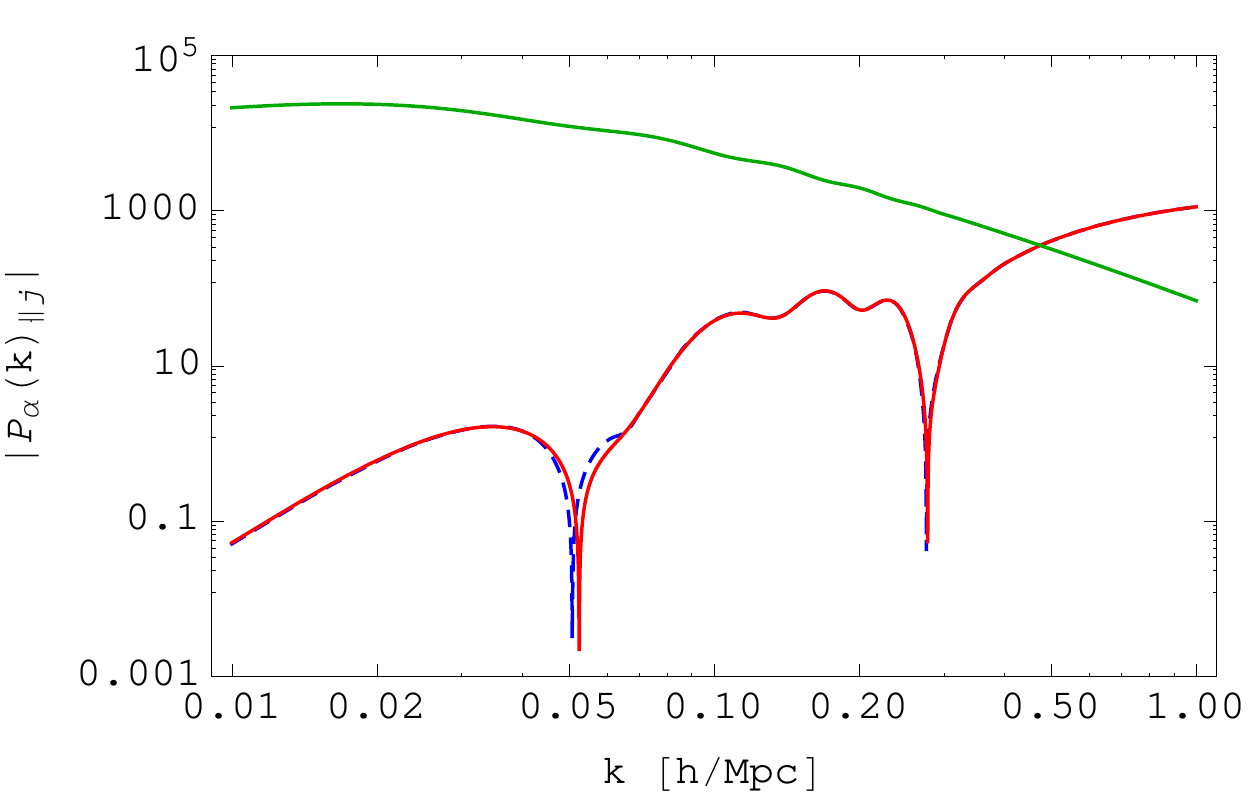}
\quad
\includegraphics[width=0.48\columnwidth]{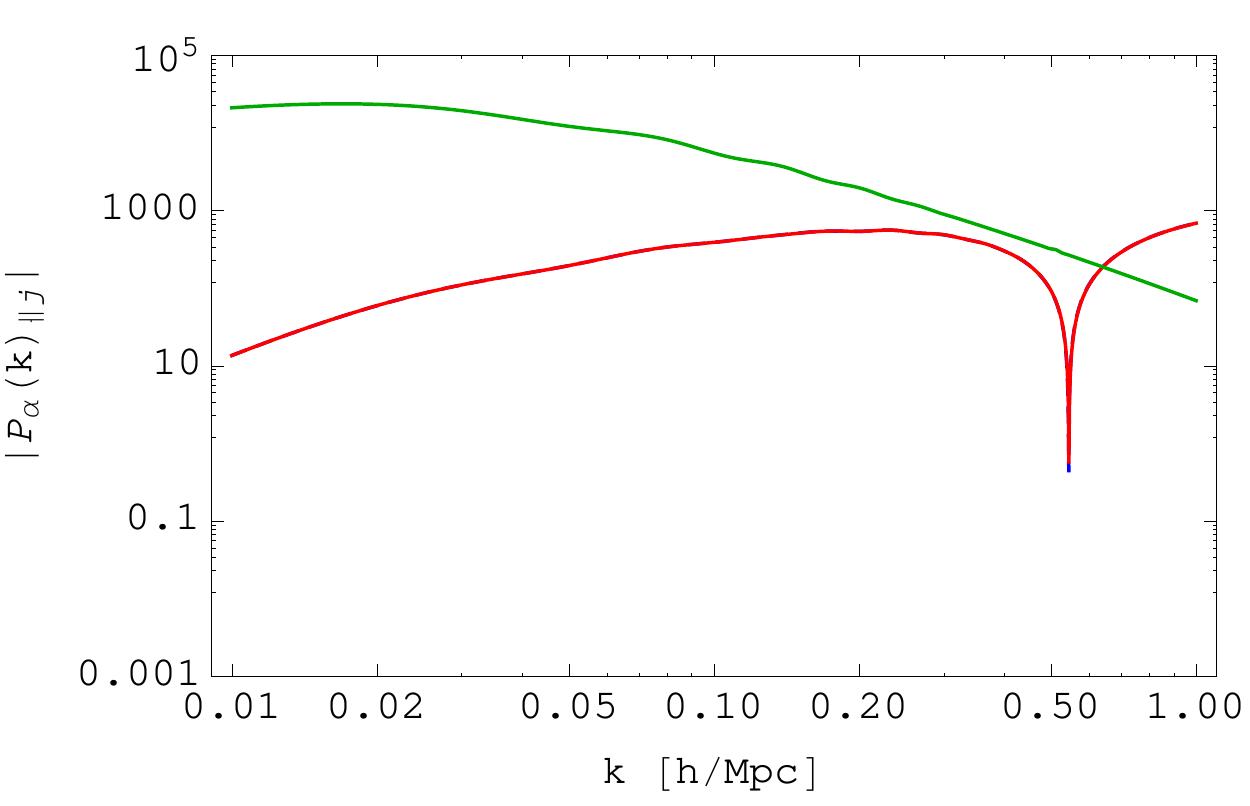}
\end{center}
\caption{
$P_{11}(k)_{\| 2}$ (green), $\ptwoloop(k)_{\| 0}$ from \codename~(dashed blue), and $\ptwoloop(k)_{\| 0}$ evaluated through full integration (red), all for the \texttt{cosmo\_5} test cosmology (see Table~\ref{tab:Cosmo_list}). We show the IR-resummed version of each term, as indicated by the subscripts.
 The $\ptwoloop$ curves in the left panel are $P_\text{2-loop}^\text{(UV-improved)}$, while the curves in the right panel are $\ptwoloopfull$. The fact that $P_\text{2-loop}^\text{(UV-improved)}\ll P_{11}$ at low $k$ makes the small deviations we observe in the left panel acceptable. On the other hand, $\ptwoloopfull$  amounts to $\sim 1\%$ of $P_{11}$ already at $k \approx 0.05 \invMpc$, leading to more demanding requirements on the precision of $\ptwoloop$ if $\ptwoloopfull$ is used.
}
\label{fig:p2loop_p11_comparison}
\end{figure*}

For the calculations above, as well as within {\codename}, we have employed the ``UV-improved'' version of $\ptwoloop$. This is defined by taking the full two-loop term, $\ptwoloopfull$, and analytically subtracting most of the leading UV contribution to the two-loop integral; this contribution has the functional form $k^2 P_{11}$, and is therefore degenerate with the $\ct$ parameter in the two-loop power spectrum prediction (see Eq.~\eqref{eq:resum_P_EFT_2loop}, and~\cite{Foreman:2015lca} for more details). The advantage of $\ptwoloop^\text{(UV-improved)}$ over $\ptwoloopfull$ is clear from Fig.~\ref{fig:p2loop_p11_comparison}, where the two versions are compared to $P_{11}$ (which is the leading term in the total prediction for the matter power spectrum), and all quantities are shown after resummation following~\cite{Senatore:2014via,Angulo:2014tfa}. In particular, the relative error on the EFTofLSS power spectrum prediction coming from the two-loop term is to a good approximation given by the difference between $\ptwoloop$ evaluated with Eq.~\eqref{eq:adj_DeltaPloop} and that obtained with the full integration, rescaled by the tree term $P_{11}$. Therefore, even though using $\ptwoloop^\text{(UV-improved)}$ we cannot achieve $\epstarget=0.5\%$ for $k \lesssim 0.35 \invMpc$, this is more than compensated by its smallness compared to $P_{11}$. Because of its larger overall amplitude, the same argument does not apply to $\ptwoloopfull$, thus demanding more stringent requirements on its precision.  For the sake of completeness, in App.~\ref{app:fullp2loop} we show how the estimate Eq.~\eqref{eq:ratioestimate} performs using $\ptwoloopfull$, and in App.~\ref{app:altestimates} we also provide a procedure to more accurately estimate $|\Delta \tilde{P}_\text{2-loop} / \ptwoloop^\text{target}|$ for this case. For the rest of the paper, $\ptwoloop$ will refer to $\ptwoloop^\text{(UV-improved)}$.

As a side remark, target cosmologies that differ from the reference cosmology only for $A_s^\text{target}$ are readily evaluated in \codename~by multiplying the reference loop integrals by a suitable factor $(A_s^\text{target}/A_s^\text{ref})^{L+1}$.

	
\subsection{Details of implementation} \label{sec:DiffImp}

\subsubsection{Inputs}

The {\codename} code implements the idea we have presented in Sec.~\ref{sec:DiffExplain}. To evaluate the integrals, we use a modification of the Copter library~\cite{Carlson:2009it} that implements the IR-safe integrands from~\cite{Carrasco:2013sva} and computes the loop integrals using Monte Carlo integration routines from the CUBA library~\cite{Hahn:2004fe}. Needless to say, the evaluation of the integrals can be performed using some of the alternative techniques presented in~\cite{Taruya:2012ut,Sherwin:2012nh,Blas:2013aba,Bertolini:2015fya,Schmittfull:2016jsw,McEwen:2016fjn}: while this might affect the evaluation time of a single integral, it will not affect the relative gain in computational cost that we obtain in computing a target cosmology with our method rather than with a direct computation~\footnote{Apart for adjustment due to how, in the various methods, the numerical error scales with the number of evaluations. Such an adjustment can be trivially performed.}. We define target cosmologies by means of the following vector of five cosmological parameters:
\begin{equation} \label{eq:refcosmo}
{\pmb\theta} \equiv \{ \omega_b, \omega_c ,  \ln(10^{10}A_s), n_s, h \}\ ,
\end{equation}
with all other cosmological parameters fixed to the reference cosmology, for which we use the current best-fit Planck parameters~\cite{Ade:2015xua} (the generalization to add additional cosmological parameters is relatively straightforward). The reference values of the five parameters listed in Eq.~\eqref{eq:refcosmo} are shown in the first row of Table~\ref{tab:Cosmo_list}.

{\codename} reads an initialization file with the following variables that specify the input cosmology: \\

\begin{lstlisting}
h = 0.6727 #Hubble parameter H_0/(100 km/s/Mpc)
Tcmb = 2.7255 #CMB temperature today 
n = 0.9645 #primordial spectral index
Omega_m = 0.313905 #total matter density
Omega_b = 0.0491685 #baryon matter density
sigma8 = 0.843107 #power spectrum normalization
tkfile = ./input/tk_planck_2015.dat #path to transfer function file
outdir = ./output/ #output directory
\end{lstlisting}

\noindent  (Note that the code takes $\Omega_{\rm m} \equiv (\omega_b+\omega_c)/h^2$ and $\Omega_b \equiv \omega_b/h^2$ instead of $\omega_c$ and $\omega_b$ as inputs, as well as $\sigma_8$ in place of $A_s$.)
The transfer function can be calculated using a Boltzmann code such as \texttt{CAMB}~\footnote{\url{http://camb.info}} or \texttt{CLASS}~\footnote{\url{http://class-code.net}}, and must be evaluated up to $k \sim 10 \invMpc$. Together with the cosmological parameters listed above, it is then used to compute the linear power spectrum that is employed in subsequent one- and two-loop integrations. In addition, the code reads in the EFTofLSS loop integrals for the reference cosmology, which have been precomputed with precision $\epsref=0.1\%$ by direct integration. By default, the required precision on the target cosmology is hard-coded and fixed to $\epstarget=0.5\%$ for all loop integrals ($\epstarget$ is named $\texttt{epsrelTar}$ in the code). For the one-loop terms, using Eq.~\eqref{eq:target_prec_modified} with the adjusted ratios set to 0.4 (see Fig.~\ref{fig:DP1loop_DP2loop}), we require a precision on the difference integral Eq.~\eqref{eq:adj_DeltaPloop} of $\epsdelta = 1.25\%$, independent of wavenumber and cosmology.

\subsubsection{Outputs}

{\codename} outputs two text files, \texttt{pk.dat} and \texttt{pXloop.dat}, containing the linear power spectrum ($P_{11}$) and the loop integrals ($\poneloop$, $\poneloopcs$, $\ptwoloop$, $\poneloop^{(\text{quad,1})}$) respectively, all evaluated at $z=0$ for the specific input cosmology. Wavenumbers are sampled sparsely in the ranges $0.01\invMpc\leqslant k  \leqslant 0.06 \invMpc$ and $0.6\invMpc \leqslant k \lesssim 10 \invMpc$, whereas we choose a denser sampling in between to accurately follow the BAO. In total, we sample each loop integral at 126 points (shown, for example, by the black points in Fig.~\ref{fig:adjusted_ratio_comparison}).


\subsubsection {\tt ResumEFT}
\label{sec:resumeft}

In a separate step, the output files from {\codename}, along with $\Omega_{\rm m}$ and desired redshift, are then used as input for the IR-resummation code \texttt{ResumEFT}~\footnote{\website}, also released with this work. \texttt{ResumEFT} implements the IR-resummation technique developed in~\cite{Senatore:2014via,Angulo:2014tfa} to incorporate the leading effect of large-scale displacements on the perturbative predictions of the EFTofLSS. 

In brief, in the Lagrangian description of large-scale structure, one finds that the effect of long-wavelength displacements on density correlation functions arises from an exponential of cumulants of the displacement field. The Eulerian approach to perturbation theory (which the computations in {\codename} are based on) expands this exponential up to a given perturbative order, assuming that the nonlinearities caused by long-wavelength displacements are no stronger than those from couplings between different modes of the density field. For cases of cosmological interest, this assumption is incorrect, and therefore it would be ideal to treat these displacements as non-perturbatively as possible. In~\cite{Senatore:2014via}, it was shown that this can be accomplished by keeping the leading (two-point) cumulant of the long displacements in the exponential mentioned earlier. In~\cite{Senatore:2014via}, with further improvements in~\cite{Angulo:2014tfa}, it was shown that these effects manifest themselves as convolutions of specific kernels $\hat{M}_{\parallel_j}$ with the terms appearing in Eulerian perturbation theory. For example, in the two-loop matter power spectrum we have
\begin{equation} \label{eq:resum2loop}
P_\text{EFT-2-loop}(k,z) =\sum_{j=0}^2 \sum_{X_j}  P_{X_j}(k,z)_{\parallel 2-j}\ ,
\end{equation}
where $\sum_{X_j}$ sums over all terms appearing at $j$-loop order in the power spectrum, $P_{X_j}(k,z)_{\parallel 2-j}$ is defined by
\begin{equation} \label{eq:resumterm}
P_{X_j}(k,z)_{\parallel 2-j} \equiv \int dk'\, \hat{M}_{\parallel_{2-j}}(k,k';z) P_{X_j}(k';z)\ ,
\end{equation}
and expressions for $\hat{M}_{\parallel_{2-j}}(k,k';z)$ can be found in~\cite{Angulo:2014tfa}.

\texttt{ResumEFT} uses the \texttt{FFTLog} algorithm~\cite{Hamilton:1999uv} to perform the integrals in Eqs.~\eqref{eq:resum2loop}-\eqref{eq:resumterm}, and outputs a text file containing all resummed EFTofLSS terms relevant to the matter power spectrum at tree, one- and two-loop levels.

\subsection{Tests and performance} \label{sec:DiffChecks}

\begin{table}[tp]
\caption{Cosmological parameters for the reference and test cosmologies employed in this work. Next to each test cosmology, we specify the number of sigmas away from the reference cosmology, as dictated by Eq.~\eqref{eq:hyperellipsoid}.}
\begin{center}
\begin{tabular}{| >{\centering\arraybackslash} m{2.5cm} || >{\centering\arraybackslash} m{1.75cm} | >{\centering\arraybackslash} m{1.75cm} | >{\centering\arraybackslash} m{1.75cm} | >{\centering\arraybackslash} m{1.75cm} | >{\centering\arraybackslash} m{1.75cm} |}
\hline 
Cosmology		&	$\omega_b$ 	&	$\omega_c$	&	$\ln(10^{10}A_s)$	&	$n_s$	&	\multirow{2}{*}{$h$} \\ [10pt]
\hline
Reference		   	&	0.02225		&	0.1198		&		3.094		&	0.9645	&		0.6727		\\
\texttt{cosmo\_1} ($3\sigma$)   	&	0.02257		&	0.1183		&		3.196		&	0.9645	&		0.6793		\\
\texttt{cosmo\_2} ($3\sigma$) 	&	0.02257		&	0.1168		&		3.026		&	0.9645	&		0.6859		\\
\texttt{cosmo\_3} ($3\sigma$)   	&	0.02193		&	0.1213		&		2.992		&	0.9645	&		0.6661		\\
\texttt{cosmo\_4} ($3\sigma$) 	 &	0.02193		&	0.1228		&		3.162		&	0.9645	&		0.6595		\\
\texttt{cosmo\_5} ($3\sigma$)  	&	0.02241		&	0.1213		&		3.026		&	0.9596	&		0.6661		\\
\texttt{cosmo\_6} ($3\sigma$)  	&	0.02209		&	0.1183		&		3.162		&	0.9694	&		0.6793		\\
\texttt{cosmo\_7} ($4\sigma$)  	&	0.02193		&	0.1183		&		2.992		&	0.9645	&		0.6727		\\
\texttt{cosmo\_8} ($4\sigma$)  	&	0.02177		&	0.1183		&		3.128		&	0.9645	&		0.6727		\\
\texttt{cosmo\_9} ($5\sigma$)  	&	0.02161		&	0.1258		&		2.924		&	0.9449	&		0.6463		\\
\texttt{cosmo\_10} ($3\sigma$)  	&	0.02177		&	0.1243		&		2.992		&	0.9498	&		0.6529		\\
\texttt{cosmo\_11} ($3\sigma$)  	&	0.02273		&	0.1153		&		3.196		&	0.9792	&		0.6925		\\
\hline

\end{tabular}
\end{center}
\label{tab:Cosmo_list}
\end{table}

We now proceed to evaluate the performance of {\codename} by comparing the code's output for various loop integrals to the results of the full integration ({\it i.e.}\ without the reference-target split described in Sec.~\ref{sec:DiffExplain}). We quantify the deviation of each test cosmology from the reference (Planck) cosmology through the following definition: an ``$n\sigma$-cosmology'' is defined as a point~${\pmb\theta}$ in parameter space separated from the reference cosmology by $n$ standard deviations, accounting for all the covariances between the cosmological parameters. More formally, the point ${\pmb\theta}$ defines an $n\sigma$-cosmology if it is contained in a thin shell around the 5-dimensional hyper-ellipsoid defined by 
\begin{equation}\label{eq:hyperellipsoid}
({\pmb\theta} - {\pmb\theta}^{\text{ref}})^{\rm T} 
\Sigma^{-1}_5 
({\pmb\theta} - {\pmb\theta}^{\text{ref}}) = \chi_5^2(p)\ .
\end{equation}
Here, $\Sigma_5^{-1}$ represents the inverse of the $5 \times 5$ covariance submatrix obtained by selecting the appropriate rows and columns from the covariance matrix for the combination ``TT+TE+EE+lowP'' in~\cite{Ade:2015xua}, giving a reasonable account for the covariances between our 5 chosen parameters when all others are marginalized over~\footnote{For easy reference, $1\sigma$ marginalized uncertainties on each parameter from Eq.~\eqref{eq:planck_cov} are given below: 
\begin{equation}
\sigma(\omega_b) = 0.00016, \quad
\sigma(\omega_c) = 0.0015, \quad
\sigma(\ln[10^{10} A_s])=0.034, \quad
\sigma(n_s)=0.0049, \quad
\sigma(h) = 0.0066\ .
\end{equation}
}:
\begin{equation}\label{eq:planck_cov}
\Sigma_5 = 
\begin{pmatrix}
    2.56 \times 10^{-8}  & -1.5 \times 10^{-7} & 1.8 \times 10^{-6} & 3.99 \times 10^{-7} & 7.95 \times 10^{-7}  \\
    -1.5 \times 10^{-7}  & 2.25 \times 10^{-6} & -1.72 \times 10^{-5} & -5.67 \times 10^{-6} & -9.62 \times 10^{-6}  \\
    1.8 \times 10^{-6}  & -1.72 \times 10^{-5} & 1.16 \times 10^{-3} & 6.54 \times 10^{-5} & 8.08 \times 10^{-5}  \\
    3.99 \times 10^{-7}  & -5.67 \times 10^{-6} & 6.54 \times 10^{-5} & 2.4 \times 10^{-5} & 2.48 \times 10^{-5}  \\
    7.95 \times 10^{-7}  & -9.62 \times 10^{-6} & 8.08 \times 10^{-5} & 2.48 \times 10^{-5} & 4.36 \times 10^{-5}  
\end{pmatrix}.
\end{equation}
In Eq.~\eqref{eq:hyperellipsoid}, $\chi_5^2(p)$ is the quantile function for probability $p$ of the chi-squared distribution with 5 degrees of freedom. Thus, for $3\sigma$, $4\sigma$ and $5\sigma$ deviations from the reference cosmology $p=0.997,0.99993,0.9999994$, respectively.

In this section, we will examine a number of selected $3\sigma$-cosmologies, given (along with the reference cosmology) in Table~\ref{tab:Cosmo_list}. All cosmologies in this table have $\Omega_k = 0$, $w = -1$, and $k_{\text{pivot}}=0.05 \, \text{Mpc}^{-1}$. Further tests for the $4\sigma$- and $5\sigma$-cosmologies in Table~\ref{tab:Cosmo_list} are shown in App.~\ref{app:morechecks}.

\begin{figure}[t]
\begin{center}
\includegraphics[width=\columnwidth]{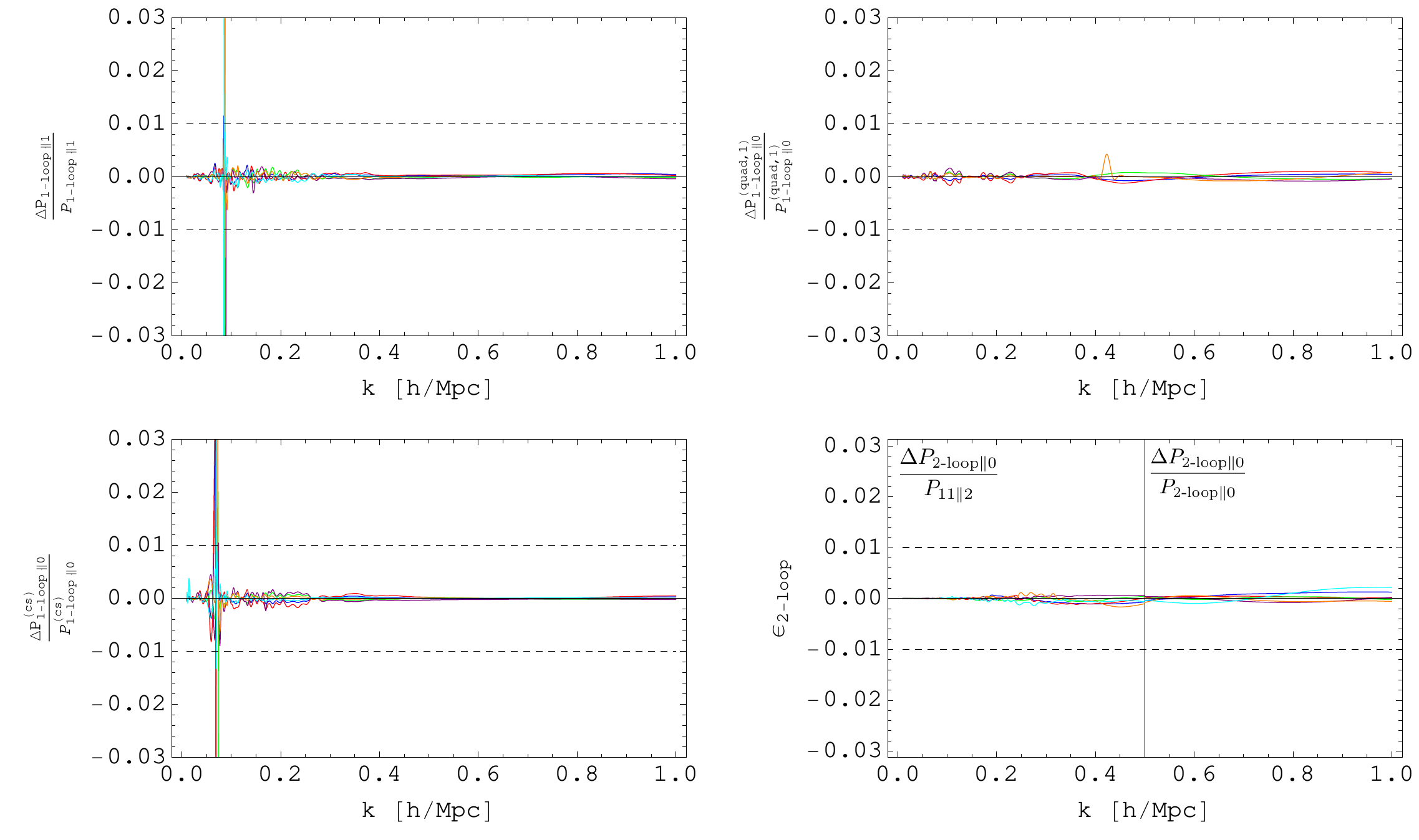}
\caption{IR-resummed {\codename} outputs for our $3\sigma$-cosmologies \texttt{cosmo\_1-6} (in order: blue, red, green, purple, orange, cyan) relative to the direct calculation of the full integrand with precision $\epsilon=0.1\%$. Here $\epsilon_{\text{target}}=0.5\%$ for all panels, and spikes are due to zero-crossing. Subscripts $\|0,\|1$ denotes the IR-resummation order as in~\cite{Senatore:2014via}. Dashed lines mark 1\% departures from direct calculations. For $\ptwoloop$ we opt for showing two quantities relevant on two disjoint scale intervals, $k < 0.5 \invMpc$ and $k > 0.5 \invMpc$. $\Delta P_{\text{2-loop} \| 0}/P_{11 \| 2}$ indicates the error on the matter power spectrum predictions, and $\Delta P_{\text{2-loop} \| 0}/P_{\text{2-loop} \| 0}$ confirms the performance of our estimates for smaller scales.}
\label{fig:fast_full_ratios_3sigma}
\end{center}
\end{figure}

We verify the integration prescription described above in Fig.~\ref{fig:fast_full_ratios_3sigma}, where we compare the outputs of {\codename} and the direct calculation of $P_{\alpha}^{\text{target}}$ for different integrals entering in the two-loop matter power spectrum prediction. Note that in all cases we have resummed the effect of large-scale displacement fields using \texttt{ResumEFT}. We find that the output for each $P_\alpha^\text{target}$ is indeed within $\sim$$0.5\%$ of the full computation over a wide range of scales, and even better than that for the one-loop integrals. A slight exception is $\ptwoloop$. For $\ptwoloop$ we divide the range of interest in two subsets, and the separation scale is chosen around where $\ptwoloop \sim P_{11}$ at $z=0$. As discussed in Sec.~\ref{sec:DiffExplain}, departures from the full integration are much larger than the required precision on scales $k < 0.35 \invMpc$ (they are about 2-3\% in this range of $k$'s). However, this error propagates to the final matter power spectrum as the ratio $\Delta \ptwoloop/P_{11}$, and Fig.~\ref{fig:fast_full_ratios_3sigma} shows that this remains within 0.2\% for $k < 0.5 \invMpc$ at $z=0$. Notably, this interval extends with redshift to increasingly smaller scales, in that $\ptwoloop \sim D(z)^6$ whereas $P_{11} \sim D(z)^2$, with $D(z)$ denoting the growth factor at redshift $z$. On scales $k > 0.5 \invMpc$, where our estimate Eq.~\eqref{eq:ratioestimate} overestimates the exact calculation, differences between \codename~and the full integration are well within the requirement $\epsdelta = 0.5\%$. Of course, overall better precision can be achieved if $\epstarget$ is adjusted to smaller values. However, within {\codename} special care must be taken to ensure that the variables encoding the maximum number of integral evaluations, i.e.\ \texttt{maxeval} and \texttt{maxeval2loop}, are properly set to ensure that the computation is not halted before the desired precision is achieved.

In Table~\ref{tab:Performance}, we show the computing time required for various evaluations, including the $4\sigma$- and $5\sigma$-cosmologies presented in App.~\ref{app:morechecks}. The integration strategy outlined in Sec.~\ref{sec:DiffExplain} reduces the computational cost by about two orders of magnitude compared to a direct computation of~$P_{\alpha}^{\text{target}}$ (though the precision requirement on $\ptwoloop$ at $k\lesssim 0.5\invMpc$ is different by a factor of a few in the direct or {\codename} computations). 

The main interest for us in this paper is to make the exploration of the various cosmologies much less computationally expensive than the direct computation. Therefore, in the context of this paper, the most important information is the relative gain with respect to the direct computation. However, one should keep in mind that both the time for obtaining the reference cosmology, as well as the running time of {\codename}, can be scaled down, probably in an obvious way, by implementing better integration techniques for the perturbation theory expressions, such as those proposed in~\cite{{Taruya:2012ut,Sherwin:2012nh,Blas:2013aba,Bertolini:2015fya,Schmittfull:2016jsw,McEwen:2016fjn}}. Furthermore, further improvements in the actual running time (wall time) can be achieved with multi-cored processors. In fact, thanks to the high degree of parallelism of the integration routines, we have verified that wall time scales down approximately as the effective number of cores. {\codename} can greatly benefit from recent advances in CPU technology, with improvements on running times that can be up to three times faster than ours. 

\ctable[
    caption = {Computational performance of {\codename} for our test cosmologies. CPU time is the amount of time used by all CPUs when executing the code, while wall time is the actual running time of the code (using a quad-core processor with hyper-threading). CPU times for the direct calculation of loop integrals correspond to a precision of $\epsilon=0.5\%$, and are to a large extent independent of the input cosmology. The ``speed-up factor'' is just the ratio of the {\codename} and direct CPU times.},
    label   = {tab:Performance},
pos =h,
mincapwidth=\textwidth,
]{| >{\centering\arraybackslash} m{2.5cm} | >{\centering\arraybackslash} m{2cm} | >{\centering\arraybackslash} m{2cm} | >{\centering\arraybackslash} m{2cm} | >{\centering\arraybackslash} m{2cm} |}{
\tnote[a]{Running on Quad-Core AMD Opteron\texttrademark Processor 2376, 2.3 GHz. The actual CPU-time on this CPU is 58 hours, although to match the Quad-Core Intel Core\texttrademark i7 performance we multiply this time by a correction factor of 0.82.}
\tnote[b]{Running on Quad-Core Intel Core\texttrademark i7, 2.4 GHz.}
}{                                                          				
\hline
Cosmology	&	CPU time direct\tmark[a]	&	CPU time {\codename}\tmark[b] 	&	Speed-up factor	&	Wall time\tmark[b]	\\ [10pt]
\hline 
\texttt{cosmo\_1} 		&	\multirow{11}{*}{48 hours}		&	28 min	&	103	&	3.6 min	\\
\texttt{cosmo\_2}		&							&	43 min	&	67	&	5.6 min	\\
\texttt{cosmo\_3}		&							&	15 min	&	192	&	2.1 min	\\
\texttt{cosmo\_4}		&							&	58 min	&	50	&	7.7 min	\\
\texttt{cosmo\_5}		&							&	7 min	&	411	&	53 sec	\\
\texttt{cosmo\_6}		&							&	8 min	&	360	&	1.1 min	\\
\texttt{cosmo\_7}		&							&	10 min	&	288	&	1.4 min	\\
\texttt{cosmo\_8}		&							&	10 min	&	288	&	1.5 min	\\
\texttt{cosmo\_9}		&							&	37 min	&	78	&	5.2 min	\\
\texttt{cosmo\_10}		&							&	29 min	&	99	&	3.8 min	\\
\texttt{cosmo\_11}		&							&	65 min	&	44	&	8.6 min	\\
\hline
}


\section{\texttt{TaylorEFT}: Taylor expansion of the loop integrals}\label{sec:Taylor}

With the latest \textit{Planck} data release our knowledge of the cosmological parameters in flat $\Lambda$CDM cosmologies has reached percent precision \cite{Ade:2015xua}. In light of this advance, along with the fact that these constraints will only improve if they are combined with information from late-time cosmological probes, it seems reasonable to restrict the EFTofLSS predictions to a sufficiently large region around our \textit{Planck}-like reference cosmology. This problem naturally lends itself to a Taylor expansion approach, where each loop integral is represented by a Taylor series up to quadratic order (or higher if necessary) in the deviation of the cosmological parameters from those of the reference cosmology. Such an approach eliminates the need to perform direct integrations to obtain the results of the loop integrals, provided that we are only concerned with cosmologies sufficiently close to the reference one, after the few cosmologies that are needed to construct the Taylor expansion have been evaluated.

	
\subsection{Details of implementation} \label{sec:TaylorImp}

\begin{figure}[tbp]
\begin{center}
\includegraphics[width=0.65\columnwidth]{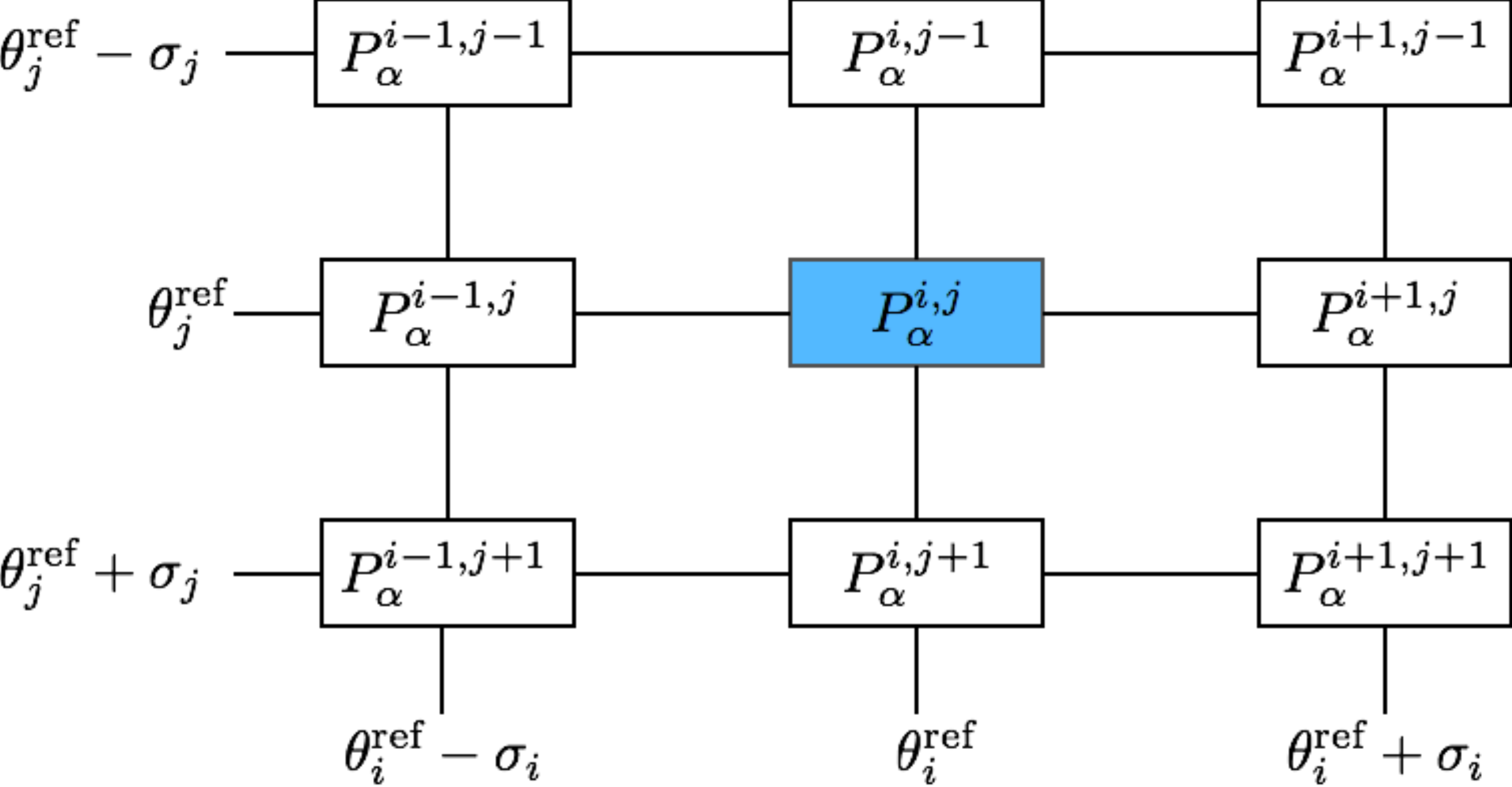}
\caption{Two-dimensional compact stencil for numerical evaluation of first and second derivatives at $\{\theta_i^{\text{ref}},\theta_j^{\text{ref}}\}$. The increments in the two parameters correspond to their \textit{Planck} standard deviations. Loop integrals $P_{\alpha}^{i,j}$ for the reference cosmology (blue box) are calculated through direct integration with 0.1\% precision. The $k$-dependence has been omitted for clarity.}
\label{fig:stencil}
\end{center}
\end{figure}

We implement this approach as follows. 
For each loop integral $P_\alpha$ we can write
\begin{equation} \label{eq:taylor_exp}
P_\alpha(k) \approx P_\alpha(k)|_{\pmb\theta^{\text{ref}}}
	+ \sum_i \Delta\theta_i \left. \frac{\partial P_{\alpha}(k)}{\partial \theta_i} 
	 \right|_{\pmb\theta = \pmb\theta^{\rm ref}}
	+ \frac{1}{2} \sum_{i,j} \Delta\theta_i \Delta\theta_j 
	\left. \frac{\partial^2 P_{\alpha}(k)}{\partial \theta_i \, \partial \theta_j} 
		\right|_{\pmb\theta = \pmb\theta^{\rm ref}},
\end{equation}
where $\Delta\theta_i \equiv \theta_i - \theta_i^{\rm ref}$. The derivatives are evaluated numerically using the compact stencil in Fig.~\ref{fig:stencil}. Considering a pair of parameters $\{\theta_i,\theta_j\}$, and fixing the remaining ones to their reference values, the derivatives at the reference cosmology are obtained with the following second order central differences:
\begin{eqnarray}
\frac{\partial P_{\alpha}}{\partial \theta_i} \biggm\lvert_{\pmb\theta^{\text{ref}}} 
	&\approx& \frac{P_{\alpha}^{i+1,j}-P_{\alpha}^{i-1,j}}{2\sigma_i}\ , \label{eq:num_deriv1} \\
\frac{\partial P_{\alpha}}{\partial \theta_j} \biggm\lvert_{\pmb\theta^{\text{ref}}} 
	&\approx& \frac{P_{\alpha}^{i,j+1}-P_{\alpha}^{i,j-1}}{2\sigma_j}\ , \label{eq:num_deriv2}\\
\frac{\partial^2P_{\alpha}}{\partial \theta_i^2} \biggm\lvert_{\pmb\theta^{\text{ref}}} 
	&\approx& \frac{P_{\alpha}^{i+1,j}-2 P_{\alpha}^{i,j} +P_{\alpha}^{i-1,j}}{\sigma_i^2}\ , \label{eq:num_deriv3}\\
\frac{\partial^2P_{\alpha}}{\partial \theta_j^2} \biggm\lvert_{\pmb\theta^{\text{ref}}} 
	&\approx& \frac{P_{\alpha}^{i,j+1}-2 P_{\alpha}^{i,j} +P_{\alpha}^{i,j-1}}{\sigma_j^2}\ , \label{eq:num_deriv4}\\
\frac{\partial^2P_{\alpha}}{\partial \theta_i \partial \theta_j} \biggm\lvert_{\pmb\theta^{\text{ref}}} 
	&\approx& \frac{P_{\alpha}^{i+1,j+1}- P_{\alpha}^{i+1,j-1} -P_{\alpha}^{i-1,j+1}+P_{\alpha}^{i-1,j-1}}{4\sigma_i\sigma_j}\ . 
	\label{eq:num_deriv5}
\end{eqnarray}

\noindent Here, for the sake of clarity, we have removed the $k$-dependence of the loop integrals, and have adopted the shorthands $P_{\alpha}^{i,j} \equiv P_{\alpha}(\theta_i^{\text{ref}},\theta_j^{\text{ref}})$, $P_{\alpha}^{i+1,j} \equiv P_{\alpha}(\theta_i^{\text{ref}}+\sigma_i,\theta_j^{\text{ref}})$, and so forth. Moreover, we have performed the finite differences over the \textit{Planck} uncertainties on each parameter, $\sigma_i$, since $\sigma_i/\theta_i^{\text{ref}} \lesssim 0.01$.  Our loop integrals depend on five cosmological parameters, and so we need 50 cosmologies (excluding the reference one) to evaluate Eqs.~\eqref{eq:num_deriv1}-\eqref{eq:num_deriv5} for all possible combinations. We have done so using a modified version of {\codename} that evaluates the $\ptwoloopfull$ 
using $\epsdelta = 0.5\%$, which for these cosmologies effectively corresponds to setting $\epstarget \approx \epsref$ (see Sec.~\ref{sec:Difference}), which is important to have sufficient control over numerical errors in the derivatives (we could have alternatively directly used $\ptwoloop$, but for the current level of precision $\ptwoloopfull$ was enough.). Note that departures from the reference cosmology along some directions in parameter space can be rather small, so much so that they are dominated by integration errors. In these cases, derivatives can be neglected, and their values are fixed to zero in the expansion in Eq.~\eqref{eq:taylor_exp}. 

The Taylor expansion scheme described above allows one to obtain the loop integrals in just a few seconds on a laptop, using a Mathematica script, \texttt{TaylorEFT}\footnote{\website}, that we have developed. We store all derivatives in text files within subdirectories organized by loop integral: \texttt{fd\_<$\theta_i$>\_<loop>.dat}, for first derivatives, \texttt{sd\_<$\theta_i$>\_<loop>.dat}, for pure second derivatives, and \texttt{smd\_<$\theta_i$>\_<$\theta_j$>\_<loop>.dat}, for mixed second derivatives. Derivatives are loaded with the dedicated \texttt{Load[]} module in the main \texttt{TaylorEFT} script. After specifying a redshift \texttt{z0} and the cosmological parameters defining \texttt{Cosmology}, the module \texttt{CalculateLoops[Cosmology]} reads off the necessary loop expressions using Eq.~\eqref{eq:taylor_exp} at $z=0$. \texttt{TaylorEFT} also calls \texttt{ResumEFT} after the EFT integrals have been obtained, applying the IR-resummation scheme described in \cite{Senatore:2014via,Angulo:2014tfa} at the chosen redshift \texttt{z0}. We shall describe in Sec.~\ref{sec:Results} how this wrapper can be used to directly obtain tree, one- and two-loop matter power spectrum predictions within the EFTofLSS. 

\subsection{Tests}

\begin{figure}[t]
\begin{center}
\includegraphics[width=\columnwidth]{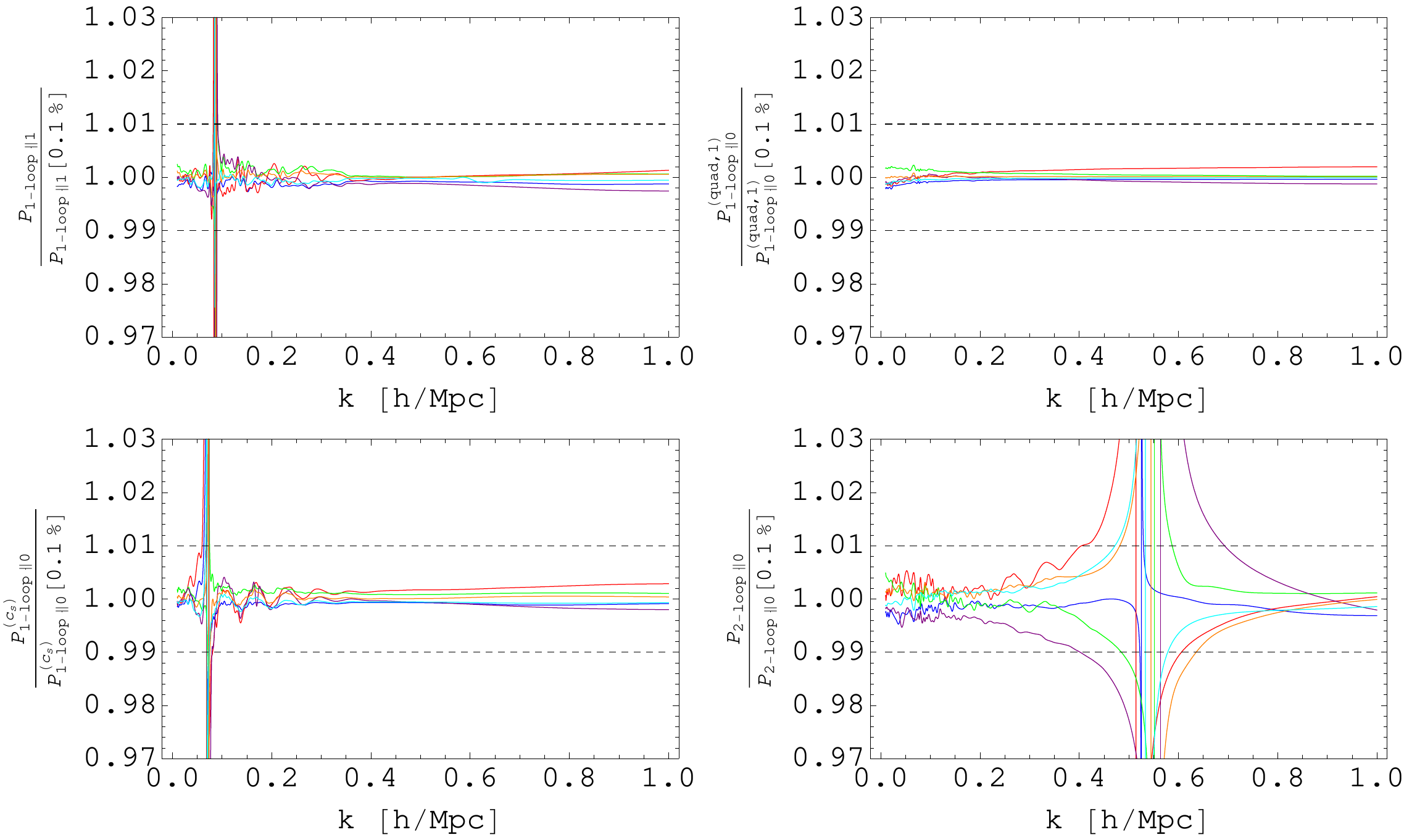}
\caption{IR-resummed \texttt{TaylorEFT} outputs for our $3\sigma$-cosmologies \texttt{cosmo\_1-6} (in order: blue, red, green, purple, orange, cyan) relative to the direct calculation of the full integrand with precision $\epsilon=0.1\%$. Spikes are due to zero-crossing and subscripts $\|0,\|1$ denotes the IR-resummation order as in~\cite{Senatore:2014via}. Dashed lines mark 1\% departures from direct calculations.}
\label{fig:taylor_full_ratios_3sigma}
\end{center}
\end{figure}

Fig.~\ref{fig:taylor_full_ratios_3sigma} shows how well the resummed EFT loop integrals at $z=0$ are approximated by the Taylor expansion~\eqref{eq:taylor_exp} for some of the $3\sigma$-cosmologies from Table~\ref{tab:Cosmo_list}. Clearly, one-loop integrals reach sub-percent precision over the entire range of scales. $\ptwoloop$ exhibits somewhat larger deviations compared to {\codename} outputs, although still inside the 1\% region for a wide range of wavenumbers. In Sec.~\ref{sec:Results}, we include two additional cosmologies (\texttt{cosmo\_10} and \texttt{cosmo\_11}) to quantify the limits of this implementation. We anticipate here that in order to obtain $P_\text{EFT-2-loop}$ within 1\% of the nonlinear power spectrum obtained from the \texttt{FrankenEmu} emulator code~\cite{Heitmann:2013bra}~\footnote{\url{http://www.hep.anl.gov/cosmology/CosmicEmu/emu.html}. This is a code that is built upon power spectrum measurements from an ensemble of dark-matter-only $N$-body simulations, and generates a power spectrum for a chosen input cosmology using a Gaussian Process emulator.}, each cosmological parameter cannot deviate from its reference value by more than $3\sigma$, i.e.\ $| \Delta \theta_i | \leqslant 3\sigma$ for any $i$. In other words, \texttt{TaylorEFT} is accurate to within 1\% only for cosmologies in a $\sim$3$\sigma$ neighborhood of our reference cosmology\footnote{{This approximately translates to a range in $\sigma_8$ of $0.78 \leqslant \sigma_8 \leqslant 0.90$ (see Fig.~\ref{fig:EFTparams_prediction}).}}. Obviously, it is possible that by increasing the order of the Taylor expansion, \texttt{TaylorEFT} can be made sufficiently accurate to a larger number of $\sigma$'s, something  that could be straightforwardly implemented. 

With this tool at hand, we are now ready to explore the cosmology dependence of the EFTofLSS parameters that incorporate the physics on nonlinear scales.


\section{The two-loop power spectrum}\label{sec:Results}

The two-loop IR-resummed matter power spectrum in the EFTofLSS is given by
\begin{align} \label{eq:resum_P_EFT_2loop}
P_\text{EFT-2-loop}(k,z) &= P_{11}(k,z)_{\|2} + P_{\text{1-loop}}(k,z)_{\|1} - 2 (2\pi) c_{s(1)}^2 \left( \frac{k^2}{k_\text{NL}^2}P_{11}(k,z) \right)_{\|1} \nonumber \\
				&+ P_{\text{2-loop}}(k,z)_{\|0} - 2(2\pi) c_{s(2)}^2 \left( \frac{k^2}{k_\text{NL}^2}P_{11}(k,z) \right)_{\|0} \nonumber \\
				&+ (2\pi) c_{s(1)}^2 P_{\text{1-loop}}^{(c_s)}(k,z)_{\|0} + (2\pi)^2 \left(c_{s(1)}^2\right)^2 \left( 1+ \frac{\zeta + \frac{5}{2}}{2 \left( \zeta + \frac{5}{4} \right)} \right) \left( \frac{k^4}{k_\text{NL}^4}P_{11}(k,z) \right)_{\|0} \nonumber  \\
				&+ (2\pi) c_1 P_{\text{1-loop}}^{\text{(quad,1)}}(k,z)_{\|0} + 2(2\pi)^2 c_4 \left( \frac{k^4}{k_\text{NL}^4}P_{11}(k,z) \right)_{\|0}.
\end{align}
Here, each term is the ``IR-resummed" version described in Sec.~\ref{sec:resumeft}, with the trailing subscripts defined by Eq.~\eqref{eq:resumterm}.
For a detailed derivation of Eq.~\eqref{eq:resum_P_EFT_2loop} see~\cite{Foreman:2015lca,Carrasco:2013mua,Carrasco:2013sva,Angulo:2014tfa}. 

The expression contains the tree-level (linear theory) power spectrum $P_{11}$, plus the one- and two-loop terms $\poneloop$ and $\ptwoloop$ that capture the mode couplings arising from the nonlinear terms in the continuity and Euler equations, in the absence of a stress tensor. However, as argued in e.g.~\cite{Baumann:2010tm,Carrasco:2012cv}, a stress tensor must be included in the description in order for it to be a sensible description of the long-wavelength dynamics. The other terms in Eq.~\eqref{eq:resum_P_EFT_2loop} are associated with the leading terms in an expansion of this stress tensor in derivatives and density and velocity fields (subleading terms in this expansion will only have relevant effects at higher wavenumbers than those we consider in this work). Specifically, the $\co$, $c_1$, and $c_4$ terms come from $\delta$, $\delta^2$, and $\d^2\delta$ terms in the stress tensor, respectively~\cite{Foreman:2015lca}.~\footnote{Other than $\delta^2$, one can also form three other quadratic terms in the stress tensor if the tidal tensor $\d_i \d_j \phi$ is used in the construction. However, it was shown in~\cite{Foreman:2015lca} that the contributions of these terms to the matter power spectrum prediction are similar enough that, at the level of precision we are concerned with, it is sufficient to only include one of them in the prediction. This conclusion was based upon comparison with the simulation used in Sec.~\ref{sec:darksky} of this paper, but based on our comparisons to other cosmologies in Sec.~\ref{sec:emulator}, the conclusion that only one quadratic term is needed seems to be robust with respect to changes in cosmological parameters.} The $\co$ parameter can be thought of as an effective speed of sound (modulo the subtleties related to renormalization discussed in e.g.~\cite{Foreman:2015uva}), $c_4$ is a kind of higher-derivative speed of sound, and $c_1$ is more generically related to nonlinearities occurring on shorter scales in the dark matter fluid (roughtly, it can be thought as a non-linear speed of sound). The scale $\knl$, defined to be approximately where perturbation theory is expected to break down, is the natural scale suppressing derivatives in the stress tensor, and therefore appears in the denominator of several terms ($\knl$ is inserted just to make the remaining parameters dimensionless, its numerical value has no physical consequences: it can be changed by a redefinition of the parameters $c_s,c_1,c_4$).

In Eq.~\eqref{eq:resum_P_EFT_2loop}, we fix $\zeta=3$ (corresponding to a specific initial assumption about the time-dependence of the free parameters) according to the discussion presented in~\cite{Foreman:2015uva}. Similarly to~\cite{Foreman:2015lca}, in this work we express the EFTofLSS parameters $c_{s(1)}^2$, $c_1$ in units of $(k_\text{NL}/2 \, h\text{Mpc}^{-1})^2$ and $c_4$ in units of $(k_\text{NL}/2 \, h\text{Mpc}^{-1})^4$, and determine $c_{s(2)}^2$ by matching the one-loop and two-loop EFTofLSS matter power spectrum at the renormalization scale $k_\text{ren} = 0.02\invMpc$. 
Using this matching procedure at such a low value of $k_\text{ren}$ fixes $\ct$ to cancel most of the contribution of the two-loop terms that is degenerate with $k^2 P_{11}$, since at low $k$ this is the dominant contribution from those terms.

We evaluate $P_{11}$ with the Boltzmann code \texttt{CLASS}~\cite{Blas:2011rf}, which within \texttt{TaylorEFT} runs through the designated module \texttt{CallCLASS[\dots]}. In Eq.~\eqref{eq:resum_P_EFT_2loop} we left implicit  both the cosmology dependence and the redshift evolution of the parameters incorporating the nonlinear physics. Below, we investigate these features using \texttt{TaylorEFT} in a neighborhood of our reference cosmology, i.e.\ for cosmologies with $| \Delta \theta_i | \leqslant 3\sigma_i$, where $\sigma_i$ is the square root of the diagonal element $\Sigma_{5,ii}$ of the covariance matrix in Eq.~\eqref{eq:planck_cov}.

\subsection{Cosmology dependence of EFT parameters}

Since  the cosmology-dependence of the EFTofLSS parameters is not predicted by the theory, we opt once again for a Taylor expansion approach that can capture the cosmology-dependence for relatively small deviations around our reference cosmology. This requires us to find the derivatives of the EFTofLSS parameters with respect to different cosmological parameters. We do so by fitting Eq.~\eqref{eq:resum_P_EFT_2loop} to \texttt{FrankenEmu} output power spectra for an ensemble of cosmologies (the details of the fits are described below), and then finding the required derivatives by fitting the Taylor expansion formula to the ensemble of fitted EFT parameters. The ensemble of cosmologies is composed of points ${\pmb \theta}$ in the parameter space associated with the nodes of a 5-dimensional cubic lattice centered at our reference cosmology, with lattice spacing $\sigma_i$ in each direction. (We also apply the additional constraint that $| \Delta \theta_i | \leqslant 3\sigma_i$.)

By testing Taylor expansions evaluated at different orders, we find that the following form of the expansion (up to third order in all cosmological parameters, plus up to fourth order in $\sigma_8$, due to its strong effect on the EFT parameters) is sufficient to guarantee 1\% precision on the final prediction for the power spectrum:
\begin{align} \label{eq:EFTparams_pred}
c_X({\pmb\theta},z) \approx c_X|_{\pmb\theta^{\text{ref}}}
	&+ \sum_i \Delta\theta_i \left. \frac{\partial c_X}{\partial \theta_i} 
	 \right|_{\pmb\theta = \pmb\theta^{\rm ref}}
	+ \frac{1}{2} \sum_{i,j} \Delta\theta_i \Delta\theta_j 
	\left. \frac{\partial^2 c_X}{\partial \theta_i \, \partial \theta_j} 
		\right|_{\pmb\theta = \pmb\theta^{\rm ref}} \nonumber \\
	&+\frac{1}{6} \sum_{i,j,k} \Delta\theta_i \Delta\theta_j \Delta\theta_k 
	\left. \frac{\partial^3 c_X}{\partial \theta_i \, \partial \theta_j \, \partial \theta_k} 
		\right|_{\pmb\theta = \pmb\theta^{\rm ref}} 
	+ \frac{1}{24}  (\Delta\sigma_8)^4
	\left. \frac{\partial^4 c_X}{\partial \sigma_8^4} 
		\right|_{\pmb\theta = \pmb\theta^{\rm ref}}\ ,
\end{align}
where $c_X$ is a placeholder for $c_{s(1)}^2$, $c_1$ or $c_4$. Note that we find it more convenient to use $\sigma_8$ in place of $A_s$ in this expansion.

We now describe the details of the fits that are used to fix the derivatives in Eq.~\eqref{eq:EFTparams_pred}. At all redshifts listed in Tab.~\ref{tab:redshift_kfit}, for each cosmology in our catalogue we extract the EFTofLSS parameters by fitting Eq.~\eqref{eq:resum_P_EFT_2loop} to the nonlinear matter power spectrum output of \texttt{FrankenEmu} up to a wavenumber $k_\text{fit}$ (also listed in Tab.~\ref{tab:redshift_kfit}) defined as the maximum wavenumber for which the parameters $c_X$ have stable values. Pushing this upper bound to increasingly larger wavenumbers results in unexpected statistical changes of the best fit parameters compensating for the omitted higher order terms   (see~\cite{Foreman:2015lca} for more details). In particular, we derive the best fits and covariances for these parameters by minimizing the chi-square function
\begin{equation} \label{eq:chi_square}
\chi^2(\mathbf{c}) = \sum_i \left[ \frac{P_\text{NL}(k_i) - P_\text{EFT-2-loop}(k_i, \mathbf{c})}{\sigma_{P}(k_i)} \right]^2,
\end{equation}
where we have explicitly included the dependence of the EFTofLSS predictions on the parameter vector $\mathbf{c} = (c_{s(1)}^2,c_1,c_4)$, and where $P_\text{NL}(k_i)$ is the emulator nonlinear power spectrum for which we assume 1\% uncorrelated gaussian errors, i.e.~$\sigma_{P}(k_i) = 0.01 \times P_\text{NL}(k_i)$, with wavenumbers $k_i$ sampled directly from the emulator's output \cite{Heitmann:2013bra}. To find the best-fit $\mathbf{c}_{(0)}$ we resort to the Levenberg-Marquardt method (see e.g.~\cite{Press:2007:NRE:1403886}), which approximates Eq.~\eqref{eq:chi_square} with a quadratic form, such that around the best fit one has
\begin{equation} \label{eq:chi_square_approx}
\Delta \chi^2 \equiv \chi^2 - \chi^2_\text{min} \approx \delta\mathbf{c} \cdot \mathcal{C}^{-1} \cdot \delta\mathbf{c},
\end{equation}
with $\mathcal{C}$ denoting the covariance matrix of the EFTofLSS parameters, and $\delta\mathbf{c} \equiv \mathbf{c} - \mathbf{c}_{(0)}$. We have validated the accuracy of this simplified approach for a sample cosmology by comparing with a full likelihood analysis (taking $-2\ln \mathcal{L} \equiv \chi^2$). 

For all redshifts in Tab.~\ref{tab:redshift_kfit}, we then fit Eq.~\eqref{eq:EFTparams_pred} to the corresponding parameter extracted with the method outlined above.
Coefficients at redshifts different than those listed in Tab.~\ref{tab:redshift_kfit} are evaluated through interpolation. 
We have implemented Eq.~\eqref{eq:EFTparams_pred} in \texttt{TaylorEFT} for $0 \leqslant z \leqslant 1$, and the user can call it through the \texttt{EFTParamsFit[\dots]} module. 

\begin{table}[tp]
\caption{Redshifts and corresponding maximum wavenumbers used in the fits of Eq.~\eqref{eq:EFTparams_pred} to the emulator power spectra. We neglect any potential variation of $k_\text{fit}$ with cosmology (expected to be small), and for each redshift we adopt the values obtained in~\cite{Foreman:2015lca}.}
\begin{center}
\begin{tabular}{| >{\centering\arraybackslash} m{1.5cm} | >{\centering\arraybackslash} m{2.5cm} |}
\hline 
$z$		&	$k_\text{fit}$ [$h\,\text{Mpc}^{-1}$] \\ 
\hline
0		&	0.33		\\
0.05   	&	0.34		\\
0.11 		&	0.36		\\
0.25   	&	0.38		\\
0.38 	 	&	0.42		\\
0.5  		&	0.46		\\
0.66  	&	0.52		\\
0.85  	&	0.52		\\
1  		&	0.52		\\
\hline

\end{tabular}
\end{center}
\label{tab:redshift_kfit}
\end{table}

\begin{figure}[t]
\begin{center}
\includegraphics[width=0.49\columnwidth]{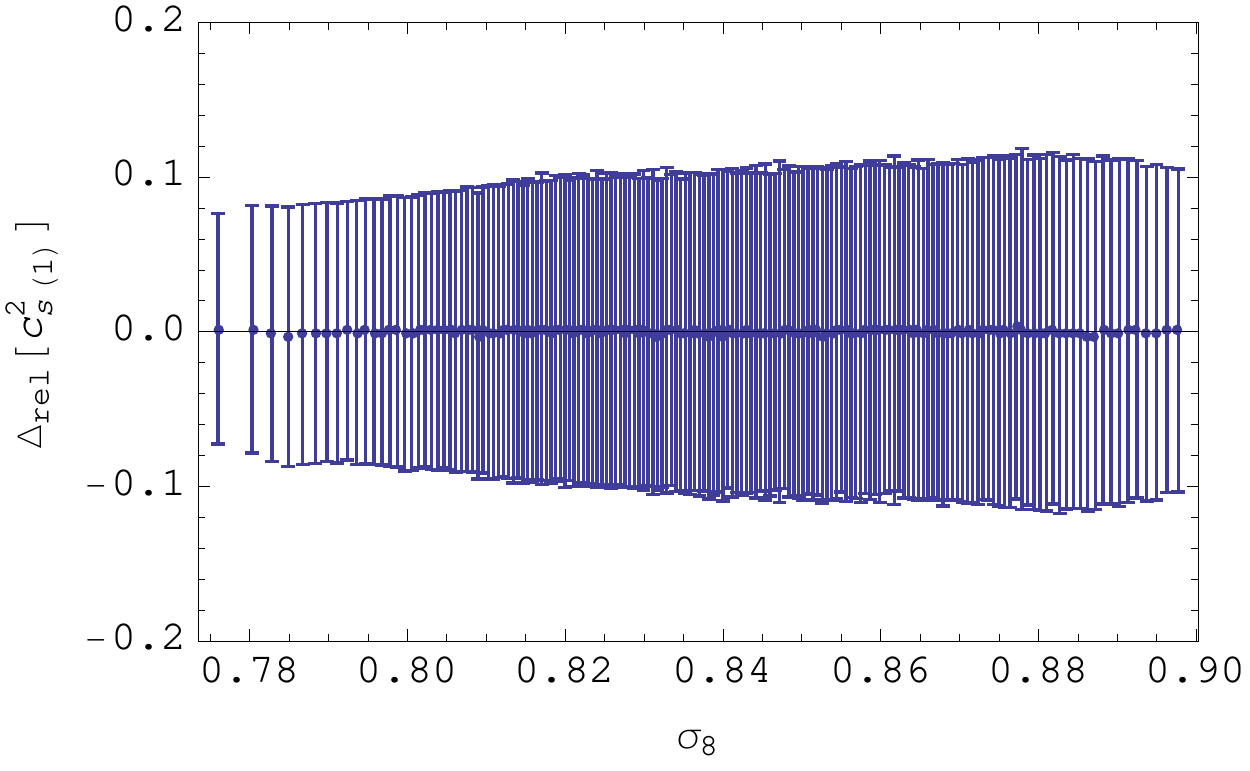} \\
\includegraphics[width=0.49\columnwidth]{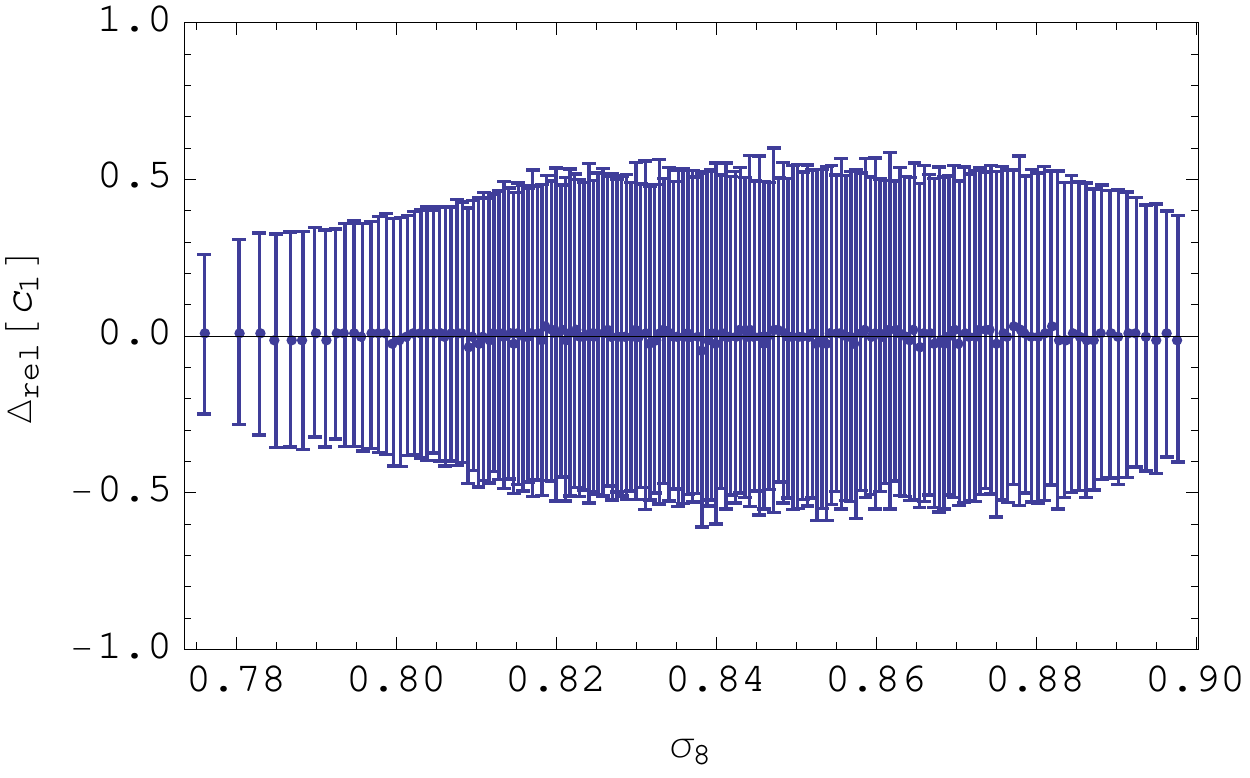} 
\includegraphics[width=0.49\columnwidth]{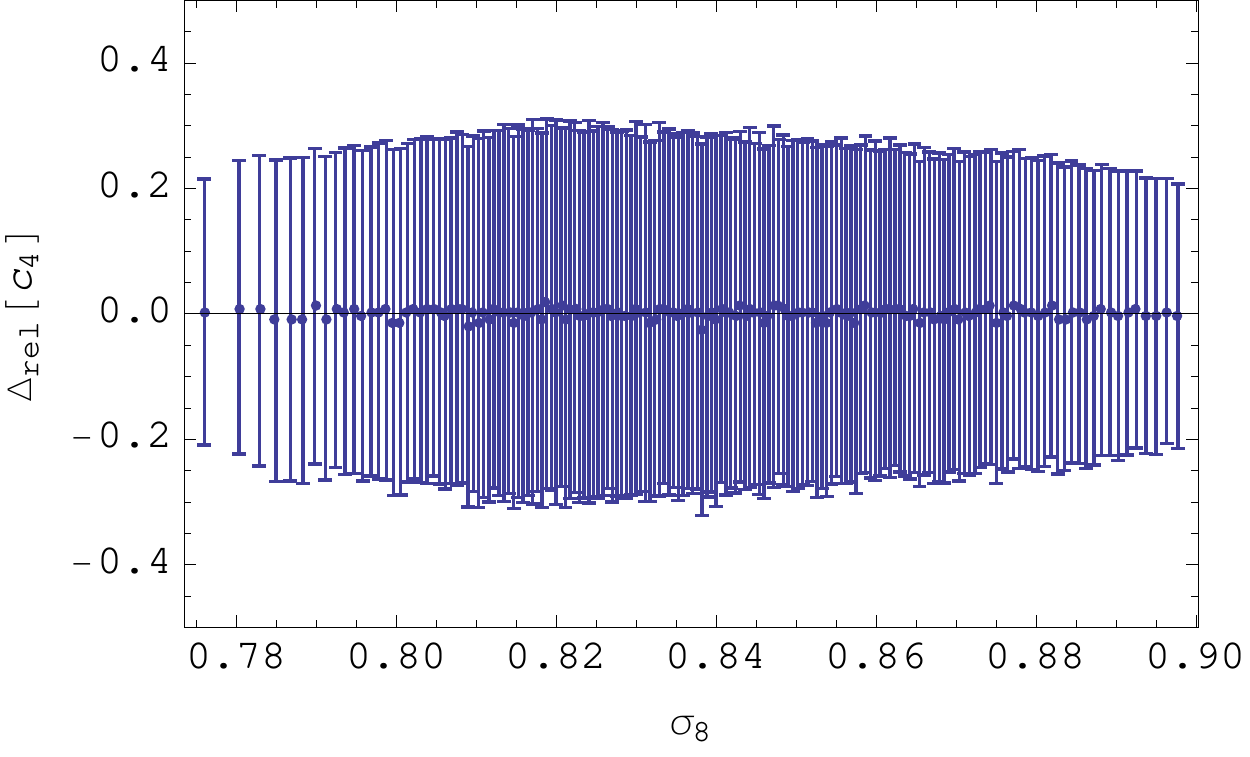} \\
\caption{Relative difference between the actual fit matching \texttt{TaylorEFT} to \texttt{FrankenEmu} and the expansion Eq.~\eqref{eq:EFTparams_pred} for the EFTofLSS parameters at $z=0$ as a function of $\sigma_8$. For clarity, each data point represents the average over 100 cosmologies sorted by $\sigma_8$, and each of them individually with cosmological parameters satisfying $| \Delta \theta_i | \leqslant 3\sigma_i$. We also average the marginalized errors assuming them independent. So the error from the mismatch between the parameters obtained directly from \texttt{TaylorEFT} and from using~\eqref{eq:EFTparams_pred}  is much smaller than the uncertainty from the fit the the numerical data.}
\label{fig:EFTparams_prediction}
\end{center}
\end{figure}

After fixing the Taylor coefficients in this way, Eq.~\eqref{eq:EFTparams_pred} reproduces the values of $\co$, $c_1$, and $c_4$ obtained from the exact fits to within 3\% for all test cosmologies,  well within the marginalized uncertainties on the exact fits: on average, $\sim$10\% for $\co$, $\sim$40\% for $c_1$, and $\sim$20\% for $c_4$ (see Fig.~\ref{fig:EFTparams_prediction}).
However, these marginalized uncertainties do not tell the whole story: it is important to account for the correlations between the EFTofLSS parameters in an assessment of Eq.~\eqref{eq:EFTparams_pred}. Using {\codename} to evaluate the loop integrals in Eq.~\eqref{eq:resum_P_EFT_2loop}, we estimate the correlation matrices (which are mostly cosmology-independent) at $z=0$ and $z=0.93$ to be
\begin{equation}
  \varrho^{(0)} = 
    \bordermatrix{ 
    	  	      & c_{s(1)}^2 & c_1 & c_4 \cr
      c_{s(1)}^2 & 1       & -0.86 & -0.80 \cr
      c_1 	      & -0.86 &    1    & 0.99 \cr
      c_4 	      & -0.80 & 0.99  & 1 },
~
  \varrho^{(0.93)} = 
    \bordermatrix{ 
    	  	      & c_{s(1)}^2 & c_1 & c_4 \cr
      c_{s(1)}^2 & 1       & -0.94 & -0.88 \cr
      c_1 	      & -0.94 &    1    & 0.99 \cr
      c_4 	      & -0.88 & 0.99  & 1 },
\end{equation}
where we have used the relation $\varrho_{ij}=\mathcal{C}_{ij}/\sigma_i\sigma_j$. Evidently, the EFTofLSS parameters display substantial degeneracies, that evolve only mildly with redshift. As discussed in~\cite{Foreman:2015lca}, this degeneracy can be expected based on theory grounds.

\begin{figure}[t]
\begin{center}
\includegraphics[width=\columnwidth]{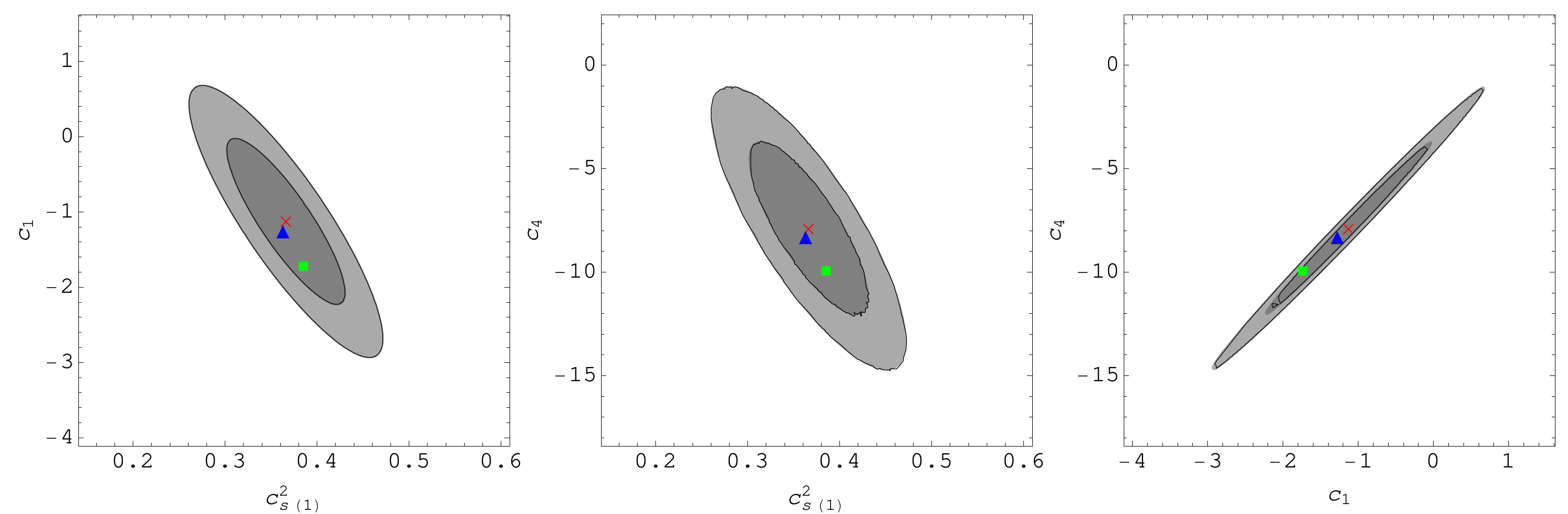}
\includegraphics[width=\columnwidth]{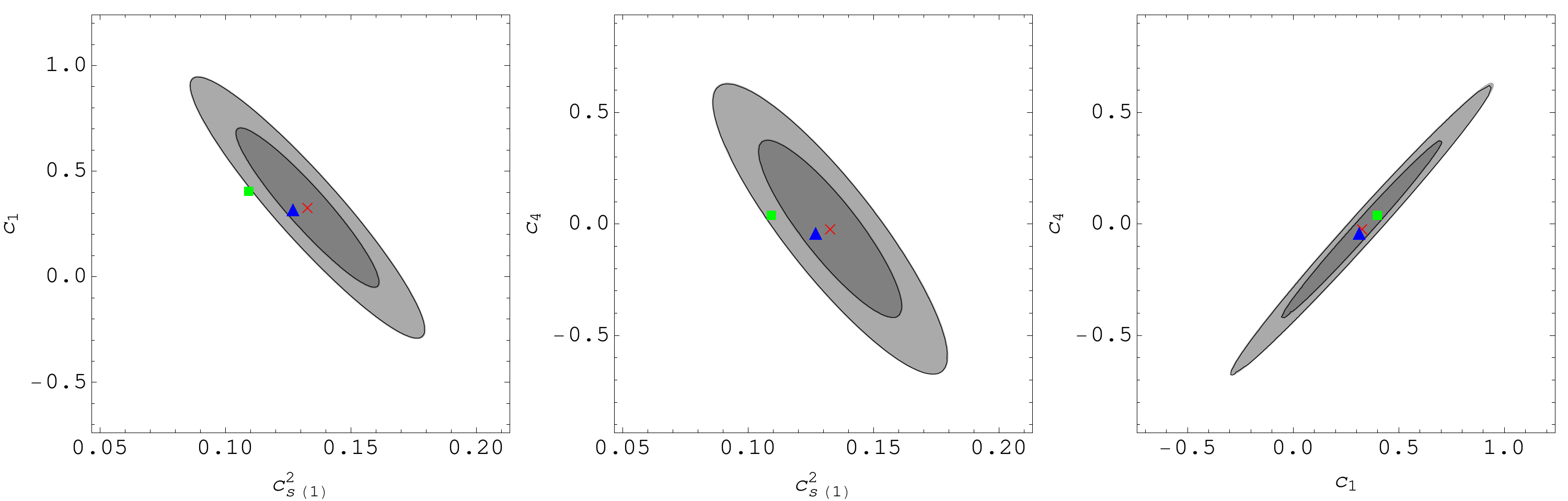}
\caption{{Levenberg-Marquardt 68.3\% (dark shadings) and 95.4\% (light shadings) confidence regions for \texttt{cosmo\_10} EFTofLSS parameters at $z=0$ (top panels) and $z=0.93$ (bottom panels). Also shown are the best fit values using {\codename} (red crosses) or \texttt{TaylorEFT} loop integrals (blue triangles), and the values given by Eq.~\eqref{eq:EFTparams_pred} (green squares). Black contours are derived from a full $\chi^2$ analysis. We find that the differences between direct fits using {\codename} or \texttt{TaylorEFT} output are much smaller than the differences from using Eq.~\eqref{eq:EFTparams_pred}, 
which can be as large as $\sim$2$\sigma$ for interpolated redshifts.}
} 
\label{fig:contours_cosmo10}
\end{center}
\end{figure}

In principle, there could be three sources of error in the parameter values, each of which should be compared to the confidence regions defined by the the full covariance matrix: (i) remainders in the Taylor expansion for $P_\alpha(k)$ in Eq.~\eqref{eq:taylor_exp}, (ii) residuals around the Taylor expansion for the EFT parameters in Eq.~\eqref{eq:EFTparams_pred}, and (iii) redshift interpolation errors in its coefficients. We have performed this comparison for a cosmology from our catalogue (\texttt{cosmo\_10}) that maximally departs from our reference cosmology, i.e.\ $|\Delta \theta_i| = 3\sigma_i$. Explicitly, we computed its parameters using three different methods: fitting Eq.~\eqref{eq:resum_P_EFT_2loop} to \texttt{FrankenEmu} using either {\codename} or \texttt{TaylorEFT} loop integrals, or directly using Eq.~\eqref{eq:EFTparams_pred}. Fig.~\ref{fig:contours_cosmo10} plots the values obtained from each method along with the Levenberg-Marquardt 68.3\% and 95.4\% confidence regions (grey shading) and the regions from the full likelihood analysis (black lines).

Differences between {\codename} and \texttt{TaylorEFT} best fitting values in this figure are due solely to (i), while differences between the {\codename} best fits and values from Eq.~\eqref{eq:EFTparams_pred} in the top panels of this figure are caused by both (i) and (ii). These differences are all negligible compared to the size of the $1\sigma$ confidence region. However, point (iii) can induce a sizable bias in the value of one or more of the $c_X$ parameters, as shown in the bottom panels of Fig.~\ref{fig:contours_cosmo10} for $\co$~(\footnote{Note that for contour plots of parameter constraints we adopt the common definitions for the 68.3\% and 95.4\% confidence regions derived from the $\chi^2$ distribution with two degrees of freedom as the surfaces enclosed by the boundaries $\Delta \chi^2 = 2.30$ and $\Delta \chi^2 = 6.17$, respectively.}), which is an interpolated redshift. {(Of course, by increasing the precision of the numerical data, the number of redshifts, and the order of the Taylor expansion, one expects this offset to be decreased)}. In Sec.~\ref{sec:comparison_data}, we will investigate the impact of this bias on the performance of $P_\text{EFT-2-loop}$ when compared to simulation data.  

\begin{figure}[t]
\begin{center}
\includegraphics[width=0.49\columnwidth]{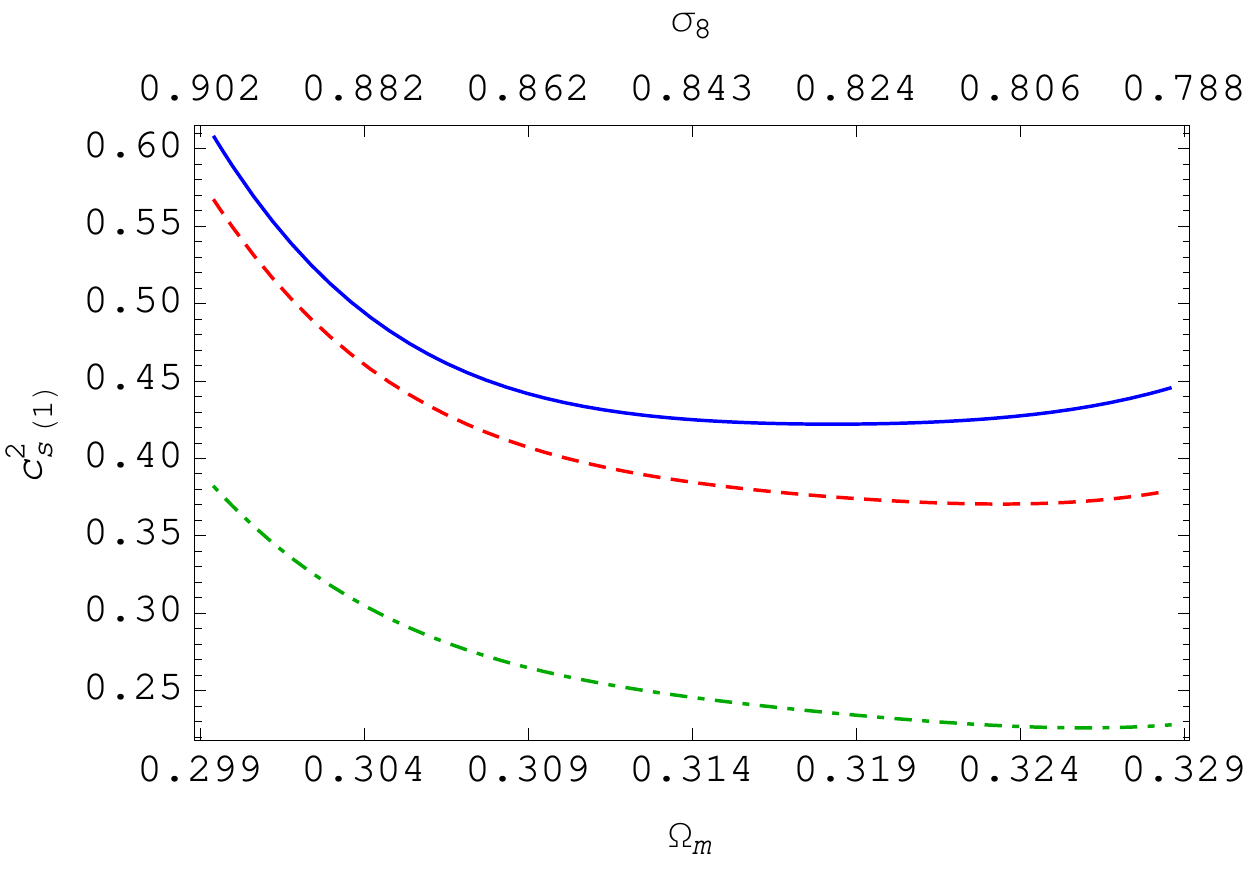} \\
\includegraphics[width=0.49\columnwidth]{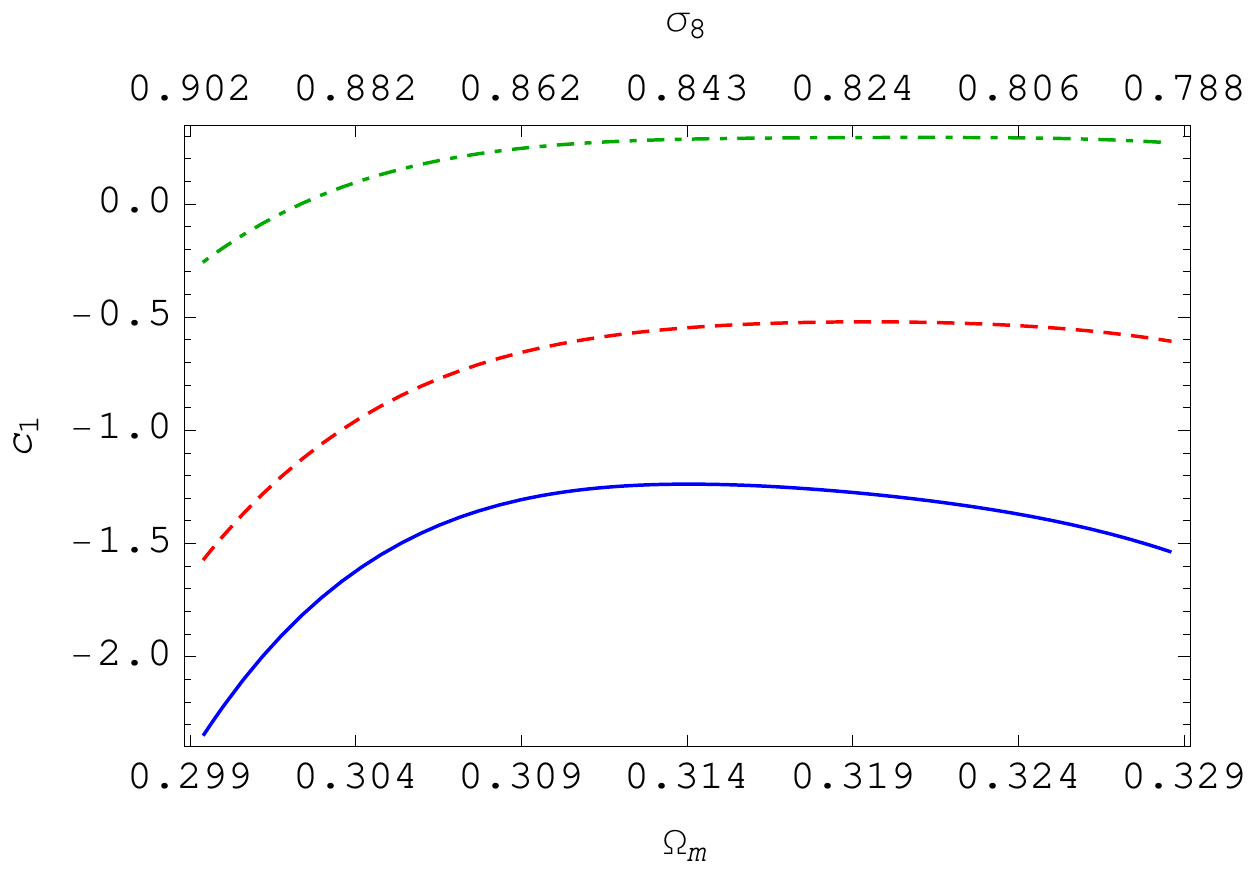} 
\includegraphics[width=0.49\columnwidth]{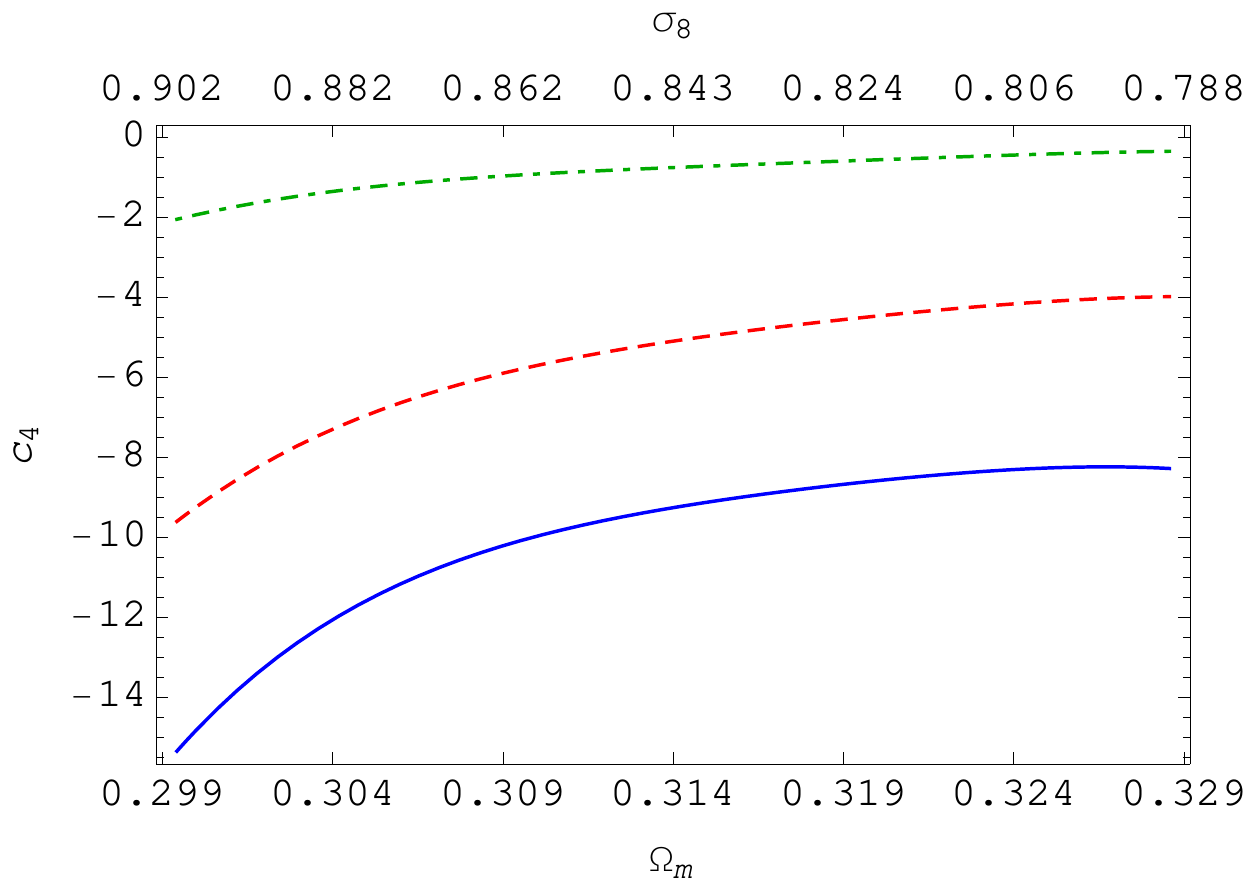} \\
\end{center}
\caption{Cosmology dependence of the EFTofLSS parameters at $z=0$ (solid blue), $z=0.11$ (dashed red) and $z=0.5$ (dash-dotted green). We show cosmologies that lie on the maximum degeneracy axis of the ellipsoid Eq.~\eqref{eq:hyperellipsoid}, up to the edge of the $p=0.997$ region. In each panel, for a given cosmology we put the corresponding $\Omega_{\rm m}$ on the bottom x-axis and $\sigma_8$ on the top x-axis. Note that absolute values of $c_{s(1)}^2$ and $c_4$ monotonically decrease with redshift regardless the cosmology, whereas $c_1$ changes sign at different redshifts for different cosmologies. The considerable relative variation of the cosmological parameters is mainly driven by the large uncertainty with which $\sigma_8$ is known.}
\label{fig:EFTparams_cosmodep}
\end{figure}

Before comparing with simulations, however, let us use Eq.~\eqref{eq:EFTparams_pred} to investigate the behavior of the EFT parameters as a function of redshift and cosmology, keeping in mind the possible bias in $c_X$ at interpolated redshifts. For this, we select the cosmologies corresponding to the maximum degeneracy axis of the ellipsoid Eq.~\eqref{eq:hyperellipsoid} with $p=0.997$. This way, all cosmological parameters vary simultaneously and monotonically, making it easier to see their impact on the EFTofLSS parameters. Furthermore, we condense the information about cosmologies in the two derived parameters, $\Omega_{\rm m}$ and $\sigma_8$, which allows us to easily present how $c_{s(1)}^2$, $c_1$ and $c_4$ change with cosmology. This cosmology dependence is shown in Fig.~\ref{fig:EFTparams_cosmodep}, at redshifts $z=0$, $z=0.11$, and $z=0.5$.  As already discussed in~\cite{Foreman:2015lca} for the cosmology used for the Dark Sky suite of simulations~\cite{Skillman:2014qca}, the absolute values of both $c_{s(1)}^2$ and $c_4$ monotonically decrease with redshift, while $c_1$ changes sign, reaches a maximum positive value and eventually approaches zero. Our new analysis suggests that the rapidity of these changes and the specific behavior of $c_1$ indeed depend on cosmology. 

\begin{figure}[t]
\begin{center}
\includegraphics[width=0.49\columnwidth]{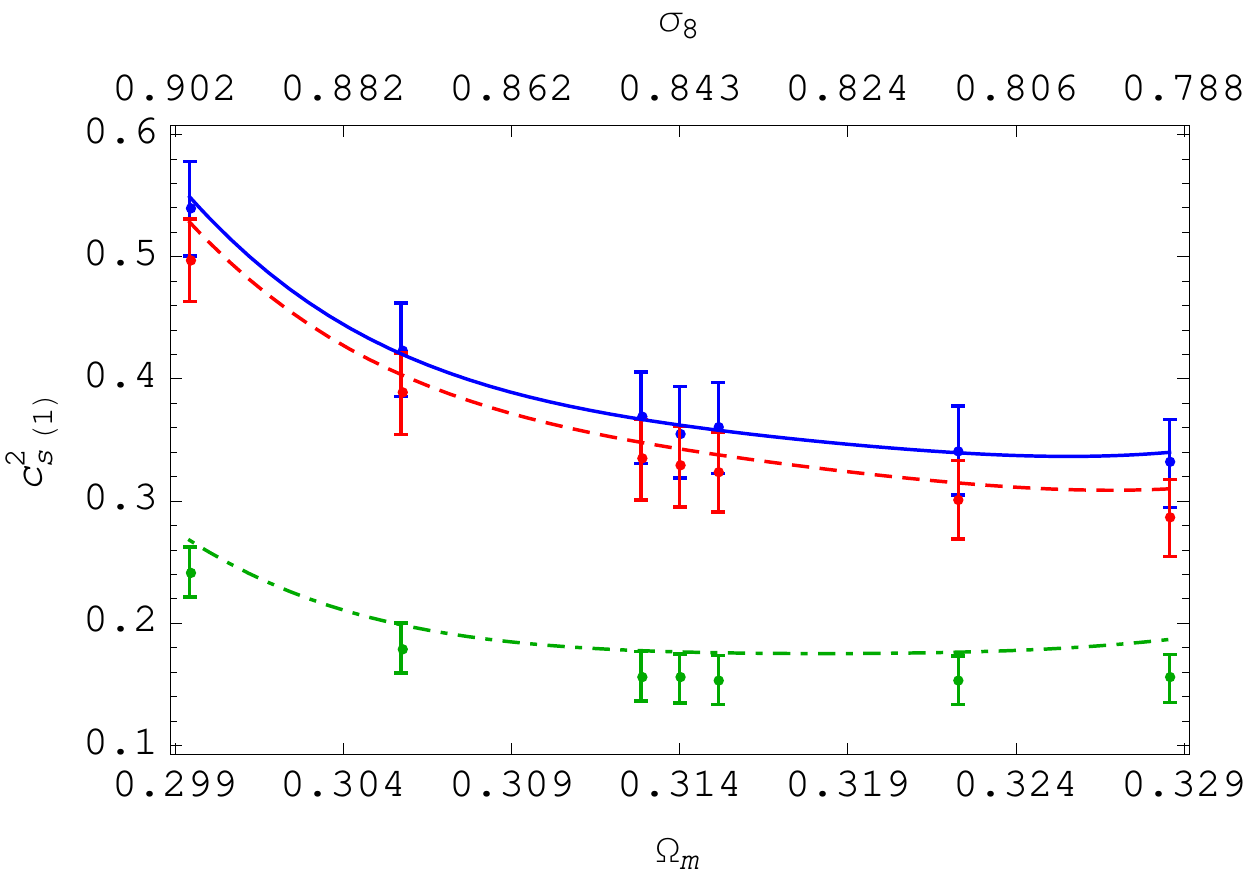} \\
\includegraphics[width=0.49\columnwidth]{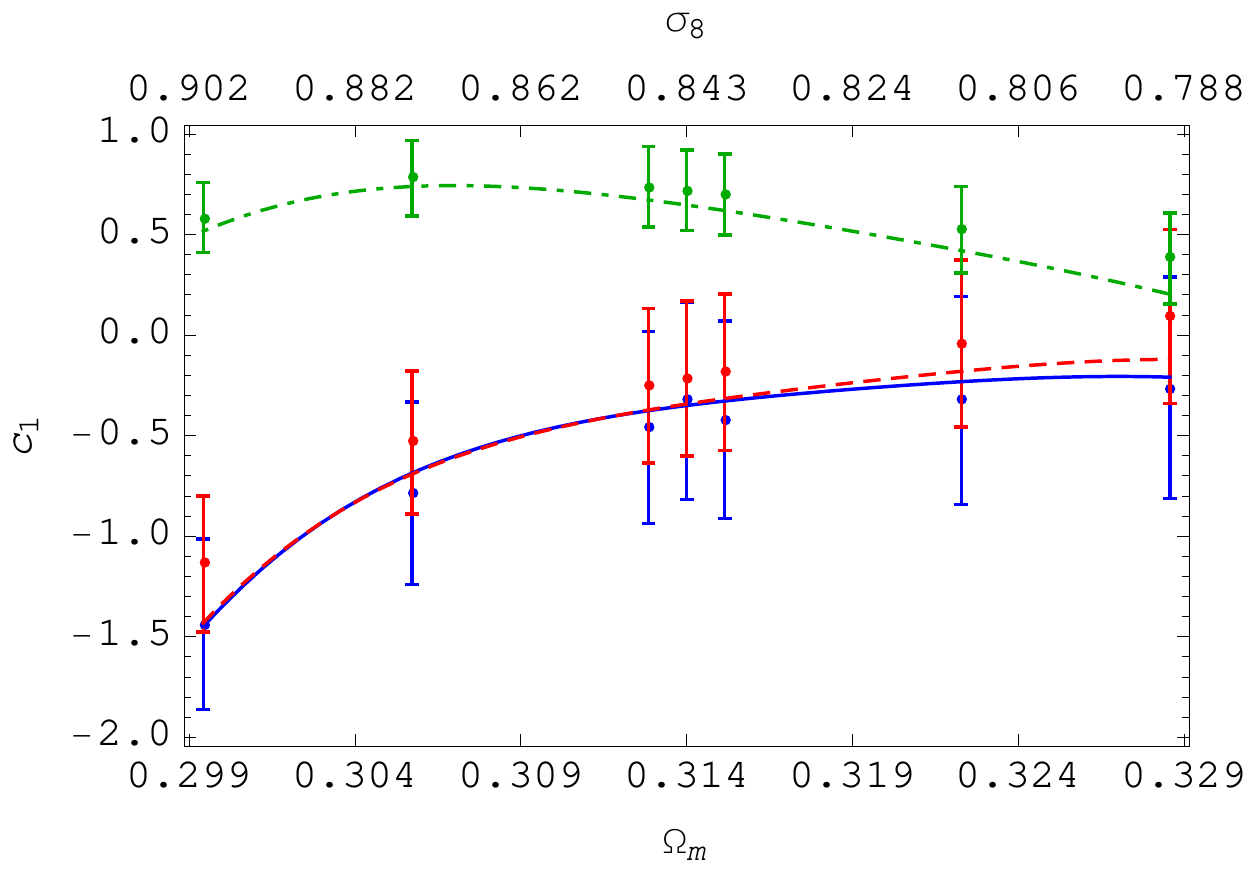} 
\includegraphics[width=0.49\columnwidth]{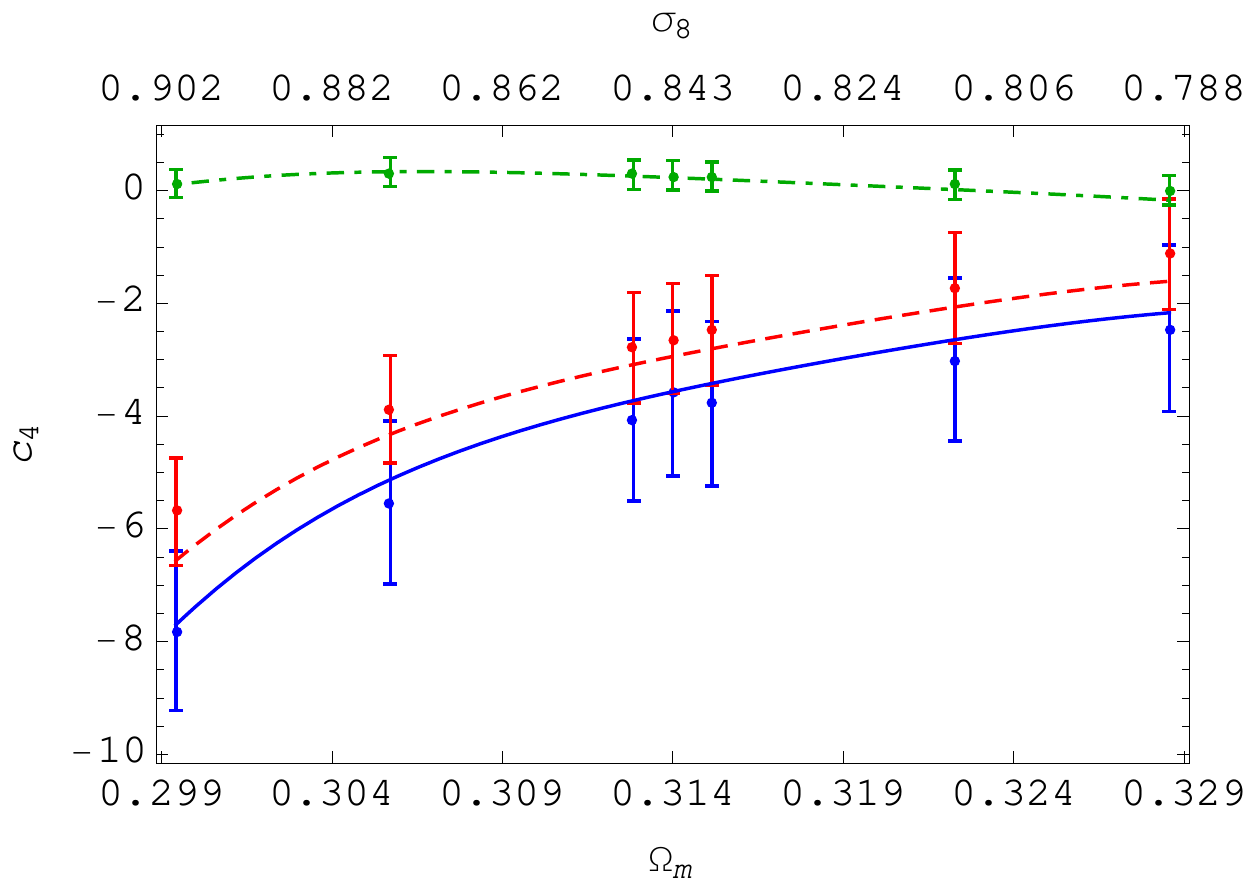} \\
\end{center}
\caption{ Crosscheck of the robustness of the Taylor expansion Eq.~\eqref{eq:EFTparams_pred} against direct fits. The selected sample of cosmologies is identical to that of Fig.~\ref{fig:EFTparams_cosmodep} with the important difference that the three redshifts showed here, $z=0.2$ (blue), $z=0.3$ (red) and $z=0.75$ (green), are not included in Table~\ref{tab:redshift_kfit}. The data points along with their associated marginalized $1\sigma$ uncertainties derive from fits to \texttt{FrankenEmu} outputs using \texttt{TaylorEFT} to compute the loop integrals, while the lines are computed from  Eq.~\eqref{eq:EFTparams_pred}. The data points located at $\Omega_{\rm m}=0.314$ represent the EFTofLSS parameters for the reference cosmology in Table~\ref{tab:Cosmo_list}, whereas $1\sigma$, $2\sigma$ and $3\sigma$ cosmologies correspond to data points progressively more distant from the $\Omega_{\rm m}=0.314$ points (e.g. the points at $\Omega_{\rm m}\approx 0.306$ and 0.322 are $2\sigma$ cosmologies).}
\label{fig:cX_Expansion_crosscheck}
\end{figure}

To further validate the performance of Eq.~\eqref{eq:EFTparams_pred} we also test it at redshifts not included in Table~\ref{tab:redshift_kfit} for $1\sigma$, $2\sigma$ as well as $3\sigma$ cosmologies. The good agreement between the Taylor expanded $c_X$ and the direct fits obtained from the comparison of the \texttt{TaylorEFT} output to \texttt{FrankenEmu} power spectra is illustrated in Fig.~\ref{fig:cX_Expansion_crosscheck}. In the future, {\codename} in tandem with higher precision cosmological simulations will enable much more detailed studies of this cosmology- and $z$-dependence.

\begin{figure}[t]
    \centering
        \includegraphics[width=0.8\textwidth]{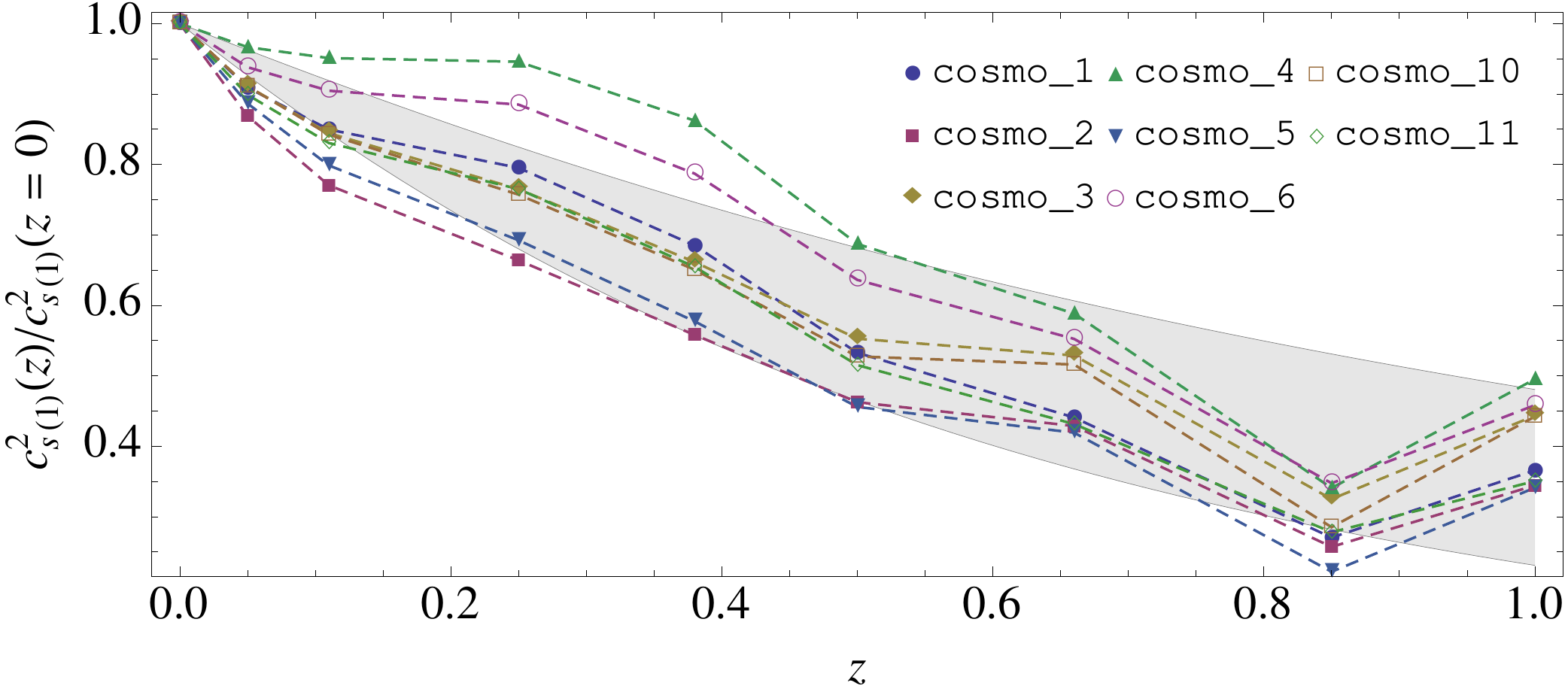}
    \caption{Redshift dependence of $\co$ for all $3\sigma$-cosmologies from Table~\ref{tab:Cosmo_list}. Each curve can be roughly approximated by a power-law in the growth factor $D(z)$; for illustration, the grey band spans the range from $D(z)^{1.5}$ to $D(z)^3$. \label{fig:cs1_z}}
\end{figure}

Fig.~\ref{fig:cs1_z} provides a more detailed picture of the redshift dependence of $\co$ for all $3\sigma$ cosmologies from Table~\ref{tab:Cosmo_list}. We find that for $z\leq 1$, $\co(z)$ scales roughly like $D(z)^p$, with $p$ ranging between 1.5 and 3 (the grey band in Fig.~\ref{fig:cs1_z}). As discussed in~\cite{Foreman:2015lca}, a sum of two power-laws of the growth factor should in principle provide a better (and more physically motivated) description of this time dependence, but we do not perform such fits here because the precision of our $\co$ measurements is limited by the precision of the emulator used for these measurements. This is because a measurement of $\co$ is essentially a measurement of the coefficient of the $k^2 P_{11}(k)$ term in Eq.~\eqref{eq:resum_P_EFT_2loop}, and a 1\% error on the total power spectrum translates into a much larger error on this coefficient ($\gtrsim 10\%$ when fitting up to the $k_{\rm max}$ we use in this work, as shown in Fig.~\ref{fig:EFTparams_prediction}).

\subsection{Comparisons with simulation data}
\label{sec:comparison_data}

\subsubsection{Emulator}
\label{sec:emulator}

\begin{figure}[t]
    \centering
    \begin{subfigure}[b]{0.45\columnwidth}
        \includegraphics[width=\columnwidth]{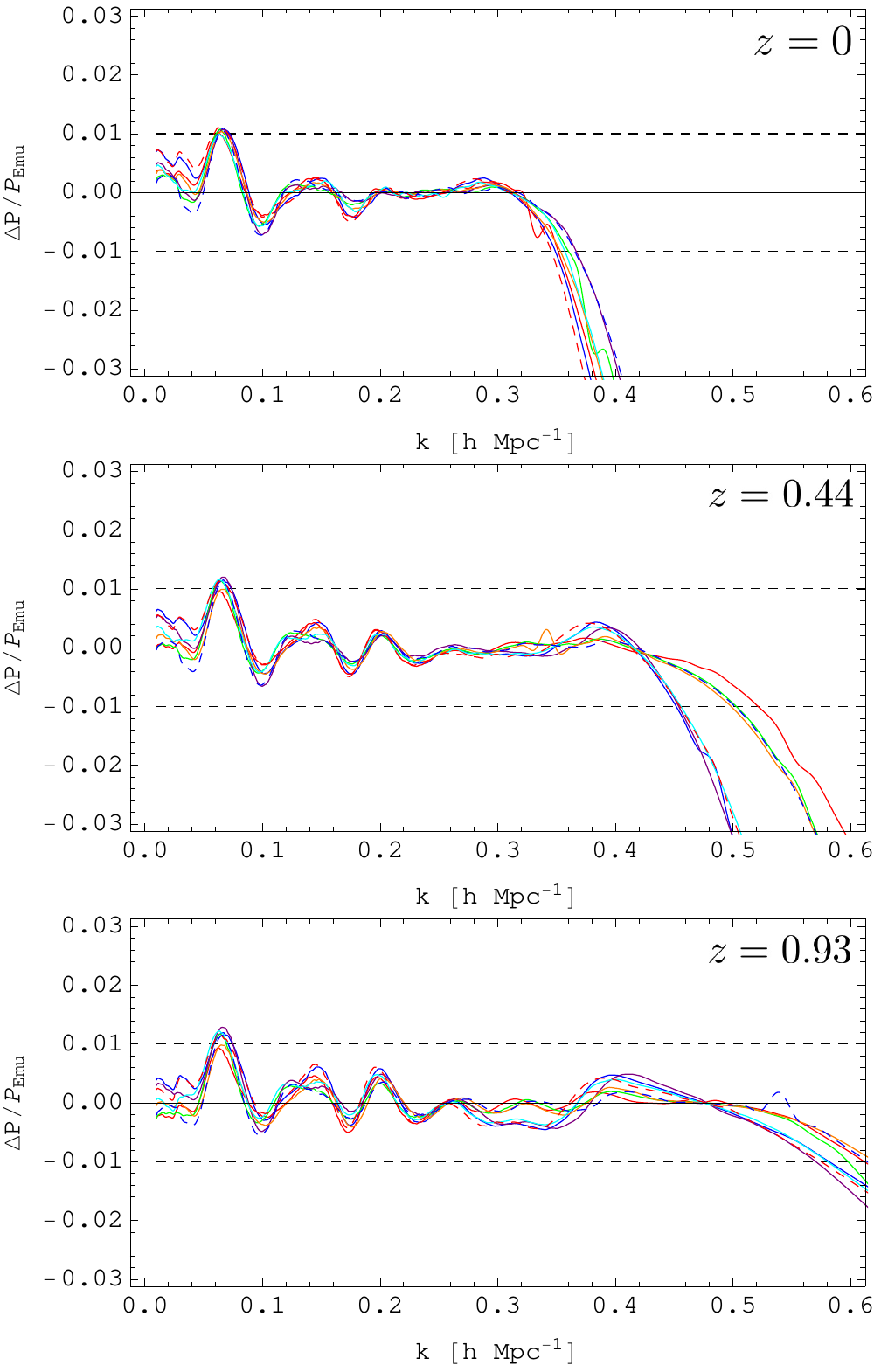}
    \end{subfigure}
    ~ 
    \begin{subfigure}[b]{0.45\columnwidth}
        \includegraphics[width=\columnwidth]{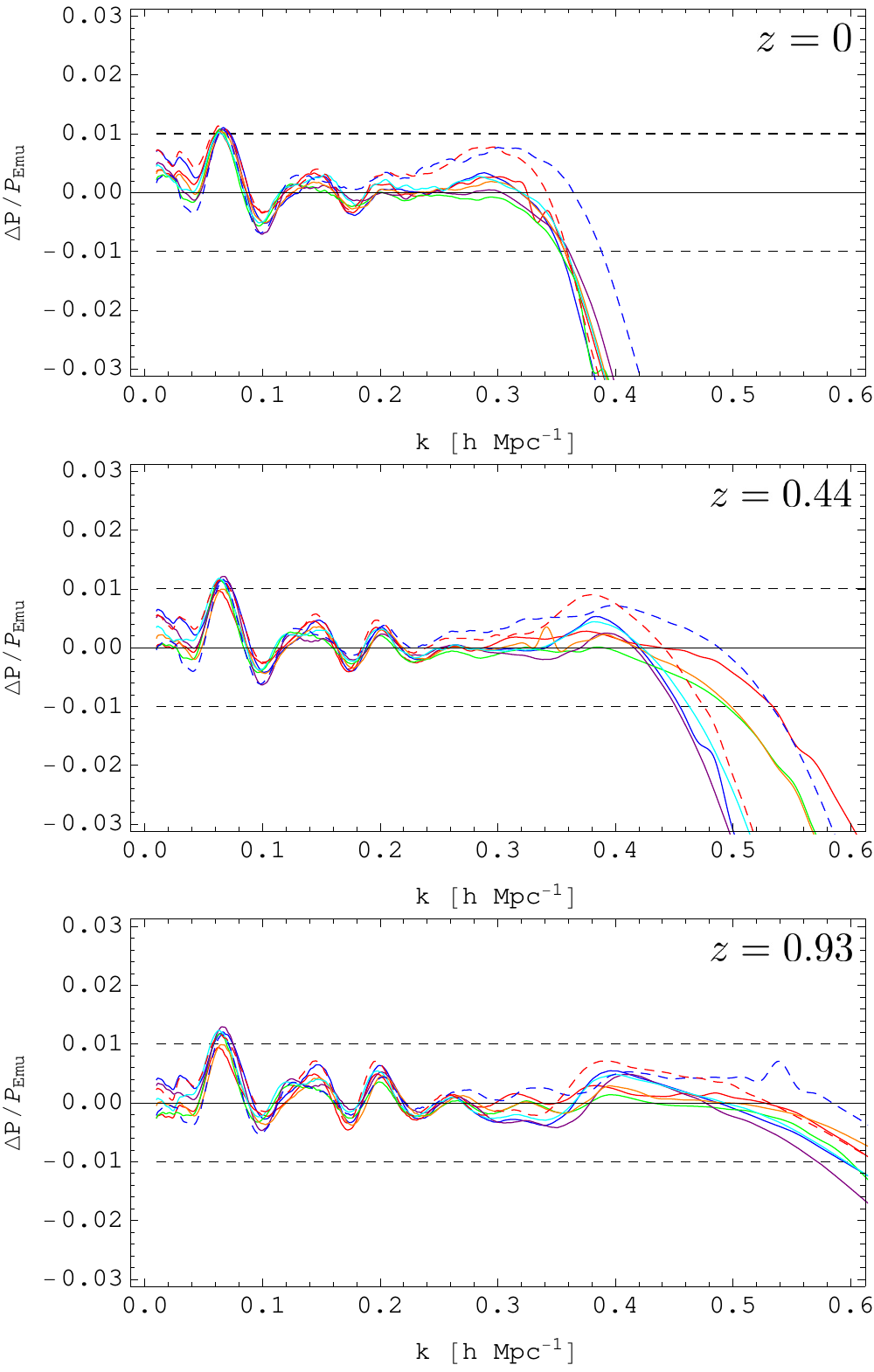}
    \end{subfigure}
    \caption{Redshift evolution of the EFTofLSS matter power spectrum predictions compared to \texttt{FrankenEmu} outputs for all of our $3\sigma$-cosmologies, \texttt{cosmo\_1-6} (in order: blue, red, green, purple, orange, cyan) and \texttt{cosmo\_10-11} (dashed blue and dashed red, respectively). Horizontal dashed lines delimit 1\% departures from the emulator power spectra. \textit{Left}: {\codename} loop integrals are used to build both the EFTofLSS power spectra and to determine the best fit EFT parameters. This column shows the excellent performance of {\codename}. \textit{Right}: predictions are constructed from {\codename} loop integrals and use best-fit EFT parameter $c_X$ values obtained with \texttt{TaylorEFT} integrals (but not using the Taylor expansion for the parameters~\eqref{eq:EFTparams_pred}). This procedure gives a sense of how the two approximations for the loop integrals affects the fitted values of the EFT parameters, and how these values affect the overall power spectrum prediction. For cosmologies with $|\Delta \theta_i| \le 3\sigma_i$, the overall power spectrum is accurate to within 1\% regardless of which loop integrals are used to determine the EFT parameters. }
\label{fig:pk_EFTexplore_BestFits}
\end{figure}

For all of our $3\sigma$-cosmologies in Tab.~\ref{tab:Cosmo_list}, we compare $P_\text{EFT-2-loop}$ to the \texttt{FrankenEmu} nonlinear power spectrum to quantify the errors introduced by the Taylor expansion of both the loop integrals and the EFTofLSS parameters discussed above. First, Fig.~\ref{fig:pk_EFTexplore_BestFits} compares the emulator output to the prediction from Eq.~\eqref{eq:resum_P_EFT_2loop}, using {\codename} to compute the loop integrals. The left panel uses values for $c_{s(1)}^2$, $c_1$ and $c_4$ that were fit to emulator output using Eq.~\eqref{eq:resum_P_EFT_2loop} with {\codename} computations, while the right panel instead uses values for $c_{s(1)}^2$, $c_1$ and $c_4$ that were obtained from fits that use \texttt{TaylorEFT} to compute the loop integrals (but not using the Taylor expansion for the parameters~\eqref{eq:EFTparams_pred}). Notably, EFTofLSS parameters extracted by means of Taylor expanded loop integrals are sufficiently close to their fiducial values obtained from {\codename} computations that the EFTofLSS power spectrum predictions are accurate within 1\% even for \texttt{cosmo\_10-11} ($|\Delta \theta_i| = 3\sigma_i$), at redshifts up to $z\sim 1$. The fits in Fig.~\ref{fig:pk_EFTexplore_BestFits} use wavenumbers up to $k_\text{fit}=0.44 \invMpc$ at $z=0.44$ and $k_\text{fit}=0.52 \invMpc$ at $z=0.93$. One should be careful in using the two-loop power spectrum results from the EFTofLSS at wavenumbers greater than roughly $k\sim 0.25h{\rm Mpc}^{-1}$ at $z=0$, and at higher wavenumbers at higher redshifts, because the theoretical errors become sizable at those scales (see~\cite{Foreman:2015lca,Baldauf:2015aha} for detailed discussions).

\begin{figure}[t]
    \centering
    \begin{subfigure}[b]{0.45\columnwidth}
        \includegraphics[width=\columnwidth]{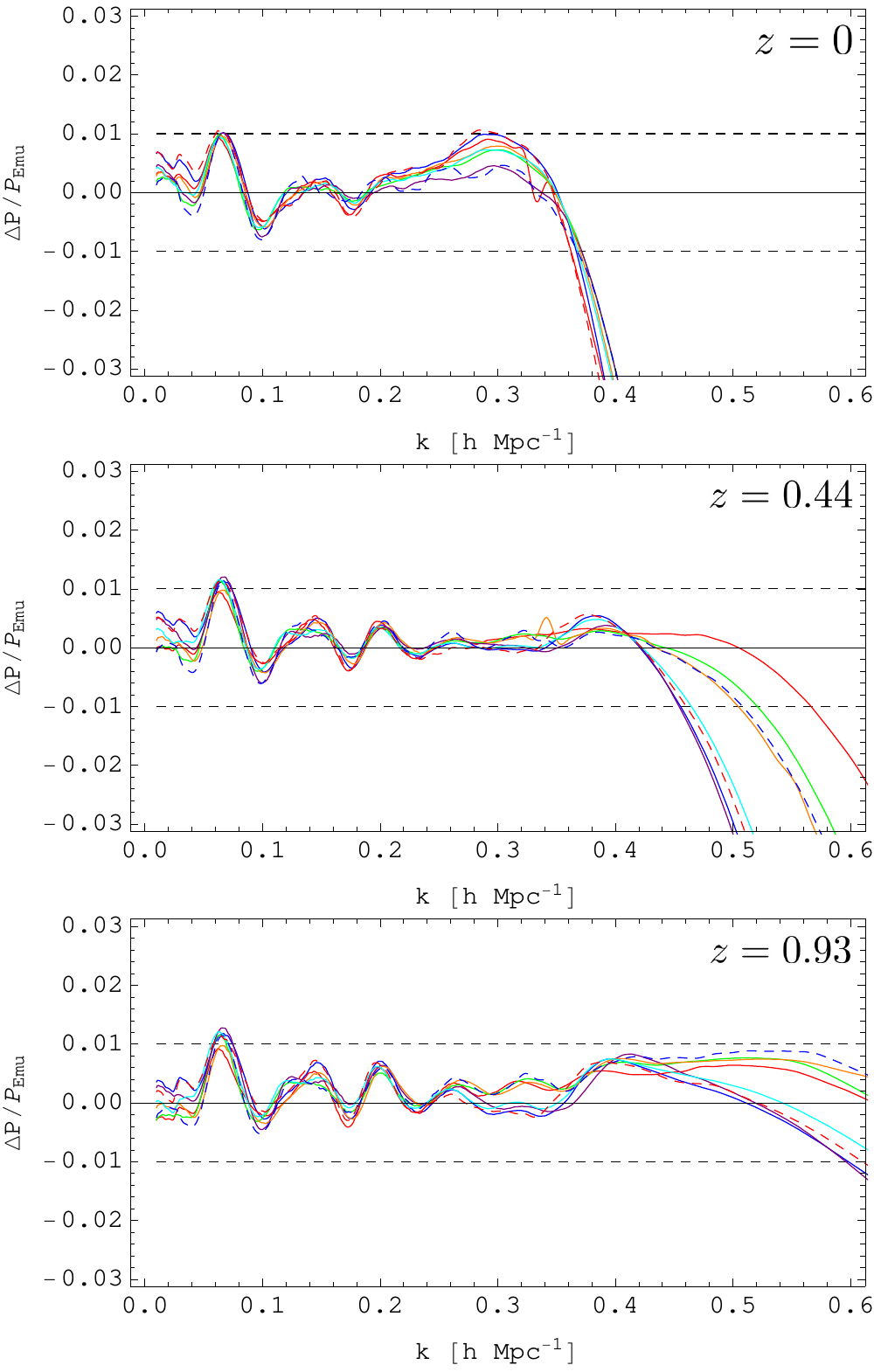}
    \end{subfigure}
    ~ 
    \begin{subfigure}[b]{0.45\columnwidth}
        \includegraphics[width=\columnwidth]{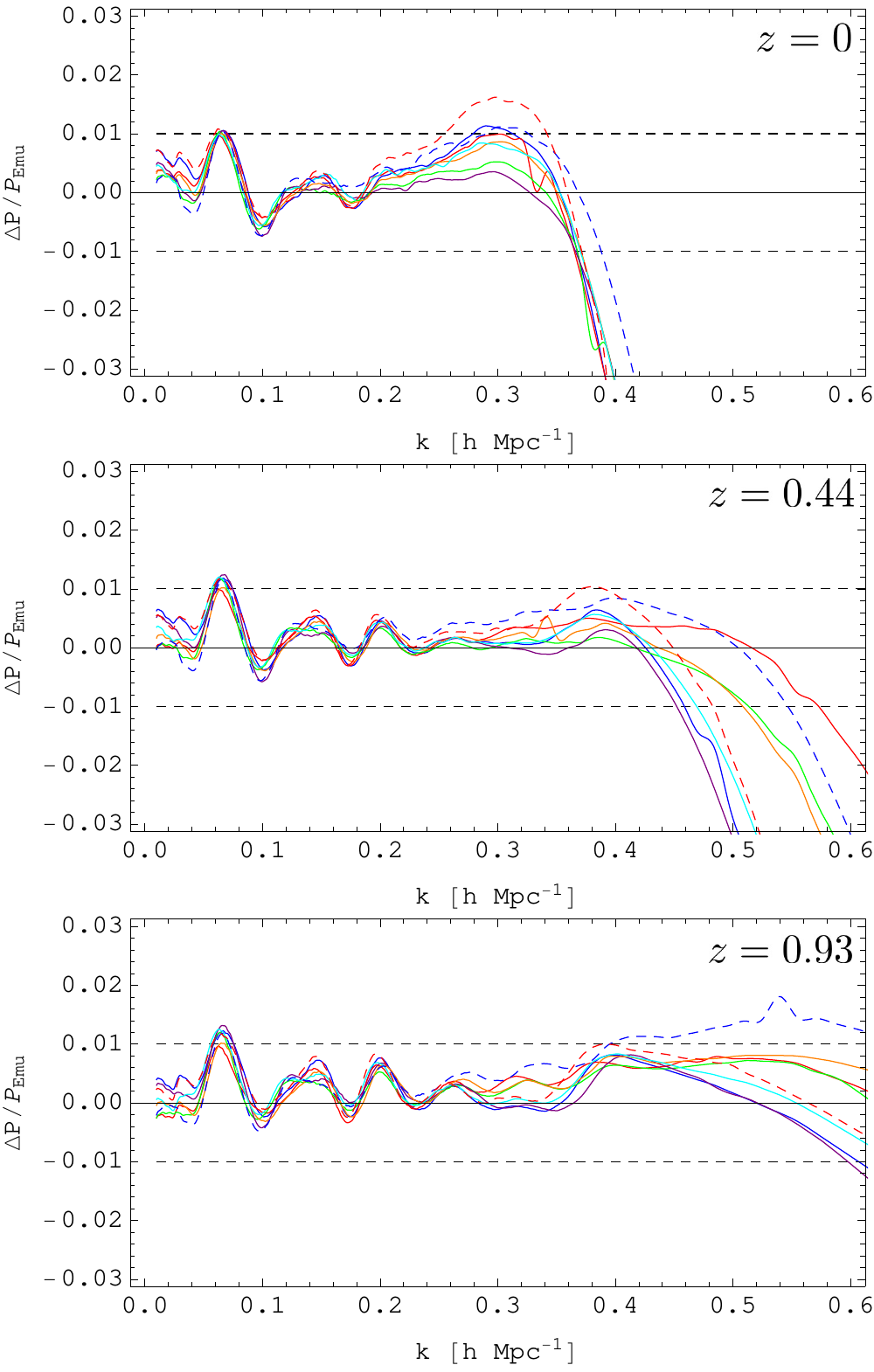}
    \end{subfigure}
    \caption{Redshift evolution of the EFTofLSS matter power spectrum predictions compared to \texttt{FrankenEmu} outputs for all of our $3\sigma$-cosmologies, \texttt{cosmo\_1-6} (in order: blue, red, green, purple, orange, cyan) and \texttt{cosmo\_10-11} (dashed blue and dashed red, respectively). Horizontal dashed lines delimit 1\% departures from the emulator power spectra.  In both columns, the EFT parameter values are estimated using Eq.~\eqref{eq:EFTparams_pred}. \textit{Left}: \texttt{TaylorEFT} is used to compute the loop integrals. This column shows the excellent performance of \texttt{TaylorEFT} combined with the predictions from \eqref{eq:EFTparams_pred}. \textit{Right}: {\codename} is used to compute the loop integrals. This procedure is clearly not the optimal one, but it gives a sense of how the parameters of the EFTofLSS obtained from \eqref{eq:EFTparams_pred} differ from those obtained from a direct fit.
        }\label{fig:pk_EFTParamsFit}
\end{figure}

In Fig.~\ref{fig:pk_EFTParamsFit}, we assess the performance of our Taylor expansion for the EFTofLSS parameters, given in Eq.~\eqref{eq:EFTparams_pred}. The left panel shows the relative differences between the emulator and the EFTofLSS two-loop power spectra constructed with the Taylor expanded loop integrals and using Eq.~\eqref{eq:EFTparams_pred} for the EFT parameters. The impact of the combination of loop integrals and $c_X$ errors on $P_\text{EFT-2-loop}$ is small enough that the EFTofLSS predictions lie within 1\% up to $k \sim k_\text{fit}$. However, using Eq.~\eqref{eq:EFTparams_pred} for the parameters but the {\codename} loop integrals in Eq.~\eqref{eq:resum_P_EFT_2loop} does not perform as well (see right panel of Fig.~\ref{fig:pk_EFTParamsFit}). This comes from the fact that the Taylor expansion of the EFTofLSS parameters has been calibrated using \texttt{TaylorEFT}, and this calibration acts to partly absorb the differences between the \texttt{TaylorEFT} and {\codename} computations of the loop integrals. Clearly, it makes more sense to use the expanded $c_X$ in the situation where the loop integrals are obtained from \texttt{TaylorEFT}~\footnote{Although this is not the recommended procedure, if one desires to do so, the expanded $c_X$ can still be used to make predictions within 2\% of the emulator power spectrum if the {\codename} loop integrals are used.}. 

Finally, we stress that it is expected that by increasing the order of the Taylor expansion both for the power spectra and for the coefficients, it should be possible to increase the accuracy of the procedure to the level of the accuracy of {\codename}. Of course the precision of the coefficients is limited by the precision of the numerical data we use to extract them. For this first release of the code, we stopped the Taylor expansion at the order given by~\eqref{eq:taylor_exp} and~\eqref{eq:EFTparams_pred}~\footnote{The reason why we can stop at second order in the Taylor expansion of $P(k)$ while we go somewhere between third and fourth order in the Taylor expansion of the EFT parameters, $c_X$'s, is simply due to the fact that when we use the Taylor expansion of $P(k)$, we let the EFT parameters be determined by the fit to the data, which allows for a partial compensation of the error in the $P(k)$ Taylor expansion. Instead, when we directly use the EFT parameters from the Taylor expansion, there is nothing left to compensate for the residual errors. This is why we go to higher order in these. Of course, intermediate procedures where one goes to cubic order both in $c_X$ and $P(k)$ could be allowed, at the cost of running more reference cosmologies (with possibly higher precision requirements).}. 

\begin{figure}[t]
\begin{center}
\includegraphics[width=\columnwidth]{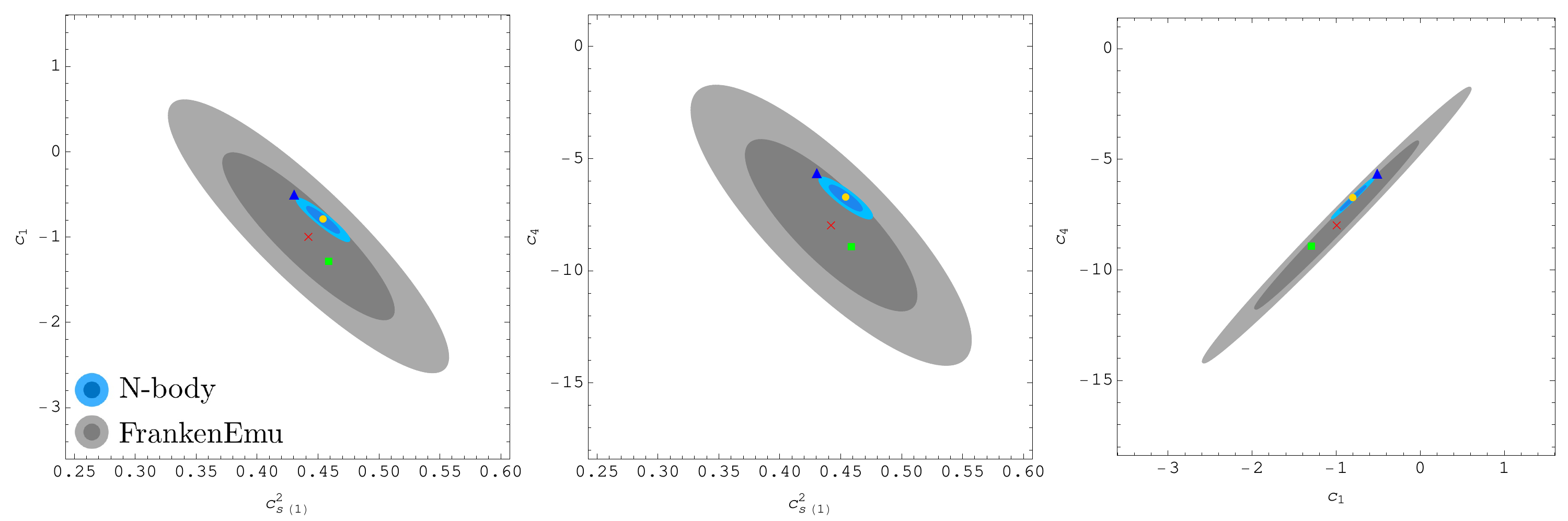}
\caption{{For the Dark Sky cosmology at $z=0$ (see main text), constraints on the EFTofLSS parameters from emulator data (gray shading) and from \texttt{ds14\_a} cosmological simulation (blue shading). Dark and light shading respectively indicate the 68.3 and 95.4\% confidence regions. Best fits obtained with {\codename} loop integrals are shown as red crosses for the emulator data and in golden circles for the simulations. Blue triangles represent best fits to the simulations derived with \texttt{TaylorEFT} loop integrals, and green squares correspond to the Taylor-expanded coefficients predictions Eq.~\eqref{eq:EFTparams_pred}.}}
\label{fig:contourts_DarkSky_z_0}
\end{center}
\end{figure}

\subsubsection{Full $N$-body}
\label{sec:darksky}

Finally, it is useful to quantify the effect of uncertainties in the nonlinear data (such as sample variance of the power spectrum measurements, or possible systematic errors in the emulator we have used for our main results) on the values of the $c_X$ parameters, and on the power spectrum predictions that make use of those values. To do so, we employ high-precision data from a dark matter-only $N$-body cosmological simulation, and compare the results to those obtained from \texttt{FrankenEmu}.
For this, we use the publicly-available redshift-zero matter power spectrum extracted from the \texttt{ds14\_a} box of the Dark Sky simulation suite~\footnote{\url{http://darksky.slac.stanford.edu}}, with $L_{\rm box}=8 \, h^{-1} \,{\rm Gpc}$, $N_{\rm part}=10240^3$, and cosmological parameters ${\pmb\theta}_{DS} = \{ 0.02214,0.11754,3.08518,0.96764,0.68806 \}$. In Fig.~\ref{fig:contourts_DarkSky_z_0} we compare the constraints on the EFTofLSS parameters obtained from fits to \texttt{FrankenEmu} with 1\% error per mode (gray shading) to those from \texttt{ds14\_a} with Gaussian errors (blue shading), where the loop integrals have been evaluated with {\codename}. Note that for these simulations, sample variance at $k_\text{fit} = 0.33 \, h\text{Mpc}^{-1}$ amounts to $\sim 0.1$-$0.2\%$. Also in this case $\epstarget=0.5\%$ along with $\sigma_\Delta=1.5 ({\rm Mpc}/h)^3$ is enough to ensure two-loop matter power spectrum predictions fall within $\sim 0.1\%$ of the data. In fact, for all of the one-loop integrals we have $\Delta P_\text{1-loop}/P_{11} \ll 0.1\%$, and the leading source of error is $\Delta P_\text{2-loop}/P_{11} \lesssim 0.1\%$ (see Sec.~\ref{sec:Difference}).
Apart from tightening the allowed region of parameter space and consistently shifting the best-fit values of the $c_X$ parameters (golden circles)\footnote{Note that the excellent agreement at $z=0$ between the emulator's output and the Dark Sky simulation illustrated in~\cite{Skillman:2014qca} would imply that the best-fit $c_X$ from {\tt FrankenEmu} are located inside the Dark Sky confidence regions. The significant offset visible in Fig.~\ref{fig:contourts_DarkSky_z_0} is due predominantly to the qualitatively different uncertainties used in Eq.~\eqref{eq:chi_square} for these two cases.}, the parameters extracted directly from the simulation power spectrum display slightly stronger degeneracies, as evident upon inspection of the following correlation matrices: 
\begin{equation}
  \varrho_{\text{sim}} = 
    \bordermatrix{ 
    	  	      & c_{s(1)}^2 & c_1 & c_4 \cr
      c_{s(1)}^2 & 1       & -0.94 & -0.89 \cr
      c_1 	      & -0.94 &    1    & 0.99 \cr
      c_4 	      & -0.89 & 0.99  & 1 },
~
  \varrho_{\text{emu}} = 
    \bordermatrix{ 
    	  	      & c_{s(1)}^2 & c_1 & c_4 \cr
      c_{s(1)}^2 & 1       & -0.89 & -0.82 \cr
      c_1 	      & -0.89 &    1    & 0.99 \cr
      c_4 	      & -0.82 & 0.99  & 1 }.
\end{equation}
We also plot the \texttt{TaylorEFT} best fit (blue triangles) resulting from the fit to the simulations and the $c_X$ predictions based on Eq.~\eqref{eq:EFTparams_pred} (green squares). Not surprisingly, the Taylor expansion of the loop integrals around our reference cosmology is good enough: the $c_X$'s are indeed shifted only mildly along the degeneracy direction. Likewise, our Taylor expanded $c_X$'s lie along the degeneracy direction of the emulator data, which have been used to calibrate that very relation. 

\begin{figure}[t]
\begin{center}
\includegraphics[width=0.7\columnwidth]{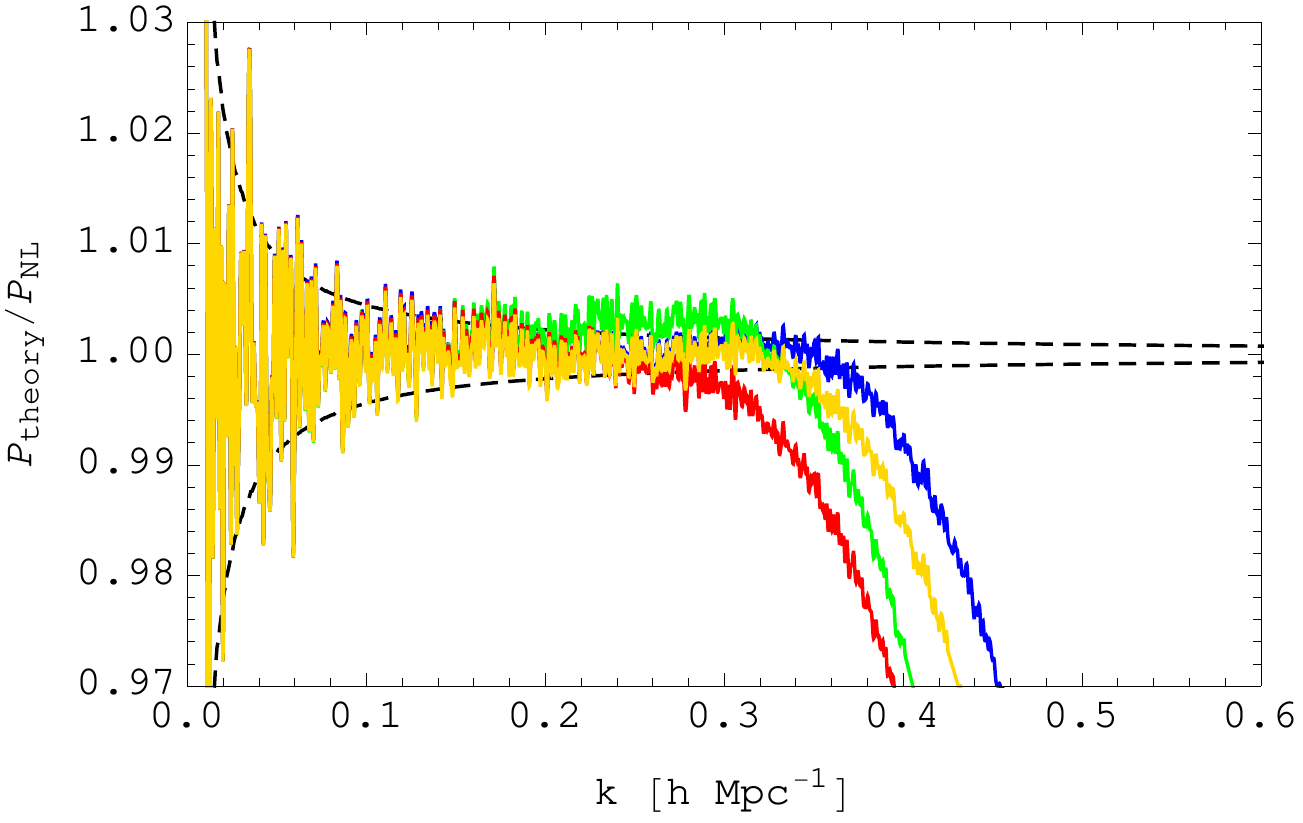}
\caption{{Comparison of two-loop EFTofLSS predictions based on {\codename} outputs and nonlinear power spectrum measured from the \texttt{ds14\_a}  Dark Sky simulation (see main text). The color coding denotes different values for the EFT parameters: best fit to \texttt{ds14\_a} power spectrum itself, using {\codename} loop integrals (gold); best fit to \texttt{ds14\_a} power spectrum itself, using \texttt{TaylorEFT} loop integrals (blue); best fit to \texttt{FrankenEmu} output for Dark Sky cosmology (red); and estimated parameter values from Eq.~\eqref{eq:EFTparams_pred} (green). Dashed lines delimit sample variance uncertainties in the simulation.}}
\label{fig:Pk_DarkSky_z_0}
\end{center}
\end{figure}

We show how the different values for the EFT parameters impact the EFTofLSS predictions in Fig.~\ref{fig:Pk_DarkSky_z_0}. We compare $P_\text{EFT-2-loop}$ to the Dark Sky power spectrum adopting the same color coding as in Fig.~\ref{fig:contourts_DarkSky_z_0} except that {\codename} integrals are used everywhere. As already shown in~\cite{Foreman:2015lca}, the direct fit to simulations produces predictions that lie within the sample variance fluctuations of the data up to $k\sim 0.34\invMpc$, and our reference {\codename} best fit (in gold) provides a similar match to the data. However, in all the other cases deviations beyond sample variance for $k < k_\text{fit}$ are the result of various inaccuracies: interpolation errors in the emulator (in red), residuals in the Taylor expanded loop integrals (in blue), and a combination of these two with other errors described above for the $c_X$ predictions (in green), which obviously cannot perform better than the red curve. In spite of this, each set of parameters corresponds to a power spectrum as close to the best answer as allowed by the method employed: at 1\% level for \texttt{FrankenEmu} and \texttt{TaylorEFT}, and within 2\% using the predicted~$c_X$ values. Finally, note that the oscillations of the theory prediction around the emulator output in Figs.~\ref{fig:pk_EFTexplore_BestFits} and~\ref{fig:pk_EFTParamsFit} are essentially absent from Fig.~\ref{fig:Pk_DarkSky_z_0}, implying that these oscillations are likely present in the emulator output itself, rather than in the EFT predictions.

Let us add two additional simple comments. In principle one could ignore simulations and match the parameters of the EFT directly to observational data (where {\codename} and \texttt{TaylorEFT} will still be valuable to obtain the functional form of the various correlation functions). Until now, large uncertainties in   observational data have hindered the pursuit of this direction. Nevertheless, this might change with planned all-sky lensing surveys, whose much smaller statistical uncertainties (and better control of systematics) could potentially have enough power to constrain the $c_X$'s at a level comparable with simulations. Of course, this will require the addition of the description of baryonic effects, as recently done in~\cite{Lewandowski:2014rca, Angulo:2015eqa}, as well as, for galaxies, of biased tracers~\cite{Angulo:2015eqa}. Moreover, a fully consistent analysis for the determination of these parameters must also account for the non-Gaussianity of the covariance matrix and potential degeneracies with cosmological parameters. On a different side, it might be interesting to compare the parameters extracted from observations to those obtained from simulations, with the purpose of reducing the parameters that are actually fit to observations, in this way minimizing the loss of information.

\section{Conclusions}
\label{sec:conclusions}

In this work, we have presented a new suite of publicly distributed codes (\codename, \texttt{ResumEFT} and \texttt{TaylorEFT}) implementing efficient algorithms to evaluate  large-scale structure observables in the EFTofLSS framework. These algorithms take advantage of the fact that we are only interested in a neighborhood of cosmological parameter space centered on a ``reference" cosmology, which we take to be the best-fit model from Planck. The main ideas of each code are as follows:
\begin{itemize}
\item Since integration is a linear operation, we can write each loop integral in perturbation theory as a sum of the integral for the reference cosmology and the integral of the {\em difference} between the integrands for the desired ``target" cosmology and the reference cosmology. If the integrals for the reference cosmology have been precomputed with high precision, then the difference integrals can be computed with much lower precision without compromising the precision of the desired result for the target cosmology. This approach has been implemented in \codename. The companion code \texttt{ResumEFT} applies the IR-resummation technique of~\cite{Senatore:2014via,Angulo:2014tfa} to the output of \codename.
\item In the same spirit, the perturbation theory integrals can be Taylor expanded in the deviation of the cosmological parameters from the reference cosmology, enabling one to simply read off the desired results from the Taylor expansion (provided that the required derivatives have been precomputed). This approach has been implemented in \texttt{TaylorEFT}.
\end{itemize}

As a proof of concept, our codes supply the two-loop IR-resummed predictions from the EFTofLSS for the dark matter power spectrum. 
These are valid up to mildly nonlinear scales ($k\lesssim 0.3\invMpc$ at redshift $z=0$, and increasingly larger wavenumbers for $z>0$), and for consistency their use should be limited to this regime. It is possible to imagine that the free parameters involved in EFTofLSS predictions can be extracted from short-distance correlations extracted from small (very roughly, $L_{\rm box} \sim 100\invMpc$) $N$-body simulations~\cite{Carrasco:2012cv}, and the cost of running these simulations and then deriving the parameters should be significantly reduced as compared to the running large-box simulations and measuring the full power spectrum on these large volumes. This is not the methodology we have adopted in this paper, where we have extracted the parameters using large volume simulations.

Our codes are based on ideas that are not specific to the matter power spectrum, and in fact they can be easily applied to other calculations of interest. It will be interesting, and in a sense essential in the light of next generation experiments, to apply these ideas to the computation of higher $N$-point matter and momentum correlation functions, following~\cite{Angulo:2014tfa,Baldauf:2014qfa,Senatore:2014via,Baldauf:2015aha,Bertolini:2015fya,Bertolini:2016bmt}, including the effects of baryons, following~\cite{Senatore:2014vja,Angulo:2015eqa}, or the correlation functions for biased tracers, following~\cite{Senatore:2014vja,Angulo:2015eqa}, or for the same quantities in redshift space, following~\cite{Senatore:2014vja,Lewandowski:2014rca}. It will also be interesting to see if similar ideas could be implemented in simulations (in the same spirit as COLA~\cite{Tassev:2013pn,Tassev:2015mia}) and in Boltzmann codes such as CMBFAST~\cite{Seljak:1996is}, CAMB~\cite{Lewis:1999bs} and CLASS~\cite{Blas:2011rf}.




\section*{Acknowledgements}

MC thanks David Rapetti for stimulating discussions.
Part of the calculations for this work utilized the Orange and Bullet computer clusters at the SLAC National Accelerator Laboratory. Further numerical computations have been performed with Wolfram $Mathematica^{\rm \tiny \textregistered}~9$. 
This work makes use of power spectrum measurements from the Dark Sky suite of simulations, and we thank the Dark Sky team for making those measurements publicly available at~\href{http://darksky.slac.stanford.edu}{http://darksky.slac.stanford.edu}.
SF is partially supported by the Natural Sciences and Engineering Research Council of Canada. 
LS is supported by DOE Early Career Award DE-FG02-12ER41854.


\appendix
\section{Alternative treatments of two-loop terms}
\label{app:detailedestimates}

\subsection{Using full version of $\ptwoloop$ instead of $\ptwoloop^\text{(UV-improved)}$}
\label{app:fullp2loop}

Instead of using the ``UV-improved'' version of $\ptwoloop$, as in~\cite{Foreman:2015lca} and Sec.~\ref{sec:DiffExplain}, we may instead use the full version of this term, as done in~\cite{Foreman:2015uva} and earlier works. As we explain below, this form of $\ptwoloop$ has the advantage of possessing fewer zero crossings than the UV-improved version. However, it also has the disadvantage that the leading UV contribution to $\ptwoloopfull$, which is absent from $\ptwoloop^\text{(UV-improved)}$, is first numerically computed and then effectively removed by a suitable choice of $\ct$, whereas $\ptwoloop^\text{(UV-improved)}$ simply does not compute this contribution in the first place.

\begin{figure}[t]
    \centering
        \includegraphics[width=0.8\columnwidth]{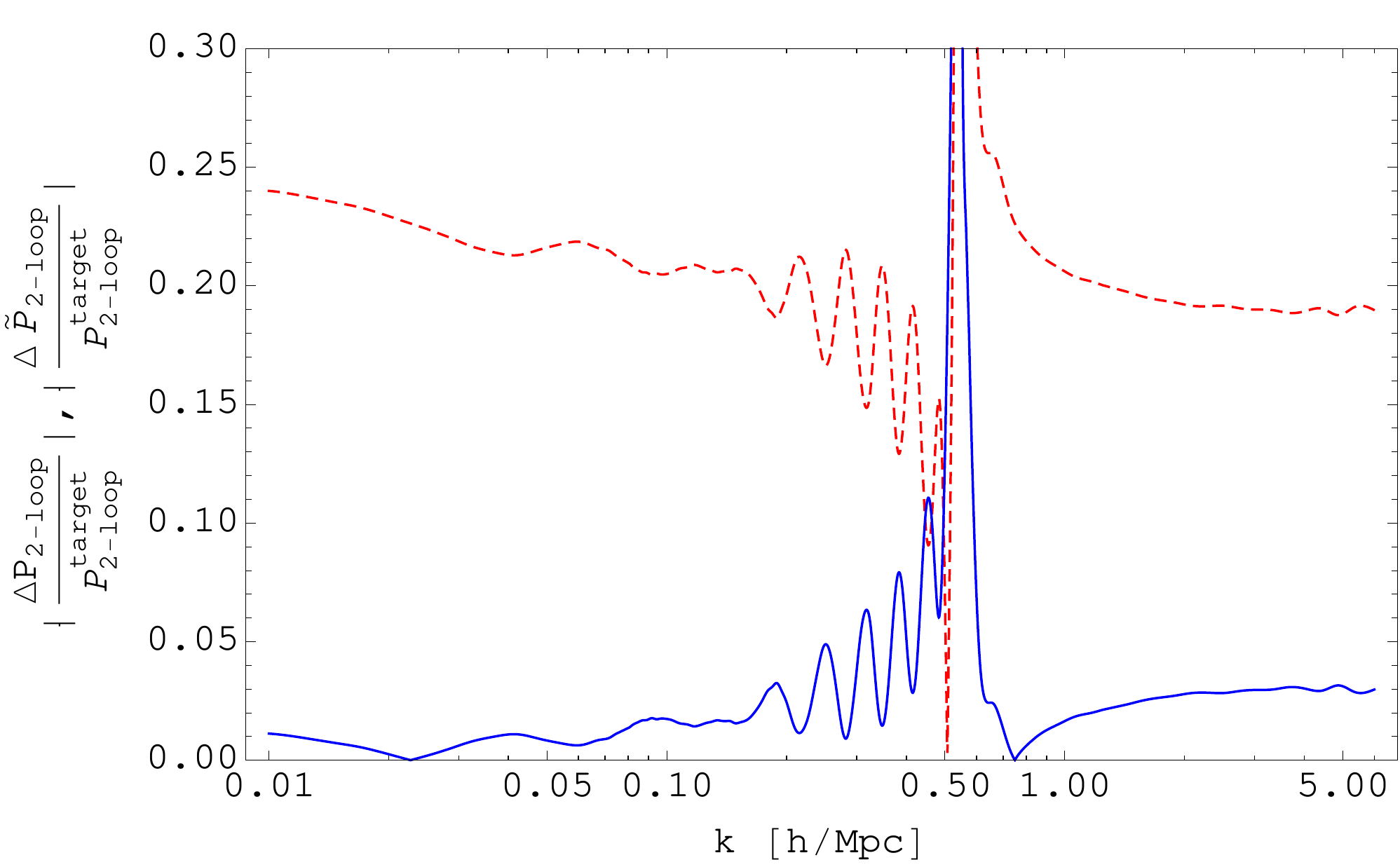}
    \caption{For the  \texttt{cosmo\_5} test cosmology, we show $\Delta \ptwoloop(k)/\ptwoloop^\text{target}$ (the red dashed line, evaluated using Eq.~\eqref{eq:diff_integration}) and $\Delta \tilde{P}_\text{2-loop}(k)/\ptwoloop^\text{target}$ (the blue solid line, evaluated using Eq.~\eqref{eq:adj_DeltaPloop}), where $\ptwoloop$ is the full calculation rather than the ``UV-improved'' version used in Sec.~\ref{sec:DiffExplain}. As with $\ptwoloop^\text{(UV-improved)}$, Eq.~\eqref{eq:adj_DeltaPloop} shows a large improvement over Eq.~\eqref{eq:diff_integration} in removing most of the difference associated with the cosmological parameter~$A_s$.
    }
\label{fig:DP2loop_nonUVsafe}
\end{figure}

As with $\ptwoloop^\text{(UV-improved)}$, we use Eq.~\eqref{eq:adj_DeltaPloop} instead of Eq.~\eqref{eq:diff_integration} when evaluating the difference between  $\ptwoloopfull$ for the target and reference cosmologies, and once again we find that the extra $A_s$ rescaling implemented in Eq.~\eqref{eq:adj_DeltaPloop} is effective in analytically removing much of the difference between the two calculations; this is shown in Fig.~\ref{fig:DP2loop_nonUVsafe}. This figure also shows that the ratio $\Delta \tilde{P}_\text{2-loop}(k)/\ptwoloop^\text{target}$ is better behaved than for the $\ptwoloop^\text{(UV-improved)}$, possessing fewer zero-crossings; this will lead to a closer agreement between the exact calculation of the ratio and the estimate described below.

The procedure for estimating $\Delta \tilde{P}_\text{2-loop}(k)/\ptwoloop^\text{target}$ is very similar to what we presented in Sec.~\ref{sec:DiffExplain}. We begin with the same expression used there, repeated below for convenience:
\begin{equation}
\label{eq:ratioestimate_appendix}
\left| \frac{\Delta \tilde{P}_{\alpha}}{P_{\alpha}^{\text{target}}} \right| \approx
	\left| 1 - \lp \frac{A_s^{\rm target}}{A_s^{\rm ref}} \rp^{L+1} 
	\lp \frac{P_{11}^{\rm ref}(k)}{P_{11}^{\rm target}(k)} \rp^{L+1} \right| .
\end{equation}
Recall that in the case of the UV-improved $\ptwoloop$, this estimate was motivated by the fact that the integrand of $\ptwoloop^\text{(UV-improved)}$ should be dominated by momenta of order $k$. In the non-UV-improved case, this is not necessarily true, as the UV contribution that is now included could cause the integral to be dominated by momenta much larger than $k$. Despite this concern, the estimate in Eq.~\eqref{eq:ratioestimate_appendix} turns out to predict the desired ratio at a level that is more than adequate for our purposes here.

\begin{figure*}[t]
\begin{center}
\includegraphics[width=0.48\columnwidth]{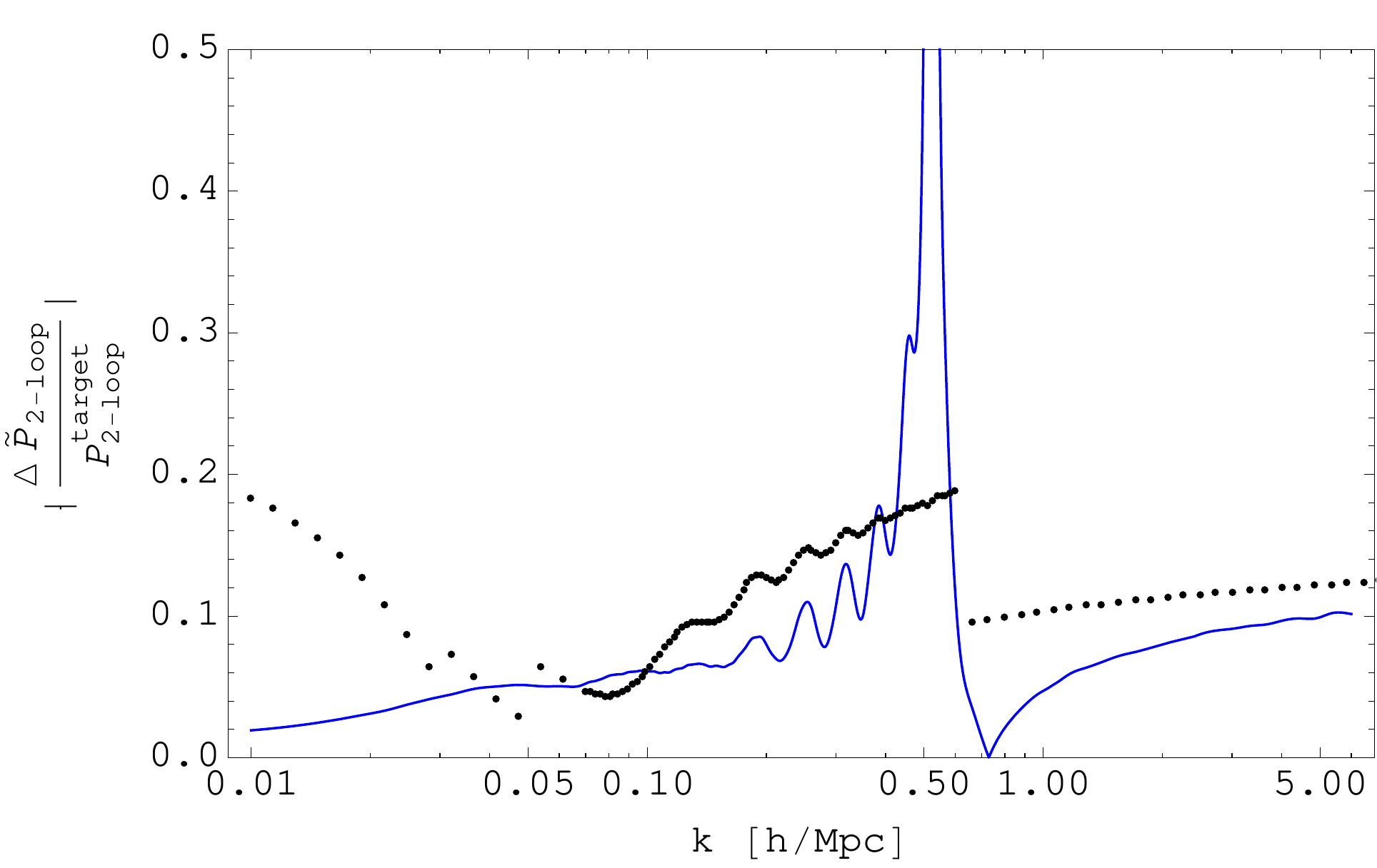}
\quad
\includegraphics[width=0.48\columnwidth]{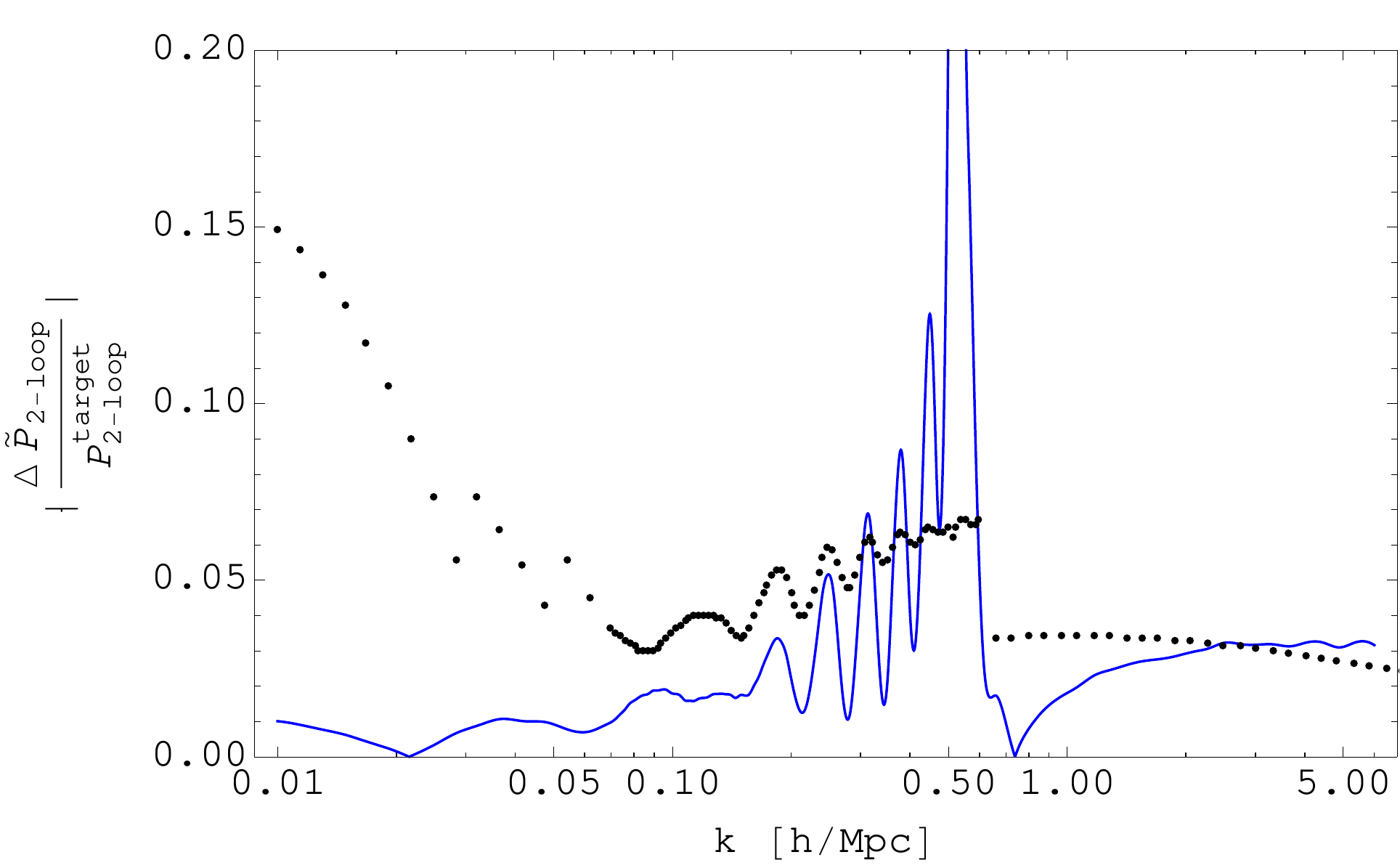}
\end{center}
\caption{Comparison of estimate (Eq.~\eqref{eq:ratioestimate}, with the modifications described in the main text; \textit{black points}) and exact calculation of $|\Delta \tilde{P}_\text{2-loop} / \ptwoloop^\text{target}|$ (\textit{blue lines}) for two test cosmologies, \texttt{cosmo\_1} (\textit{left}) and \texttt{cosmo\_5} (\textit{right}), given in Table~\ref{tab:Cosmo_list}. On average, the estimate slightly over-predicts the exact calculation, but this only means that the precision requested for the integration of $\Delta \tilde{P}_\text{2-loop}$ is slightly more conservative than necessary. Also, the estimate has the desirable feature of automatically limiting the precision requested close to the zero-crossing of $\ptwoloopfull(k)$, by setting a ceiling on the value of $|\Delta \tilde{P}_\text{2-loop} / \ptwoloop^\text{target}|$ in Eq.~\eqref{eq:ratioestimate}.
}
\label{fig:adjusted_ratio_comparison_appendix}
\end{figure*}

As before, this estimate on its own is slightly sub-optimal, so we implement the following extra steps:
\begin{enumerate}
\item We apply the smoothing given in Eq.~\eqref{eq:adj_ratio_smooth} over the range $0.03 \invMpc < k < 0.5\invMpc$.
\item Since Eq.~\eqref{eq:ratioestimate_appendix} occasionally under-predicts the exact ratio at low wavenumbers, we multiply the estimate by 2 for $k<0.6\invMpc$.
\end{enumerate}
In Fig~\ref{fig:adjusted_ratio_comparison_appendix}, we compare this estimate to the exact calculation of $|\Delta \tilde{P}_\text{2-loop} / \ptwoloop^\text{target}|$ in two test cases. The estimate is slightly higher than the exact calculation overall, again leading to slightly more conservative precision requirements on the integral evaluation than strictly necessary. The zero-crossing around $k\sim 0.5\invMpc$ is also nicely handled by the estimate, which again imposes a ceiling on the value of $|\Delta \tilde{P}_\text{2-loop} / \ptwoloop^\text{target}|$ in the relevant region.

To provide more detailed proof that the estimate described above is suitable for our purposes, in App.~\ref{app:altestimates} we provide a procedure to more accurately estimate $|\Delta \tilde{P}_\text{2-loop} / \ptwoloop^\text{target}|$ for the non-UV-improved case. The resulting runtimes of the {\codename} code are not significantly different whether we use these more detailed estimates or the estimates presented in this section. Overall, both estimates typically determine $\epsdelta$ to be $\sim$5-10 times larger than $\epstarget$, clearly a significant improvement over performing an exact calculation using $\epstarget$ at all wavenumbers.

\begin{figure}[t]
    \centering
        \includegraphics[width=0.6\columnwidth]{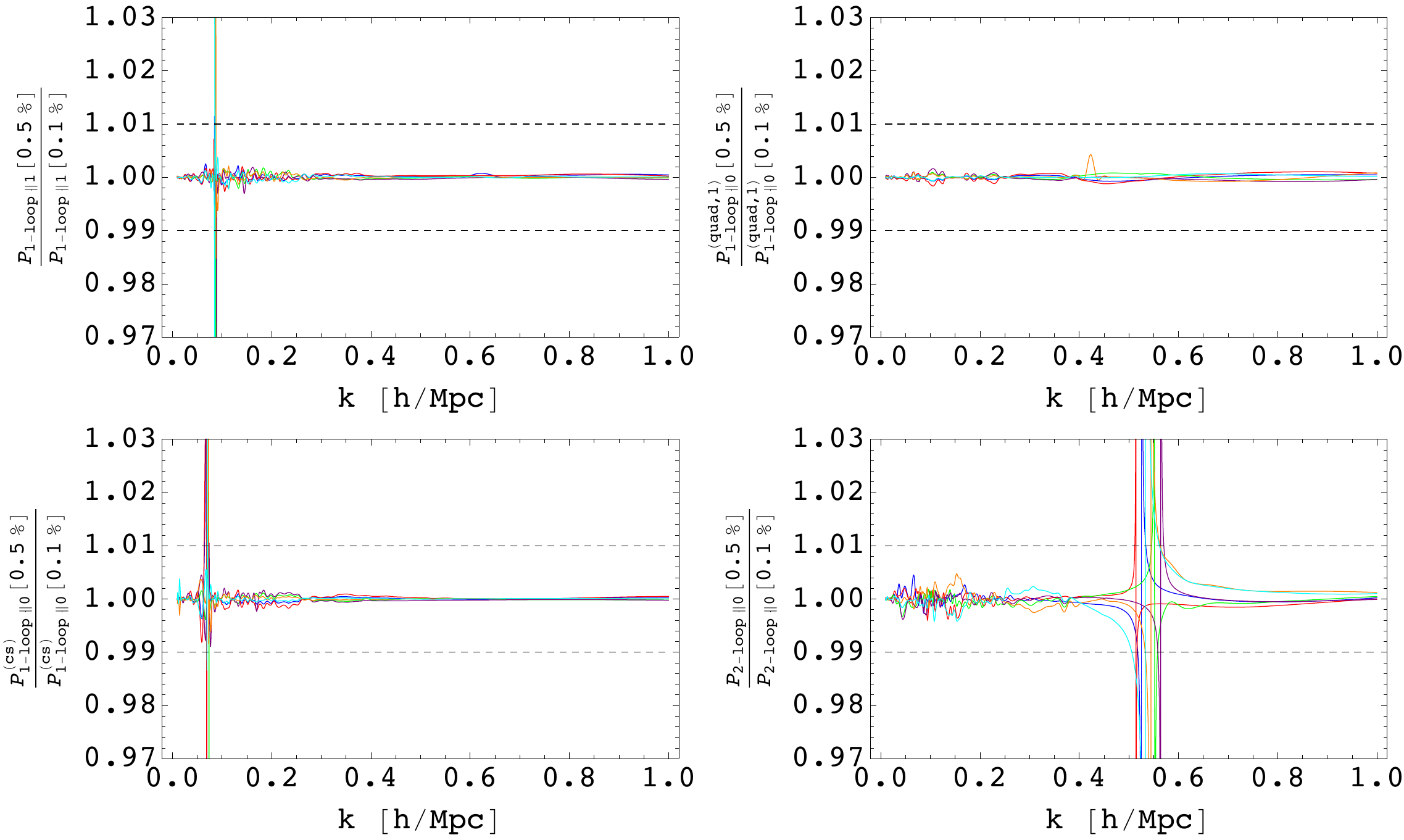}
    \caption{For all of the $3\sigma$-cosmologies from Table~\ref{tab:Cosmo_list}, we show {$\ptwoloopfull$} evaluated with $\epstarget$ determined using the procedure in App.~\ref{app:fullp2loop}, normalized to {$\ptwoloopfull$} computed using $\epstarget=0.1\%$. Aside from the region close to the zero crossing at $k\sim0.5\invMpc$, the faster computation is within $\sim$0.5\% of the more precise computation over the entire wavenumber range of interest.
    }
\label{fig:p2loop_vs_exact_appendix}
\end{figure}

Finally, in Fig.~\ref{fig:p2loop_vs_exact_appendix}, we compare the exact ($\epstarget=0.1\%$) $\ptwoloopfull$ computation with the computation where $\epstarget$ is set using Eq.~\eqref{eq:target_prec_modified} and the procedure we have just described. Aside from the region close to the zero crossing at $k\sim0.5\invMpc$, the faster computation is within $\sim$0.5\% of the more precise computation for at least $k<1\invMpc$. 

\subsection{Alternative estimates for determining integration precision for $\ptwoloop$}
\label{app:altestimates}

Recall that, in order to set the the integration precision $\epsdelta$ used by computations in {\codename}, we make use of Eq.~\eqref{eq:target_prec_modified}, repeated below for convenience:

\begin{equation}
\label{eq:appendixratio}
\epsdelta \approx \left| \frac{\Delta \tilde{P}_{\alpha}}{P_{\alpha}^{\text{target}}} \right|^{-1} \epstarget\ ,
\end{equation}

In the main text of the paper, we use Eq.~\eqref{eq:ratioestimate} to approximate $|\Delta\tilde{P}_\alpha/P_\alpha^\text{target}|$. In this appendix, we describe a more detailed estimate, and compare its performance to that of Eq.~\eqref{eq:ratioestimate}.

In~\cite{Foreman:2015uva}, it was found that $\ptwoloopfull$ is roughly described by the following:
\begin{equation}\label{eq:p2loop_est}
\ptwoloopfull(k)  \approx  
\alpha \frac{\poneloop^2(k)}{P_{11}(k)} + 2(2\pi) \ct \frac{k^2}{k_\text{NL}^2} P_{11}(k)\ .
\end{equation}
As stated in \cite{Foreman:2015uva}, for $\alpha=0.3$ the estimate is within a factor of 2 of exact calculation for $k \lesssim 0.7\invMpc$, with the accuracy increasing at lower $k$. We have verified that this value of $\alpha$ works well for all of our test cosmologies described below, which suggests $\alpha$ is only weakly dependent on cosmology and can be fixed for our purposes.

Note that, as extensively explained in \cite{Carrasco:2013mua,Foreman:2015uva}, $\ct$ can be completely determined from theory alone: it will take the form of a function of the free parameters of the EFTofLSS, but determination of this function does not rely on access nonlinear data. However, to make use of the estimate~(\ref{eq:p2loop_est}), we require prior knowledge of $\ct$ as a function of cosmology. Therefore, our strategy will be to pre-calibrate $\ct$ by fitting approximate predictions to the output of simulations, and then feed these pre-calibrated values back into the estimates used in the {\codename} algorithm. Here and throughout, we express $\ct$ in $(\knl/2\invMpc)^2$ units.

\begin{figure*}[t]
\begin{center}
\includegraphics[width=0.48\columnwidth]{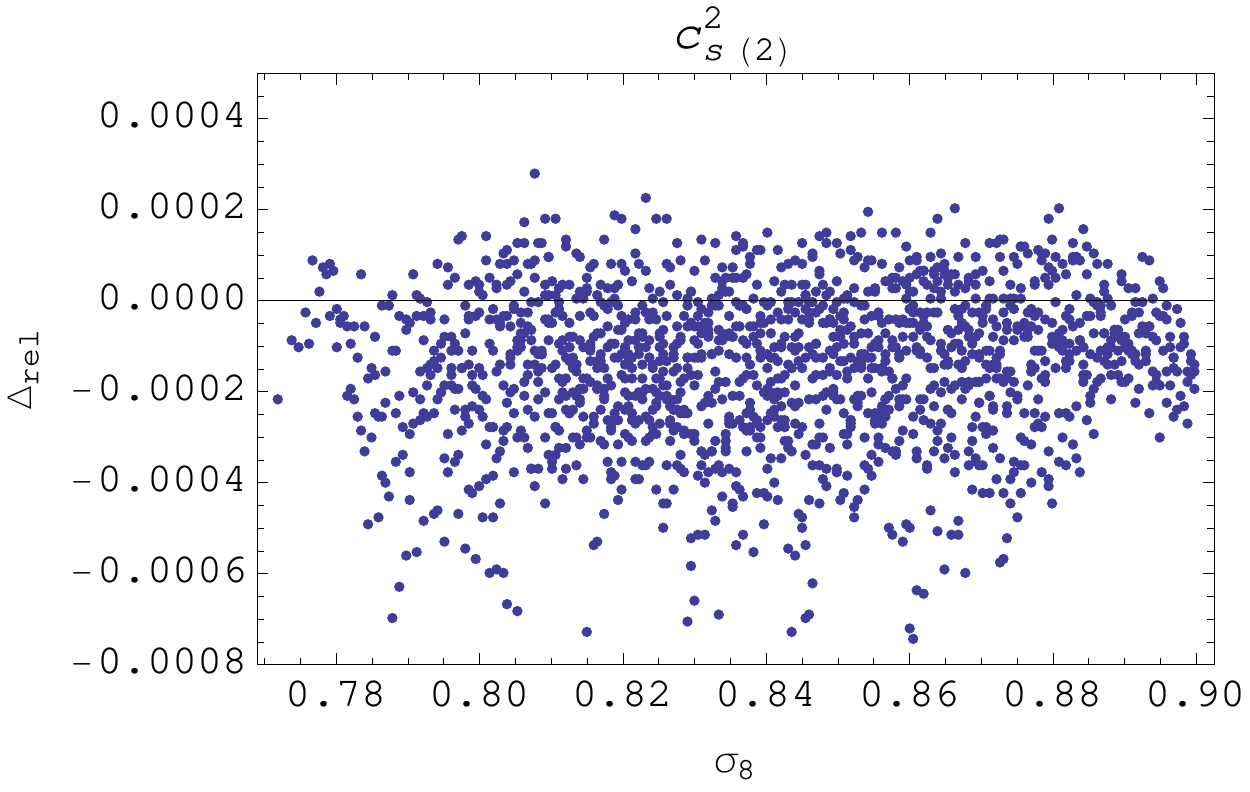}
\quad
\includegraphics[width=0.48\columnwidth]{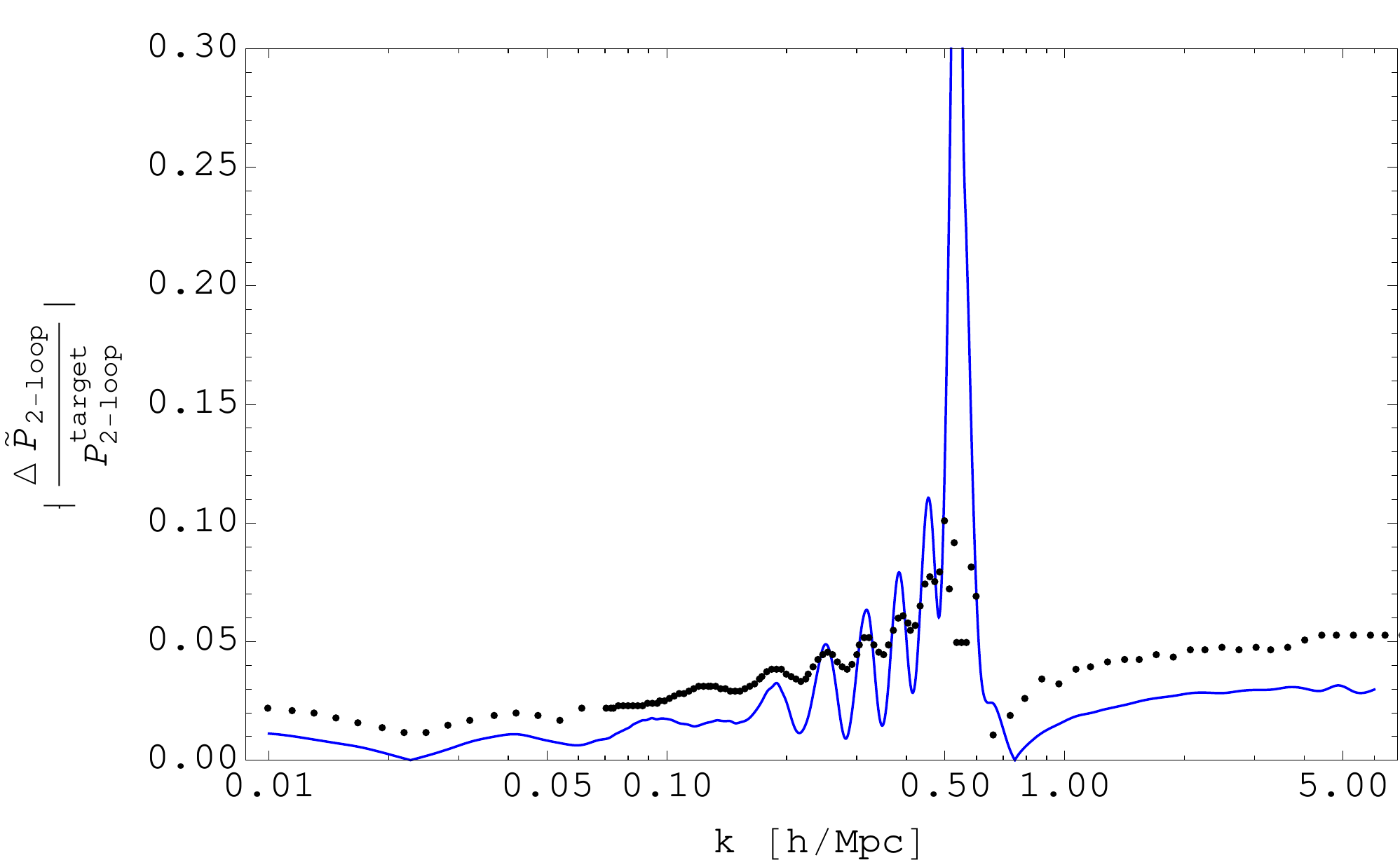}
\end{center}
\caption{\textit{Left:} Relative departure of the estimated $\ct$ values from the predictions of Eq.~\eqref{eq:cs2_pred} for a large sample of cosmologies in the $4\sigma$ 5-cube centered at our reference cosmology. In all cases, the agreement between the two is excellent (within 0.1\%). \textit{Right:} For the \texttt{cosmo\_5} test cosmology (given in Table~\ref{tab:Cosmo_list}), comparison between the exact adjusted relative ratio (blue) and its estimate (dots) based on Eq.~\eqref{eq:p2loop_est} together with the $\ct$ prediction of Eq.~\eqref{eq:cs2_pred}.}
\label{fig:DP2loop_cfr_cs2}
\end{figure*}

We model $\ct$ at $z=0$ with a Taylor expansion in the five cosmological parameters. Specifically, we use
\begin{align} \nonumber
\ct (\pmb\theta) &= \ct(\pmb\theta^{\rm ref}) 
	+ \sum_i \Delta\theta_i 
		\left. \frac{\d\ct(\pmb\theta)}{\d\theta_i} \right|_{\pmb\theta = \pmb\theta^{\rm ref}}
	+ \frac{1}{2} \sum_{i,j} \Delta\theta_i \Delta\theta_j 
		\left. \frac{\d^2 \ct(\pmb\theta)}{\d\theta_i \, \d\theta_j} \right|_{\pmb\theta = \pmb\theta^{\rm ref}}  \\
&\quad + \frac{1}{6} \sum_{i,j,k} \Delta\theta_i \Delta\theta_j \Delta\theta_k 
		\left. \frac{\d^3\ct(\pmb\theta)}{\d\theta_i \, \d\theta_j \, \d\theta_k} 
		\right|_{\pmb\theta = \pmb\theta^{\rm ref}}\ ,
\label{eq:cs2_pred}
\end{align}
where $\Delta\theta_i \equiv \theta_i - \theta_i^{\rm ref}$ and $i,j,k$ run over the 5 parameters in~\eqref{eq:refcosmo}.
We determine the derivatives in Eq.~\eqref{eq:cs2_pred} by fitting to a large sample of cosmologies whose parameters lie within a $4\sigma$ 5-cube centered at our reference cosmology. That is, for each of these cosmologies, we obtain the nonlinear power spectrum from the \texttt{FrankenEmu} emulator, calculate the corresponding two-loop EFT prediction using \texttt{TaylorEFT} (described in Sec.~\ref{sec:Taylor}), and extract the corresponding value of $\ct$ by fitting the prediction to the nonlinear spectrum. The values of the derivatives in Eq.~\eqref{eq:cs2_pred} are obtained by fitting this equation to the resulting~$\ct$ values as a function of $\pmb\theta$. The left panel of Fig.~\ref{fig:DP2loop_cfr_cs2} shows that, after fixing the values of the derivatives, the accuracy of Eq.~\eqref{eq:cs2_pred} is within 0.1\% for all these cosmologies.  

To compensate for fitting errors and expansion errors in \texttt{TaylorEFT}, we add a constant additive shift of~0.01 to the estimation for $\Delta \tilde{P}_{\alpha}/P_{\alpha}^\text{target}$ that follows from Eq.~\eqref{eq:p2loop_est}, clearly visible at low $k$ in the right panel of Fig.~\ref{fig:DP2loop_cfr_cs2}. Interestingly enough, Eq.~\eqref{eq:p2loop_est} can predict the zero-crossing $k_c$ of the adjusted ratio~\eqref{eq:appendixratio} fairly well; for all cosmologies of interest, the zero-crossing is in the range $\delta k_{c} \equiv (0.5,0.6) \invMpc$. We use this, together with our specific $k$ sampling, to place an upper limit on the absolute value of $\Delta \tilde{P}_{\alpha}/P_{\alpha}^{\text{target}}$ used by the code by imposing that at most six points within $\delta k_{c}$ can be above the corresponding threshold. For these wavenumbers, we use an arbitrary small value $r_{c} = 0.05$ for $\Delta \tilde{P}_\alpha/P_{\alpha}^\text{target}$ in Eq.~\eqref{eq:appendixratio}, effectively reducing integration times by increasing the required relative precision to $\epsdelta = 10\%$ wherever $\Delta \tilde{P}_{\text{2-loop}}$ is negligible compared to $(A_s^{\text{target}}/A_s^{\text{ref}})^3 \ptwoloop^{\text{ref}}$.
 
As with Eq.~\eqref{eq:ratioestimate},  Eq.~\eqref{eq:p2loop_est} systematically predicts BAO in antiphase with the exact calculations. Therefore, for $ 0.055 \invMpc < k < k_c $ we smooth the estimated adjusted ratios with the top-hat window function from Eq.~\eqref{eq:adj_ratio_smooth}. When Eq.~\eqref{eq:adj_ratio_smooth} returns a ratio larger than the upper limit, we replace it by its unsmoothed value. Finally, for $k > 4.4 \invMpc$, our estimation for $\Delta \tilde{P}_{\alpha}/P_{\alpha}^{\text{target}}$ always approaches a constant in the $k$-range we are interested in, hence for these wavenumbers we simply use its value at $k = 4.4 \invMpc$.

In Table~\ref{tab:Performance_detailed_estimates}, we compare the wall times for evaluation of each of our 11 test cosmologies using the simpler estimates from the main text or the estimates from this appendix. In most cases, we find comparable performance, while the few larger discrepancies are caused by because the simpler estimates are more conservative in some cases.

\ctable[
    caption = {Computational performance of {\codename} for our test cosmologies, when either the estimates from Sec.~\ref{sec:DiffExplain} or App.~\ref{app:detailedestimates} are used to set the integration precision $\epsdelta$. In most cases, the running times are almost identical, indicating that the estimates from Sec.~\ref{sec:DiffExplain} may be used without a significant loss of computational efficiency. },
    label   = {tab:Performance_detailed_estimates},
pos =h,
mincapwidth=\textwidth
]{| >{\centering\arraybackslash} m{2.5cm} | >{\centering\arraybackslash} m{2cm} | >{\centering\arraybackslash} m{2cm} | >{\centering\arraybackslash} m{2cm} | >{\centering\arraybackslash} m{2cm} |}{

}{                                                          				
\hline
Cosmology	&	Wall time, simple estimates	& Wall time, detailed estimates	\\ [10pt]
\hline 
\texttt{cosmo\_1} 	&	5.2 min	& 5.3 min	\\
\texttt{cosmo\_2}	&	10.9 min	& 10.8 min	\\
\texttt{cosmo\_3}	&	3.3 min	& 3.3 min	\\
\texttt{cosmo\_4}	&	11 min	& 7.3 min	\\
\texttt{cosmo\_5}	&	1.4 min	& 1.3 min	\\
\texttt{cosmo\_6}	&	2.3 min	& 1.4 min	\\
\texttt{cosmo\_7}	&	2.0 min	& 1.6 min	\\
\texttt{cosmo\_8}	&	1.7 min	& 1.8 min	\\
\texttt{cosmo\_9}	&	14 min	& 9.0 min	\\
\texttt{cosmo\_10}	&	9 min	& 6 min	\\
\texttt{cosmo\_11}	&	15 min	& 11 min	\\
\hline
}


\section{{\codename}: additional checks}\label{app:morechecks}

\begin{figure}[t]
\includegraphics[width=0.48\columnwidth]{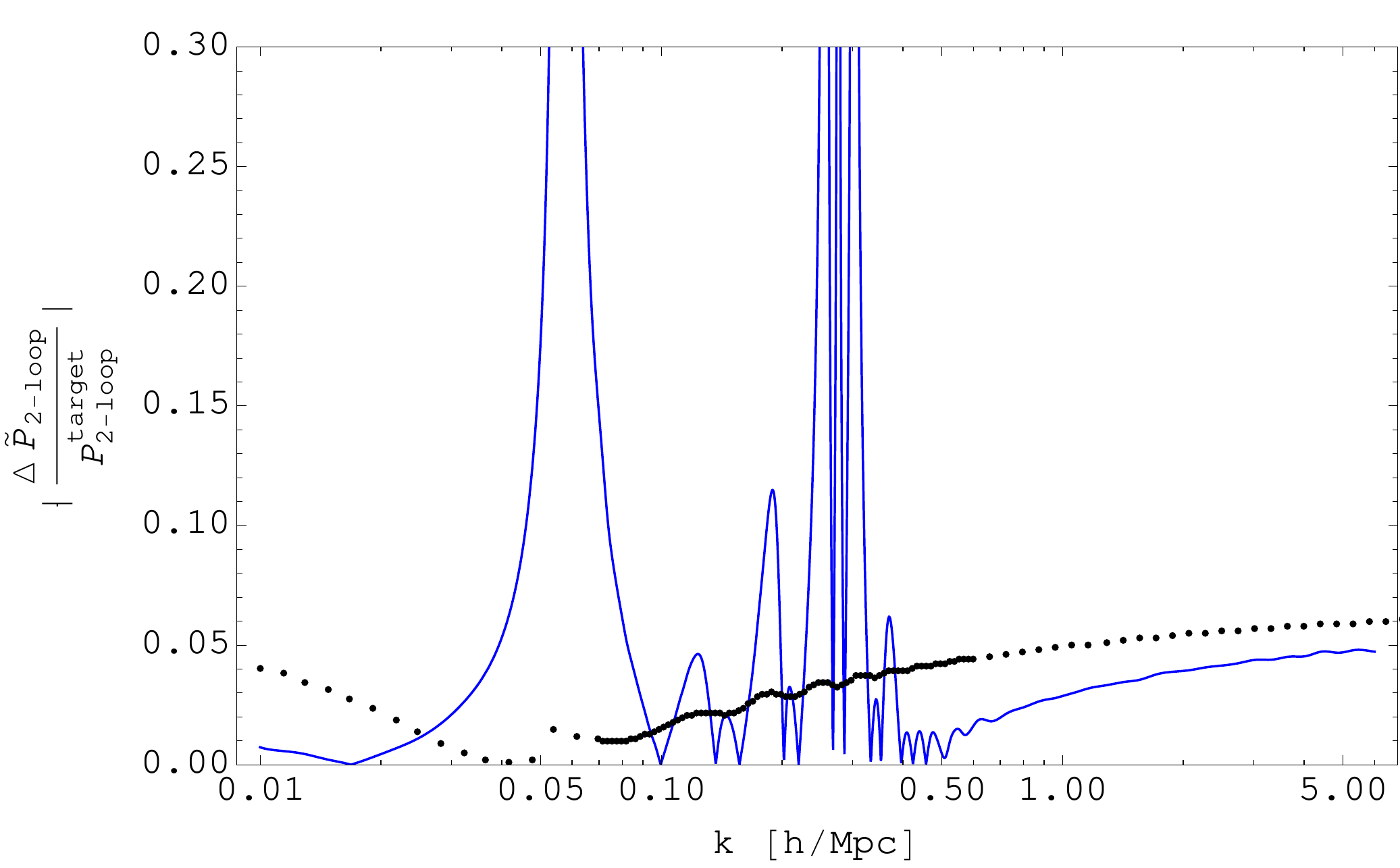}
\quad
\includegraphics[width=0.48\columnwidth]{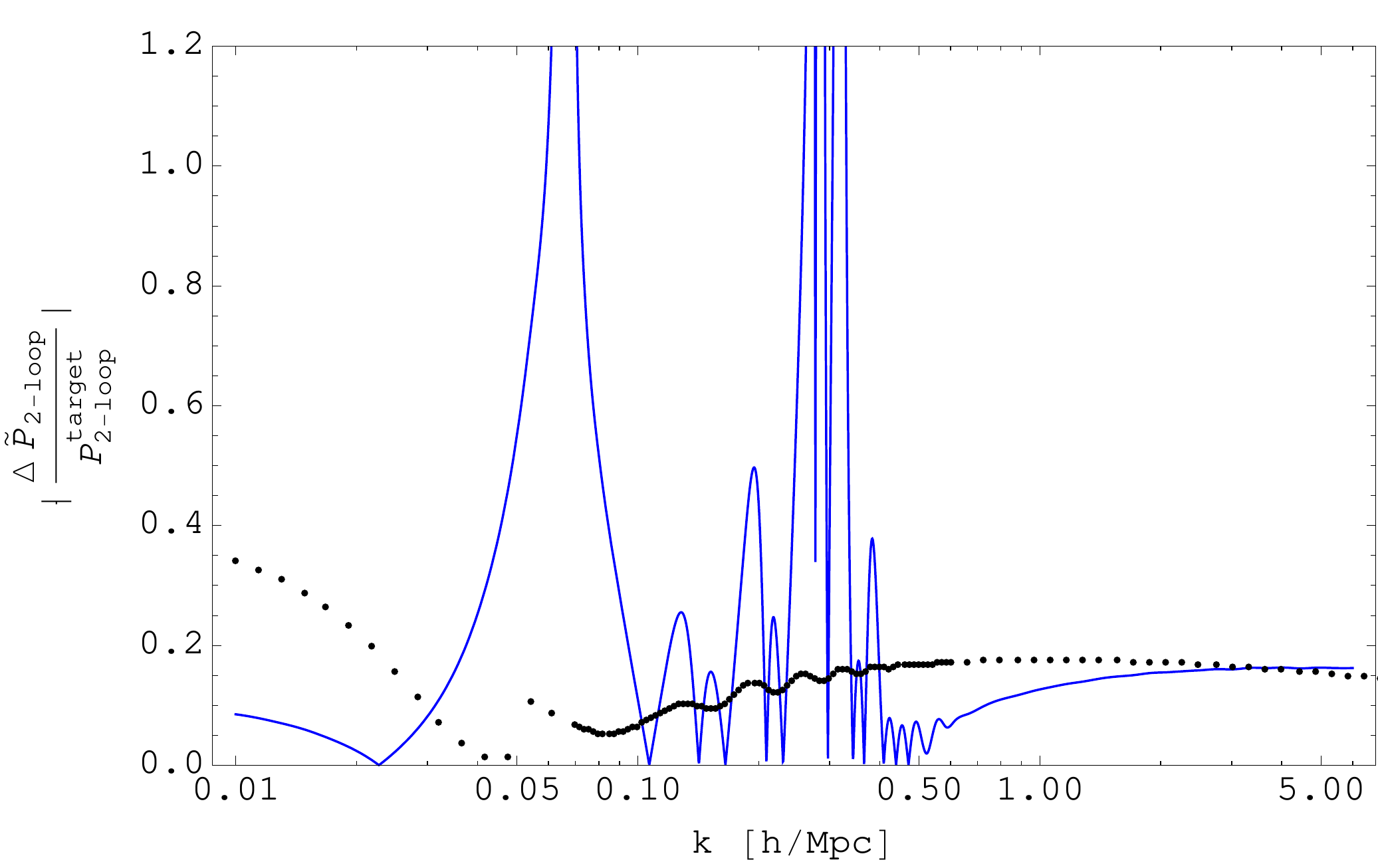}
\caption{{Comparison between the exact adjusted relative ratio (blue) and its estimate (dots) based on Eq. \eqref{eq:ratioestimate} for \texttt{cosmo\_8} (left) and \texttt{cosmo\_9} (right).}}
\label{fig:DP2loop_ratios_cosmo8_cosmo9}
\end{figure}

\begin{figure}[t]
\begin{center}
\includegraphics[width=\columnwidth]{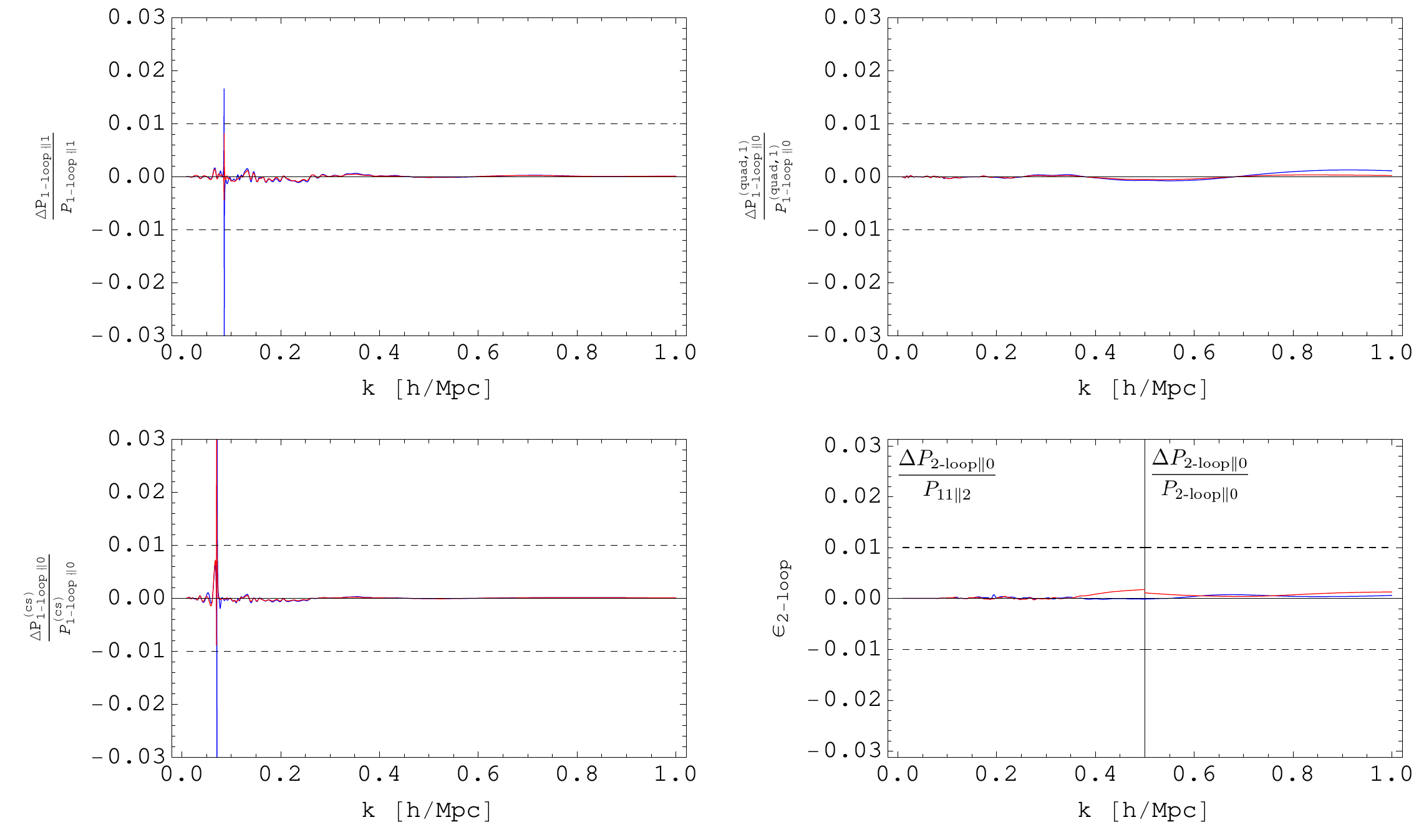}
\caption{Resummed {\codename} outputs for our $4\sigma$-cosmologies \texttt{cosmo\_7-8} (blue and red, respectively) relative to the direct calculation of the full integrand with precision $\epsilon=0.1\%$. We used $\epsilon_{\text{target}}=0.5\%$, and spikes indicate zero-crossing. Subscripts $\|0,\|1$ denotes the resummation order as in~\cite{Senatore:2014via}. Dashed lines mark 1\% departures from direct calculations. {As in Fig.~\ref{fig:fast_full_ratios_3sigma}, for $\ptwoloop$ we show $\Delta P_{\text{2-loop} \| 0}/P_{11 \| 2}$ for $k < 0.5 \invMpc$, which propagates directly to the matter power spectrum predictions, and $\Delta P_{\text{2-loop} \| 0}/P_{\text{2-loop} \| 0}$ for $k > 0.5 \invMpc$, which proves the goodness of our estimates for smaller scales.}}
\label{fig:fast_full_ratios_4sigma}
\end{center}
\end{figure}

\begin{figure}[t]
\begin{center}
\includegraphics[width=\columnwidth]{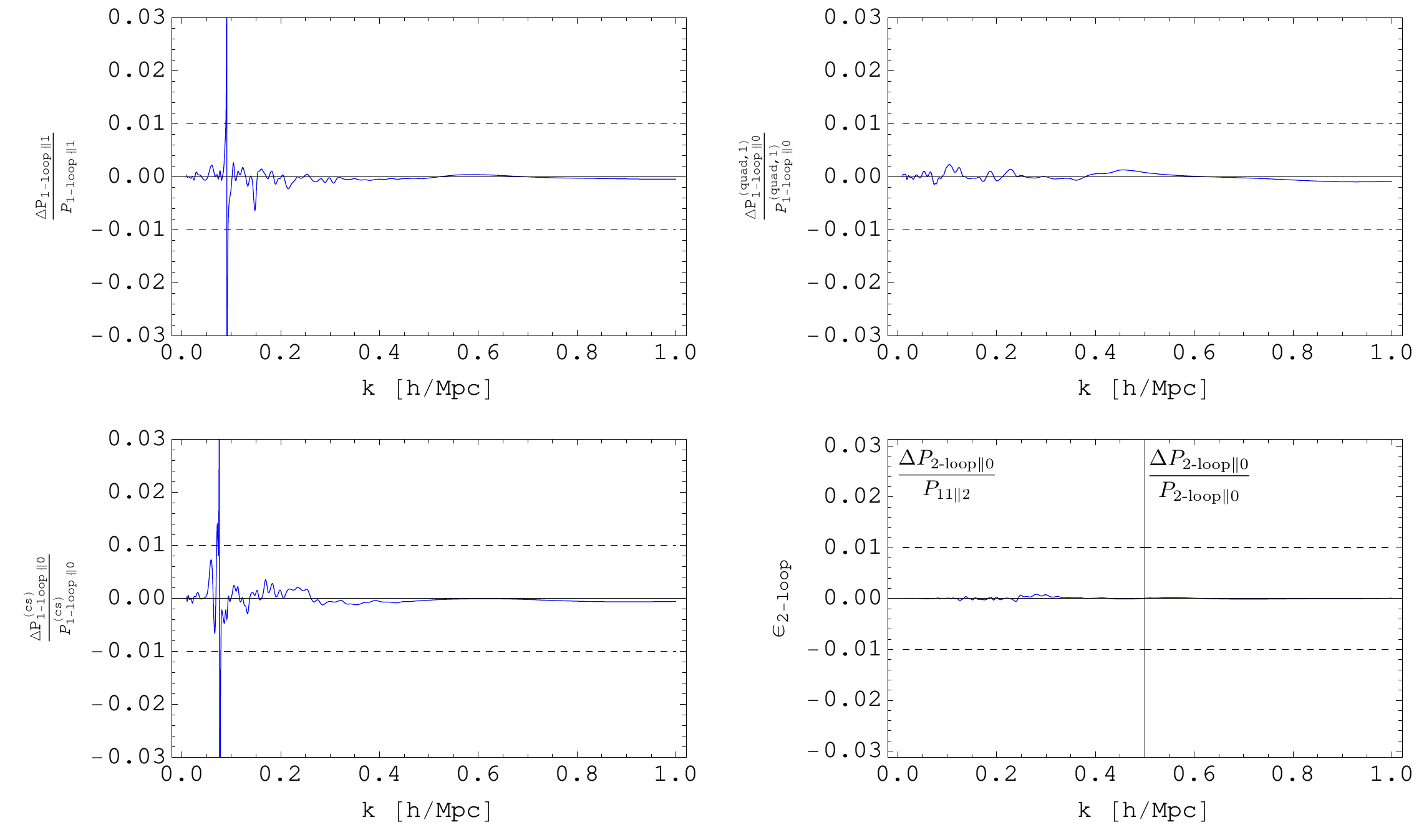}
\caption{Resummed {\codename} outputs for our $5\sigma$-cosmology \texttt{cosmo\_9} relative to the direct calculation of the full integrand with precision $\epsilon=0.1\%$. We used $\epsilon_{\text{target}}=0.5\%$, and spikes indicate zero-crossing. Subscripts $\|0,\|1$ denotes the resummation order as in~\cite{Senatore:2014via}. Dashed lines mark 1\% departures from direct calculations. {As in Fig.~\ref{fig:fast_full_ratios_3sigma}, for $\ptwoloop$ we show $\Delta P_{\text{2-loop} \| 0}/P_{11 \| 2}$ for $k < 0.5 \invMpc$, which propagates directly to the matter power spectrum predictions, and $\Delta P_{\text{2-loop} \| 0}/P_{\text{2-loop} \| 0}$ for $k > 0.5 \invMpc$, which verifies the performance of our estimates for smaller scales.}}
\label{fig:fast_full_ratios_5sigma}
\end{center}
\end{figure}

In this appendix, we repeat the tests from Sec.~\ref{sec:DiffChecks} for the $4\sigma$- and $5\sigma$-cosmologies from Table.~\ref{tab:Cosmo_list}. Here, we also use $\epstarget = 0.5\%$, and first check that the estimates Eq.~\eqref{eq:ratioestimate} are a good approximation of the two-loop modified ratios in Fig.~\ref{fig:DP2loop_ratios_cosmo8_cosmo9}. In Figs.~\ref{fig:fast_full_ratios_4sigma}-\ref{fig:fast_full_ratios_5sigma} we then compare the resummed {\codename} calculations with the equivalent outputs obtained through direct integration, verifying that {\codename} achieves the expected precision.

\vfill
\bibliographystyle{JHEP}
\bibliography{references}

\end{document}